\documentclass[12pt]{iopart}
\usepackage{amsfonts}
\usepackage{amssymb}
\expandafter\let\csname equation*\endcsname=\relax
\expandafter\let\csname endequation*\endcsname=\relax
\usepackage{amsmath}
\usepackage{graphicx}

\begin{document}
\title[Spectral density of the Ising model: Gaussian and multi-Gaussian approximations]{Spectral density of the quantum Ising model in two fields: Gaussian and multi-Gaussian approximations}
\author{Y.Y. Atas, E. Bogomolny}
\address{Univ. Paris Sud, CNRS, LPTMS, UMR8626, Orsay F-91405, France}
\eads{\mailto{yasar.atas@lptms.u-psud.fr}, \mailto{eugene.bogomolny@lptms.u-psud.fr}}
\begin{abstract}
Spectral density of quantum Ising model in two fields for large but finite number of spins $N$, is discussed in detail. When all coupling constants are of the same order, spectral densities in the bulk are well approximated by a Gaussian function which is typical behaviour for many-body models with short-range interactions. The main part of the paper is devoted to the investigation of a different characteristic case when spectral densities have peaks related with strong degeneracies of unperturbed states in certain limits of coupling constants. In the strict limit $N\to\infty$  peaks overlap and disappear but for values of $N$ accessible in numerical calculations  they often strongly influence spectral densities and other quantities as well.  A simple method is developed which permits to find general approximation formulae for multi-peak structure of spectral density in good agreement with numerics.       
\end{abstract}

\section{Introduction}

Spectral density is a fundamental and one of the simplest characteristics of quantum systems. In general, spectral density is system dependent. Nevertheless, there exist a few 'universal' densities which appear in numerous different models in the limit of large number degrees of freedom. 

Probably the best known example of such universality is the Wigner semicircle law \cite{wigner_1}-\cite{mehta} 
\begin{equation} 
\rho_W(E;N)=\left \{\begin{array}{cc}\dfrac{1}{\pi N \sigma^2 }\sqrt{2N\sigma^2 -E^2},&|E|<\sigma \sqrt{2N} \\0,&|E|>\sigma \sqrt{2N}\end{array}\right . \
\label{semi_circle}
\end{equation}  
Here $N$ is the matrix dimension and $\sigma$ is a constant. 

The semicircle law is well investigated in mathematics. Initially it appears as limiting spectral density of random matrices with independent elements \cite{wigner_1} but later it was proved that it describes densities  of a large variety of random matrix ensembles (e.g. in free probability \cite{free}).  In physics its importance is limited as there is  practically no physical models with such density. 

Instead, there exist models whose spectral density is well described by a different universal law, namely by the Gaussian function
\begin{equation}
\rho_{G}(E;N)=\frac{1}{\sqrt{2\pi N\sigma^2}}\exp \left (-\frac{E^2}{2N\sigma^2}\right )\ .
\label{Gaussian}
\end{equation}
A typical and the most investigated example is the shell model with $N$ fermions with two-body interaction occupying $M$ one-particle states with $M\gg N$  \cite{french_1}-\cite{bohigas_2} but one can argue that the Gaussian density is universal behaviour (under certain conditions) for any $N$-body model with short-range interaction.

We stress that the both formulae \eqref{semi_circle} and \eqref{Gaussian} (with corresponding rescaling)  are valid only in the bulk, namely when energy is scaled as follows 
\begin{equation}
E=\sqrt{N}e
\label{bulk}
\end{equation}  
with fixed $e$ and $N\to\infty$. 

For larger energies (e.g. $E\sim N\, e$) and, in particular close to the ground state, the spectral densities have a different forms, usually  exponential  or  stretched exponential. For example, for standard Gaussian ensembles of random matrices at the left of zero point of the semicircle law, $E_0=-\sqrt{2N}\sigma$, the density for $E<E_0$ is  given by $\rho(E)\sim N^{1/6}\exp \left [-\mathrm{const.}\, N^{1/6}(E_0-E)\right ]$ \cite{tail}. For the shell model with equally spaced one-particle states, the density close to the ground state $E_{gs}$ increases as follows: $\rho(E)\sim \exp \left [\mathrm{const.}\, \sqrt{N(E-E_{gs})}\right ]$ \cite{bethe_1}, \cite{bethe_2}.      

Though there exist many physical models with the Gaussian density \eqref{Gaussian}, this subject is not well investigated due to the fact that in solid state physics the number of particles is so large that the behaviour of the density in the bulk very far from the ground state is meaningless. 

Recently the situation has changed. First, modern experiments can be performed with a small number of particles. Second, progress in numerical calculations permits to obtain the full solution  of quantum  models with a few tenths of particles (but not much more). Consequently, the investigation of such models at large but finite values of $N$ becomes of interest.

The purpose of this paper is to discuss in detail spectral densities of one-dimensional spin chains. As a typical example we consider the quantum spin-$\tfrac{1}{2}$ Ising model in transverse and longitudinal fields whose Hamiltonian is  
\begin{equation}
\mathcal{H}=-\sum_{n=1}^N \sigma_n^{x}\sigma_{n+1}^x-\lambda  \sum_{n=1}^N \sigma_n^z-\alpha \sum_{n=1}^N\sigma_n^x \ ,
\label{ising_2}
\end{equation}
where $\sigma^{x,y,z}$ are usual Pauli matrices. Parameter $\lambda$ determines the strength of the transverse field and parameter $\alpha$ fixes the longitudinal field.  We refer to   $\lambda$ and $\alpha$ as coupling constants. For simplicity we assume periodic boundary conditions. 

The purpose of the paper is twofold. First we confirm that in the thermodynamic  limit $N\to\infty$ the spectral densities of the Ising model when all coupling constants are of the order of unity  attain the Gaussian form \eqref{Gaussian} even at relatively small number of spins $N$. Second, we discuss  different types of corrections at large but finite $N$.
The simplest is power series corrections to \eqref{Gaussian} in close analogy with  usual corrections to the central limit theorem. The second kind of corrections is more drastic as it manifests as pronounced peaks in the spectral density. These peaks are related with strong degeneracies of eigenvalues in certain limits of coupling constants. Though  this phenomenon leads to spectral densities quite different from the smooth Gaussian shape  prediction~\eqref{Gaussian},  it seems that it has  not been discussed in detail in the literature.  The investigation of such corrections constitutes the main part of the paper. We develop a simple method which permits to obtain approximate formulae describing well  multi-peaks spectral densities in  good agreement with numerical calculations for different values of coupling constants. Limiting densities are represented as a sum of different Gaussian functions (multi-Gaussians) whose parameters are calculated analytically from the Hamiltonian without the full solution of the problem.   

The plan of the paper is the following. In Section~\ref{integrable} the Ising model in transverse field (with $\alpha=0$) is considered. As its spectrum is known, all calculations are straightforward and can be performed analytically.  In Section~\ref{two_fields}  the non-integrable case of the Ising model in two fields is discussed. 
Similar to the  integrable case we demonstrate the existence of Gaussian and multi-Gaussian regimes for this model.   The calculations are slightly different for large values of coupling constants and for small values of $\lambda$.  The summary of results is given in Section~\ref{conclusion}. Details of calculations are presented in Appendices.           
   

\section{Ising model in transverse field}\label{integrable}

The quantum Ising model in transverse field  \cite{pfeuty} is one of the most investigated one-dimensional spin-$\tfrac{1}{2}$ models. In particular, it serves as the standard model of quantum phase transition \cite{sachdev}.

This model  is determined by the Hamiltonian \eqref{ising_2} with $\alpha=0$
\begin{equation}
\mathcal{H}=-\sum_{n=1}^{N}\sigma_n^{x}\sigma_{n+1}^x-\lambda \sum_{n=1}^{N}\sigma_n^z\ .
\label{H_Ising}
\end{equation}
The quantum  Ising model in transverse field  is integrable and its spectrum are calculated explicitly by the Jordan-Wigner transformation \cite{lieb}, \cite{mattis}. The result corresponds to the fermionic filling of one-particle levels 
\begin{equation}
E_{\vec{n}}=\sum_{j=1}^N  e_j\left (n_j-\tfrac{1}{2}\right )\ ,
\label{energy_exact}
\end{equation} 
where $n_j=0,1$ are number of fermions in state $j$, $e_j=e(\phi_j)$ with
\begin{equation}
e(\phi)=2\sqrt{1-2\lambda \cos \phi +\lambda^2}\ ,
\label{one_particle_levels}
\end{equation}
and phase $\phi_j$ equals $\pi k_j/N$  with integer momenta $k_j$.

More precisely,   for $|\lambda|<1$  the spectrum consists of two sub-spectra with only even number of excitations, $\sum_{j=1}^N n_j=$ even integer, but with the both even  and odd momenta, $k_j=2m_j$ and $k_j=2m_j+1$ for $m_j=0,\ldots, N-1$. For   $|\lambda|>1$ any numbers of excitations are possible but if $\sum_{j=1}^N n_j=$ even integer one has to choose odd $k_j=2m_j+1$ and when the total number of excitations is odd, i.e $\sum_{j=1}^N n_j=$ odd integer, $k_j$ is even integer, $k_j=2m_j$. For $|\lambda|=1$ there exists one zero mode and the both expressions give the same result. For completeness,  in \ref{ap_0} the derivation of this result is briefly reminded. 
 

\subsection*{Exact asymptotic spectral density}

The existence of the exact spectrum rends  the calculation of the spectral density straightforward (see e.g. \cite{lieb}, \cite{mattis}).  Consider first the case $|\lambda|>1$ and let us calculate the canonical sum of energies \eqref{energy_exact}
\begin{equation}
Z(\beta)=\frac{1}{2^N}\sum_{\vec{n}}\mathrm{e}^{-\beta \,E_{\vec{n} }}= \prod_j \cosh \left ( \tfrac{1}{2} \beta e_j \right )=\exp \left [ \sum_j \ln  \cosh \left ( \tfrac{1}{2} \beta e_j \right )\right ]\ .
\end{equation}
For large $N$ the sum in the exponent can be substituted by the integral over phases $\phi$
\begin{equation}
Z(\beta)\approx  \exp \left [\frac{N}{2\pi} \int_0^{2\pi}\mathrm{d}\phi \, \ln  \cosh \left (\beta \sqrt{1-2\lambda \cos \phi +\lambda^2}\right )  \right ]\ .
\end{equation}
The knowledge of the canonical sum permits to calculate the spectral density using the Laplace transform
\begin{equation}
\rho(E)\equiv \sum_{\vec{n}}\delta(E-E_{\vec{n}})=\frac{1}{2\pi\mathrm{i}}\int_{c-\mathrm{i}\infty}^{c+\mathrm{i}\infty}\mathrm{e}^{\beta\, E}Z(\beta)\mathrm{d}\beta 
\underset{N\to \infty}{\longrightarrow} \frac{1}{2\pi\mathrm{i}}\int_{c-\mathrm{i}\infty}^{c+\mathrm{i}\infty}\mathrm{e}^{NS(\beta\, e)}\mathrm{d}\beta\ ,
\end{equation}
where  $e=E/N$ is the energy per spin and the entropy 
\begin{equation}
S(\beta ,e)=e \beta +\frac{1}{2\pi} \int_0^{2\pi}\mathrm{d}\phi \, \ln  \cosh \left (\beta \sqrt{1-2\lambda \cos \phi +\lambda^2}\right ) \ .
\end{equation}
When $N\to\infty$ it is legitimate to perform the integration over $\beta$ by  the saddle point method. The saddle point, $\beta_{\mathrm{sp}}$, is determined from the equation $\partial S(\beta,e)/\partial \beta|_{\beta=\beta_{\mathrm{sp}}}=0$, or
\begin{equation}
e=-\frac{1}{2\pi}\int_0^{2\pi}\tanh \left (\beta_{\mathrm{sp}} \sqrt{1-2\lambda \cos \phi +\lambda^2} \right ) \sqrt{1-2\lambda \cos \phi +\lambda^2}\,  \mathrm{d}\phi\ ,
\label{saddle_point_ising}
\end{equation} 
and the spectral density in this approximation takes the form
\begin{equation}
\rho(e)=A\mathrm{e}^{NS(e)}\ .
\label{exact_density_ising}
\end{equation}
Here $S(e)$ plays the role of the entropy per spin and is given by following expression
\begin{equation}
S(e)=e \beta_{\mathrm{sp}}+\frac{1}{2\pi} \int_0^{2\pi}\mathrm{d}\phi \, \ln  \cosh \left (\beta_{\mathrm{sp}} \sqrt{1-2\lambda \cos \phi +\lambda^2}\right ) \ .
\label{entropy_ising} 
\end{equation} 
The pre-factor $A$ is determined by  the second derivative of $S(\beta,e)$ calculated at the saddle point, 
\begin{equation}
\frac{N}{A^2}=\int_0^{2\pi} \frac{(1-2\lambda \cos \phi +\lambda^2)\, \mathrm{d}\phi }{\cosh^2 \left ( \beta_{\mathrm{sp}} \sqrt{1-2\lambda \cos \phi +\lambda^2}\right ) } \ .
\label{pre_factor_ising}
\end{equation}
Of course, the saddle point method is equivalent to the standard thermodynamic relations between the free energy, energy, and entropy. 
In calculations it is convenient to plot $\rho$ as function of $\beta_{\mathrm{sp}}$ versus $e(\beta_{\mathrm{sp}})$ from \eqref{saddle_point_ising}.

For $|\lambda|<1$ one has to take into account the restriction that the total number of excitations over the two vacua with odd and even momenta should be even. It can be done e.g. by calculating first a slightly modified canonical sum depending on parameter $x$ controlling the total number of excitations,
\begin{equation}
Z(\beta,x)=\frac{1}{2^N}\sum_{\vec{n}}\mathrm{e}^{-\beta \,E_{\vec{n} }}\ x^{\, \sum_j n_j}= \prod_j \left [ \frac{1+x}{2}\cosh(\beta e_j/2)+\frac{1-x}{2}\sinh(\beta e_j/2) \right ]\ .
\label{generation_function}
\end{equation}
The canonical sum of energies with even number of excitations is simply
\begin{eqnarray}
Z_{+}(\beta)&=&\tfrac{1}{2}(Z(\beta,1)+Z(\beta,-1))\\
&=&\frac{1}{2} \left \{ 
\exp \left [ \sum_j \ln  \cosh \left ( \tfrac{1}{2} \beta e_j \right )\right ]+\exp \left [ \sum_j \ln  \sinh \left ( \tfrac{1}{2} \beta e_j \right ) \right ]
\right \} \ .
\nonumber
\end{eqnarray}
For $|\lambda|<1$ there is two sub-spectra with different parity of momenta. In limit $N\to\infty$ in the both cases one can change the summation over momenta to the integration over $\phi$ from $0$ to $2\pi$. Therefore  one gets exactly as above 
\begin{eqnarray}
Z(\beta)&\approx &  \exp \left [\frac{N}{2\pi} \int_0^{2\pi}\mathrm{d}\phi \, \ln  \cosh \left (\beta \sqrt{1-2\lambda \cos \phi +\lambda^2}\right )  \right ]
\nonumber\\
&+& 
\exp \left [\frac{N}{2\pi} \int_0^{2\pi}\mathrm{d}\phi \, \ln  \sinh \left (\beta \sqrt{1-2\lambda \cos \phi +\lambda^2}\right )  \right ]\ .
\end{eqnarray}
At finite $y$,  $\cosh y>\sinh y$,  the second term in the bulk is negligible at large $N$ and for all $\lambda$ in the bulk the spectral density is given by \eqref{exact_density_ising}-\eqref{pre_factor_ising} with $\beta_{\mathrm{sp}}$ determined from \eqref{saddle_point_ising}. 


\subsection*{Gaussian approximation for the density}

\begin{figure}
\begin{center}
\includegraphics[width=.3\linewidth, angle=-90, clip]{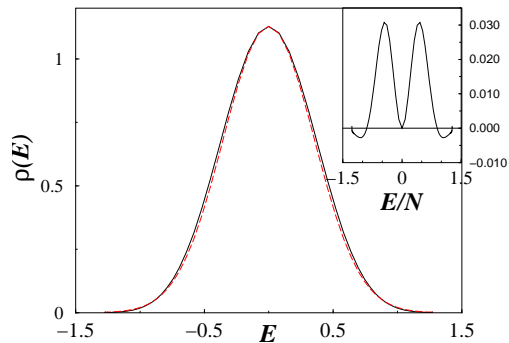}
\end{center}
\caption{Density of states for the Ising model in transverse field with $\lambda=1$ and $N=16$. Solid black line is the saddle point approximation \eqref{exact_density_ising}. Dashed red line indicates the Gaussian approximation \eqref{Ising_Gaussian}. Insert: the difference between them. }
\label{difference}
\end{figure}

\begin{figure}
\begin{minipage}{.33\linewidth}
\begin{center}
\includegraphics[width=.85\linewidth, angle=-90,clip]{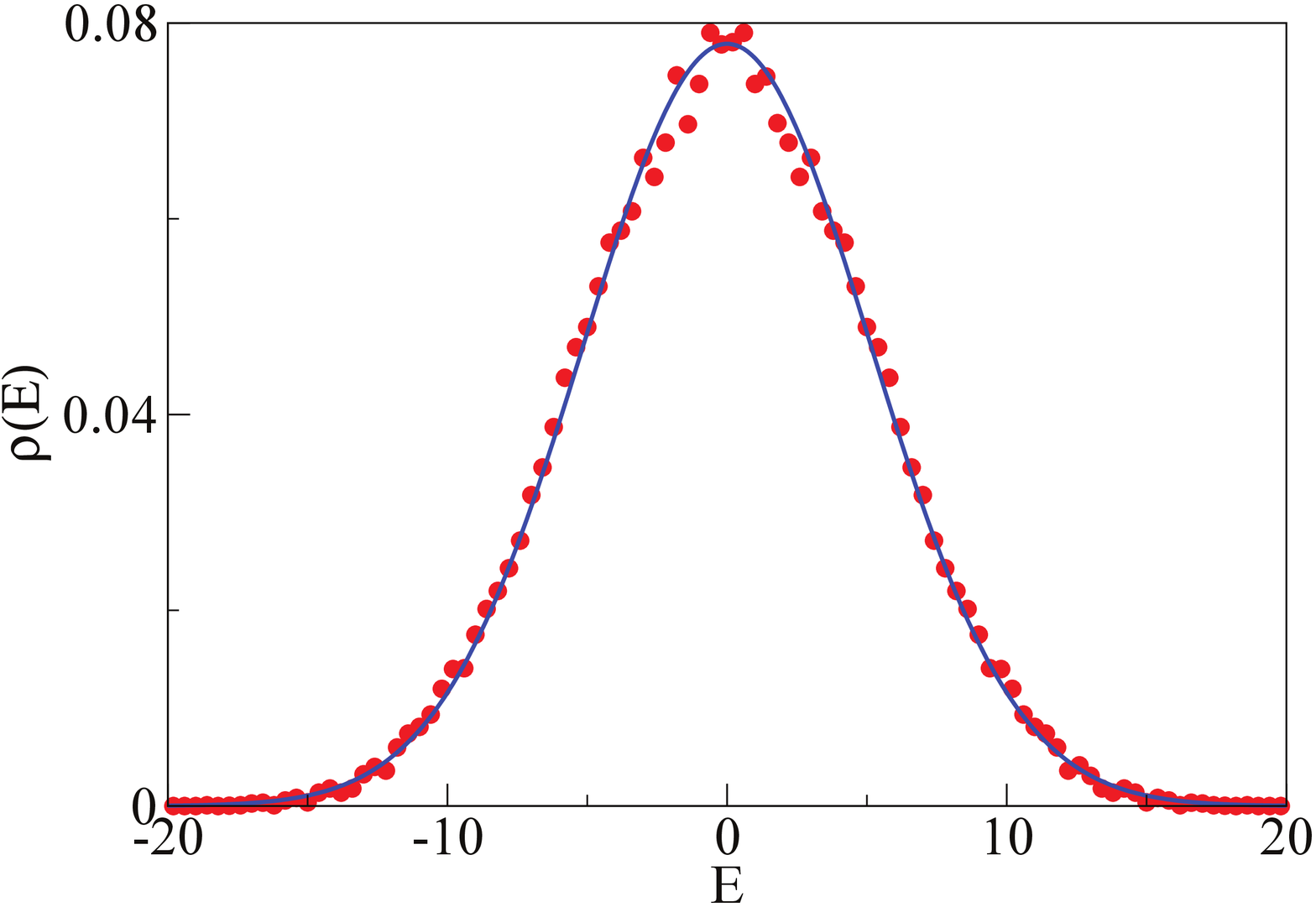}\\
(a)
\end{center}
\end{minipage}
\begin{minipage}{.33\linewidth}
\begin{center}
\includegraphics[width=.85\linewidth, angle=-90,clip]{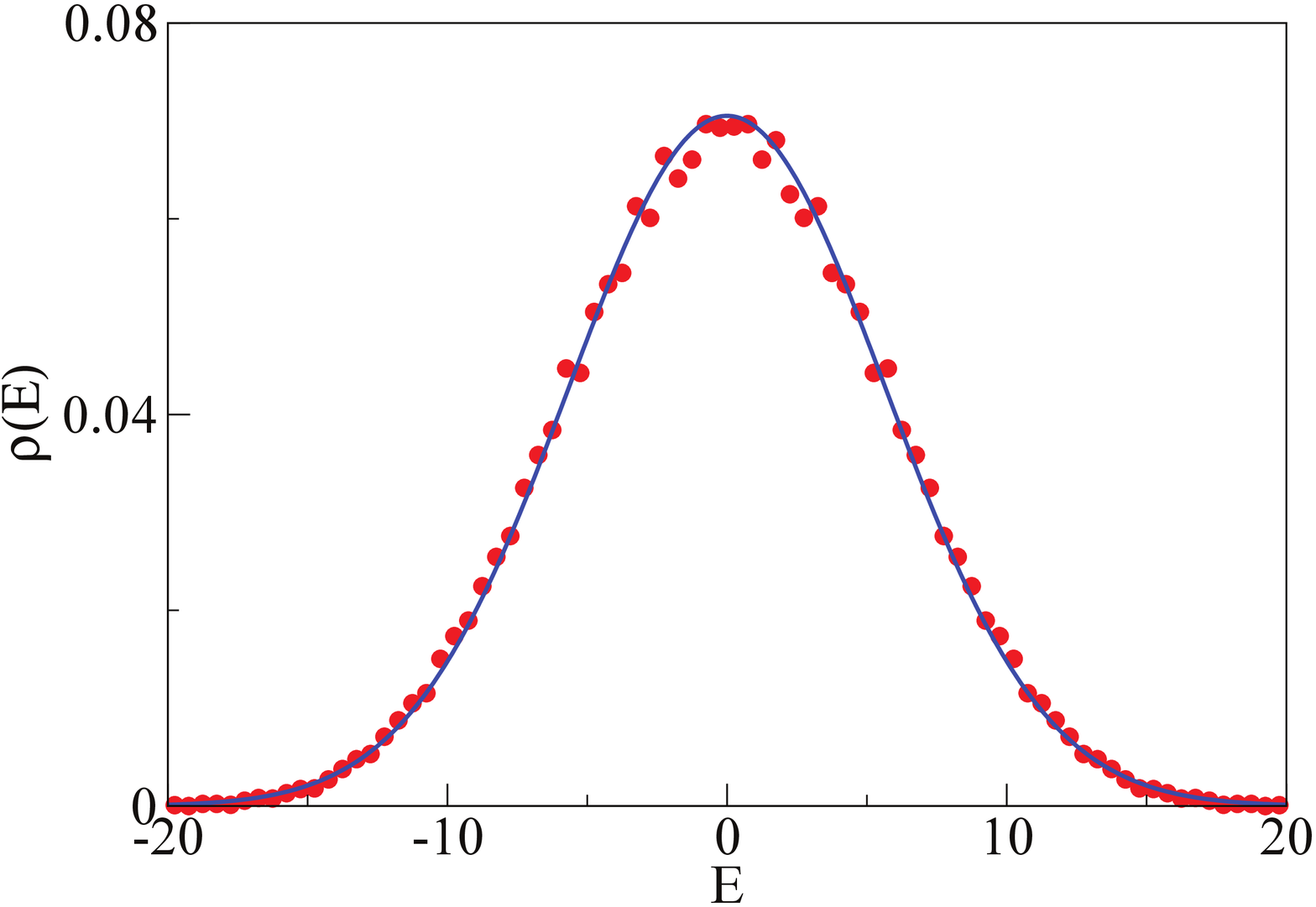}\\
(b)
\end{center}
\end{minipage}
\begin{minipage}{.3\linewidth}
\begin{center}
\includegraphics[width=.94\linewidth, angle=-90,clip]{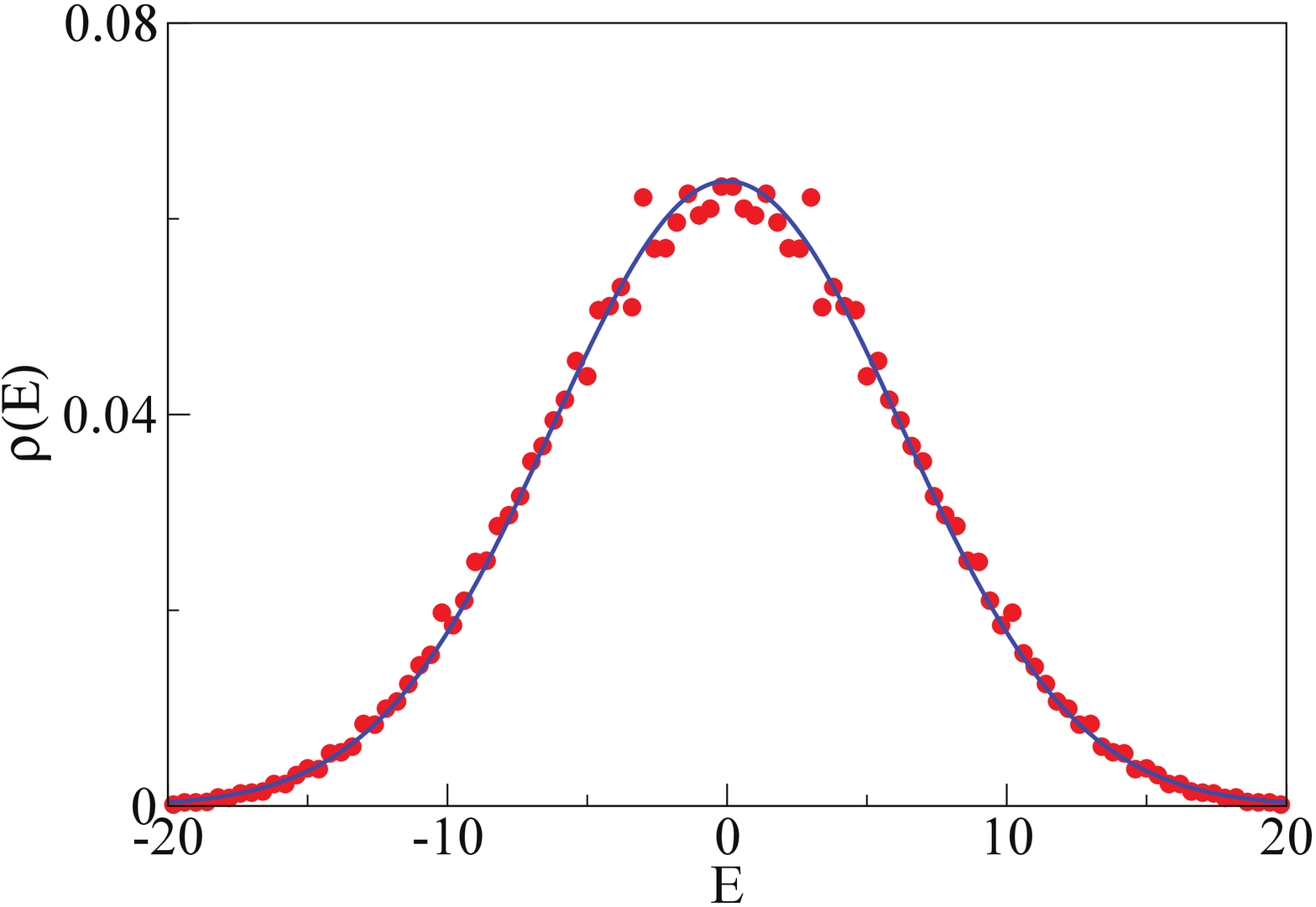}\\
(c)
\end{center}
\end{minipage}
\caption{Spectral density for the quantum Ising model in transverse field with $N=16$:
(a) $\lambda=0.8$, (b) $\lambda=1$, (c) $\lambda=1.2$. Red circles represent numerical calculated densities. Blue thick lines indicate the Gaussian approximation \eqref{Ising_Gaussian}.  }
\label{fig_exact_Gaussian_Ising}
\end{figure}

It is instructive to find the approximate behaviour of the above spectral density in the bulk (i.e. close to the maximum) as in \eqref{bulk}. In this case the dominant contribution is given by vicinity of point $\beta=0$ and 
\begin{equation}
\rho_G(e)\approx \sqrt{\frac{N}{2\pi (1+\lambda^2)}}\exp \left (-\frac{N e^2}{2(1+\lambda^2)} \right )\ .
\label{Ising_Gaussian}
\end{equation}
This is exactly the Gaussian prediction \eqref{Gaussian} for the rescaled variable $e=E/N$. As the true energy is the sum of individual ones (cf. \eqref{energy_exact}), it is plain that in this case the origin of the Gaussian behaviour for the spectral density is the same as in the usual central limit theorem. Indeed, the quantity 
\begin{equation}
E_{\vec{n}}=\sum_{j=1}^N  e_j\left (n_j-\tfrac{1}{2}\right ), \quad \vec{n}=(n_{1},\dots,n_{N})
\end{equation}
can be considered as the sum of $N$ random variables $n_j=0,1$ (or equivalently as a realization of a random walk) and the central limit theorem states that probability distribution of $\lbrace E_{\vec{n}} \rbrace$ is the Gaussian with parameters determined by the first and the second moments of $\lbrace E_{\vec{n}} \rbrace$ which gives \eqref{Ising_Gaussian}.  Of course, the same result can be obtained  directly from the Hamiltonian without reference to the exact solution. It is sufficient to calculate the first traces of powers of the Hamiltonian \eqref{H_Ising}. Denote
\begin{equation}
\langle \mathcal{H}^k\rangle=\frac{1}{2^N}\mathrm{Tr}\, \mathcal{H}^k\ .
\label{traces}
\end{equation}
Simple calculations give that for Ising model in the transverse field
\begin{equation}
\langle \mathcal{H}^{2k+1}\rangle=0,\qquad \langle \mathcal{H}^2\rangle=N(1+\lambda^2),\qquad \langle \mathcal{H}^4\rangle=3N^2(1+\lambda^2)^2-N(2+8\lambda^2+2\lambda^4)\ . 
\label{Ising_moments}
\end{equation}  
Often the Gaussian approximation \eqref{Ising_Gaussian} is practically indistinguishable in the bulk from the exact saddle point formula \eqref{exact_density_ising}.  For example, in figure~\ref{difference} the both expressions are present and it is clear that the difference is hardly visible. 
The Gaussian approximation is also in a good agreement with the results of direct numerical calculations for values of coupling constant close to $1$ as indicated in  figure~\ref{fig_exact_Gaussian_Ising}. 

It is also of interest to find the behaviour of the spectral density close to the ground state. This limit corresponds to $\beta\to\infty$ so that ground state energy is 
\begin{equation}
e_{\mathrm{gs}}=-\frac{1}{2\pi}\int_0^{2\pi}\sqrt{1-2\lambda \cos \phi +\lambda^2}\,  \mathrm{d}\phi \ .
\end{equation}
Expanding the above expressions for large $\beta$ one can find the corresponding formulae. We omit the details and present only  the results for  $\lambda=1$ 
\begin{equation}
\rho(E)=2^{-N } (E-E_{\mathrm{gs}})^{-3/4} \left ( 8\sqrt{6\pi}N  \right )^{-1/2}\exp \left (\sqrt{\frac{\pi N(E-E_{\mathrm{gs}})}{6}}\,\right ) \ .
\label{asymptotic_lambda_1}
\end{equation}
When $N$ is not too big, the region of applicability of \eqref{asymptotic_lambda_1} is small and even the uniform approximation is of little interest. 


\subsection*{Multi-Gaussian approximation}

\begin{figure}
\begin{minipage}{.48\linewidth}
\begin{center}
\includegraphics[width=.8\linewidth, angle=-90,clip]{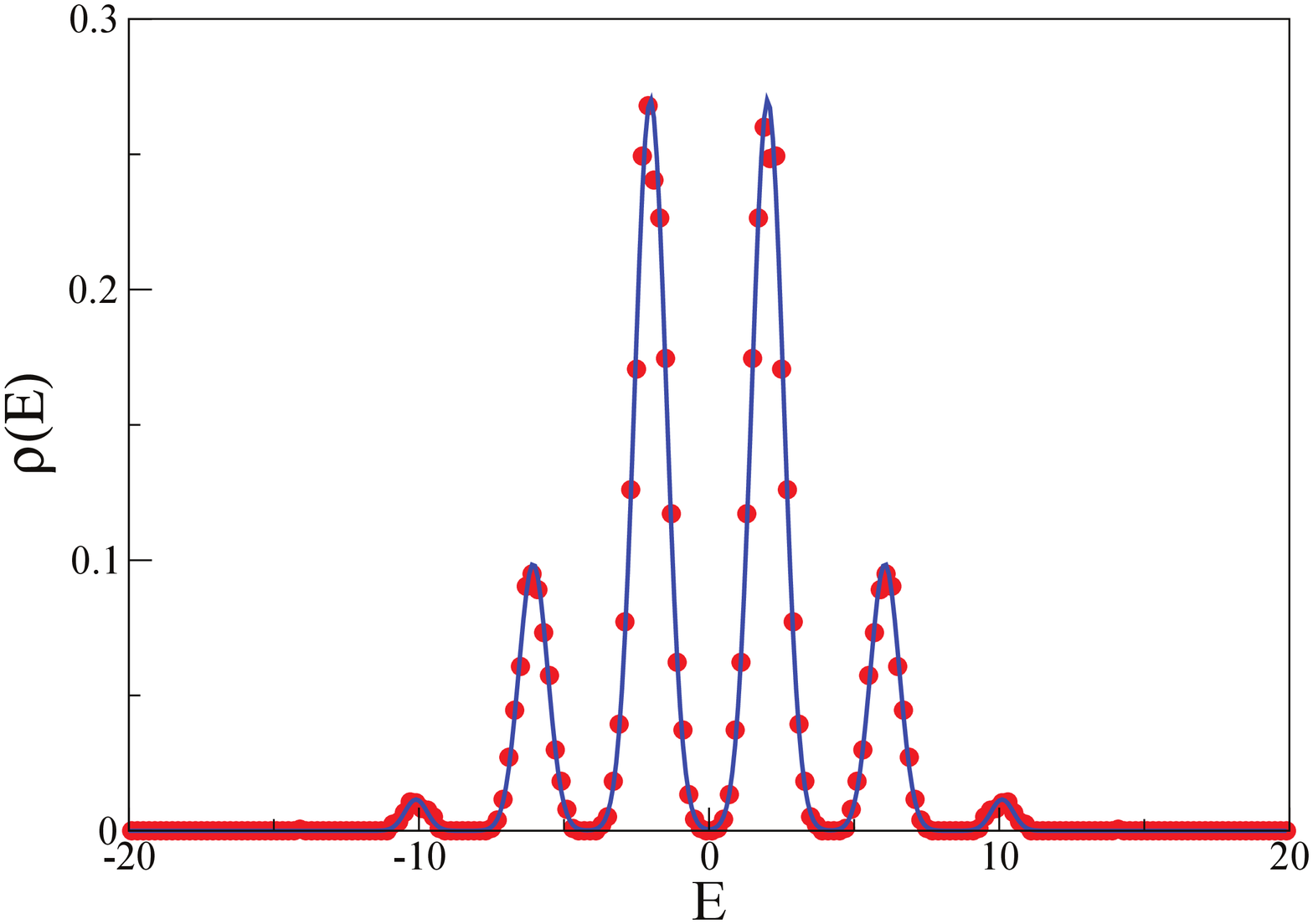}\\
(a)
\end{center}
\end{minipage}
\begin{minipage}{.48\linewidth}
\begin{center}
\includegraphics[width=.8\linewidth, angle=-90,clip]{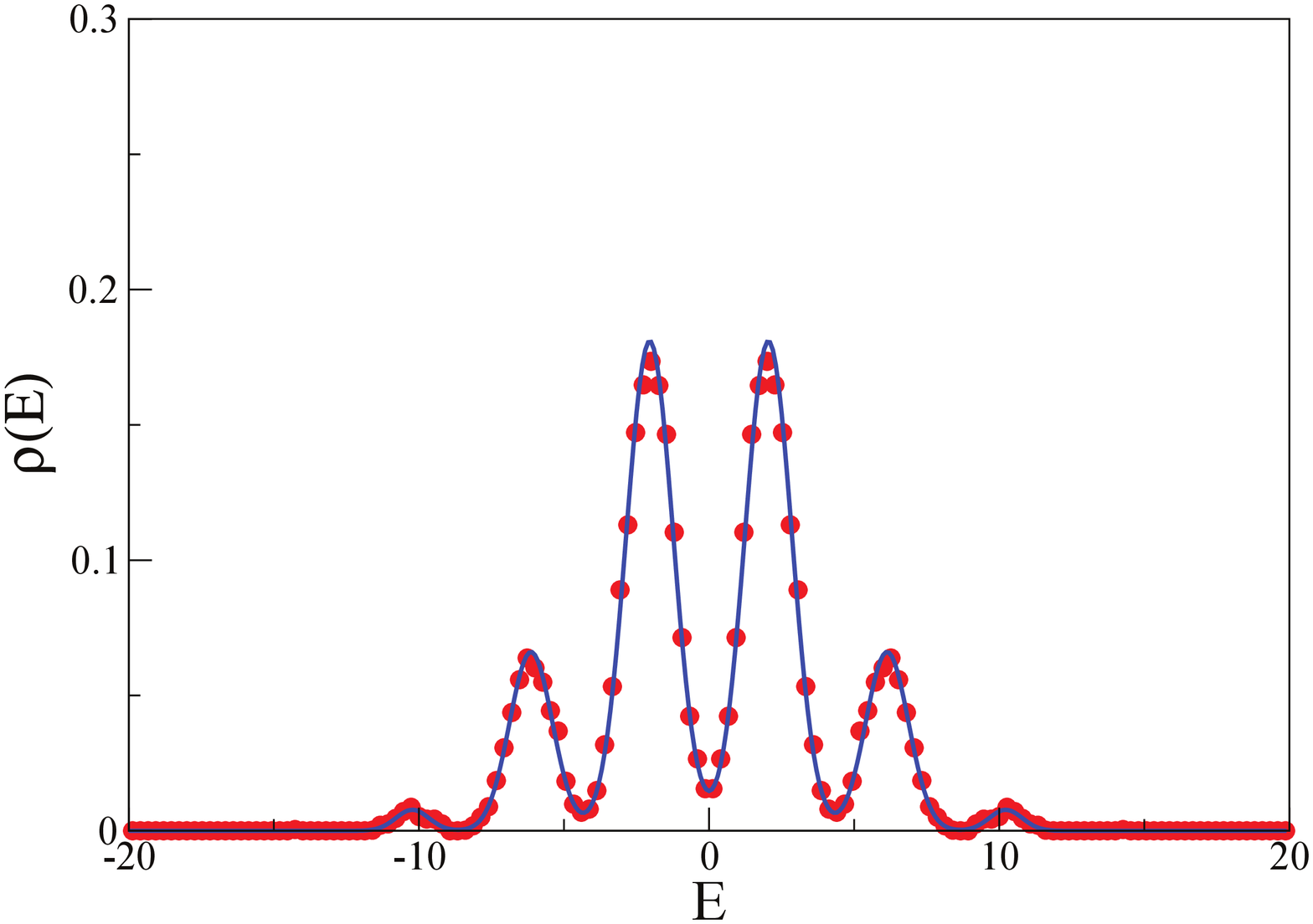}\\
(b)
\end{center}
\end{minipage}

\begin{minipage}{.48\linewidth}
\begin{center}
\includegraphics[width=.8\linewidth, angle=-90,clip]{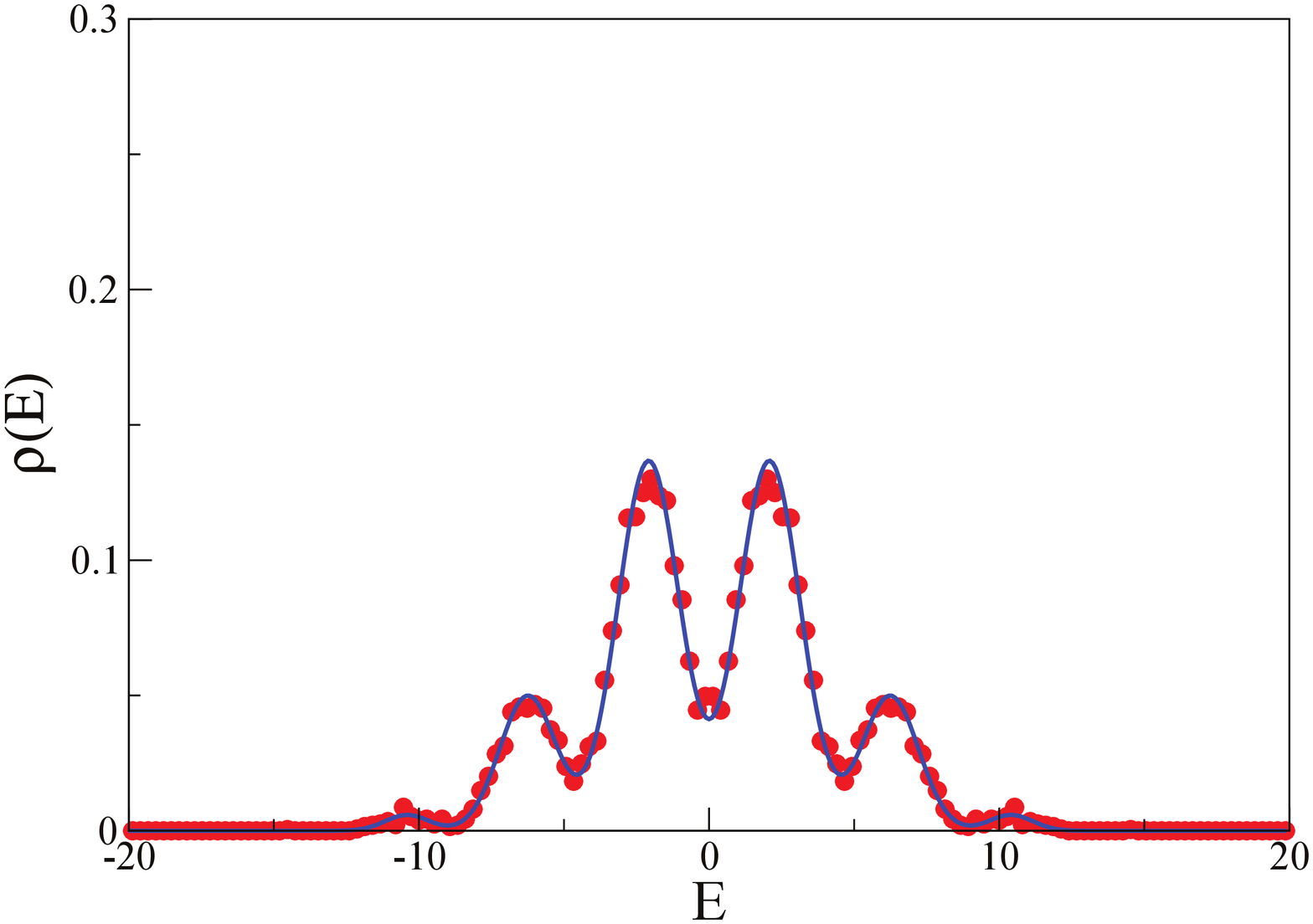}\\
(c)
\end{center}
\end{minipage}
\begin{minipage}{.48\linewidth}
\begin{center}
\includegraphics[width=.8\linewidth, angle=-90,clip]{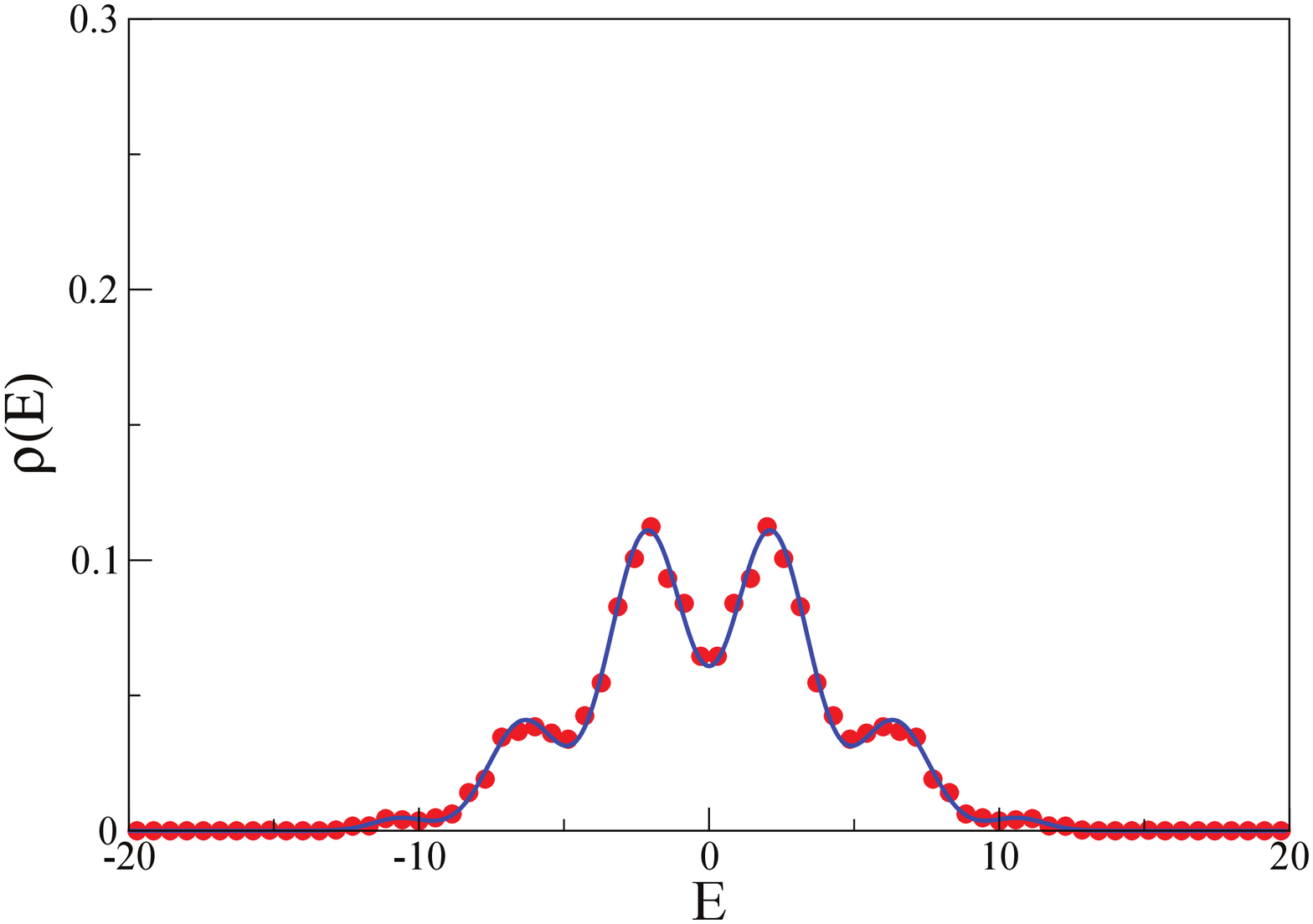}\\
(d)
\end{center}
\end{minipage}
\caption{Spectral density for the quantum Ising model in transverse field with $N=14$:
  (a) $\lambda=0.2$, (b) $\lambda=0.3$, (c) $\lambda=0.4$. (d) $\lambda=0.5$. Red  circles  are numerically calculated densities. Blue thick lines indicate the multi-Gaussian approximation \eqref{rho_lambda_less_1}.  }
\label{Ising_less_1}
\end{figure}

\begin{figure}
\begin{minipage}{.48\linewidth}
\begin{center}
\includegraphics[width=.8\linewidth, angle=-90,clip]{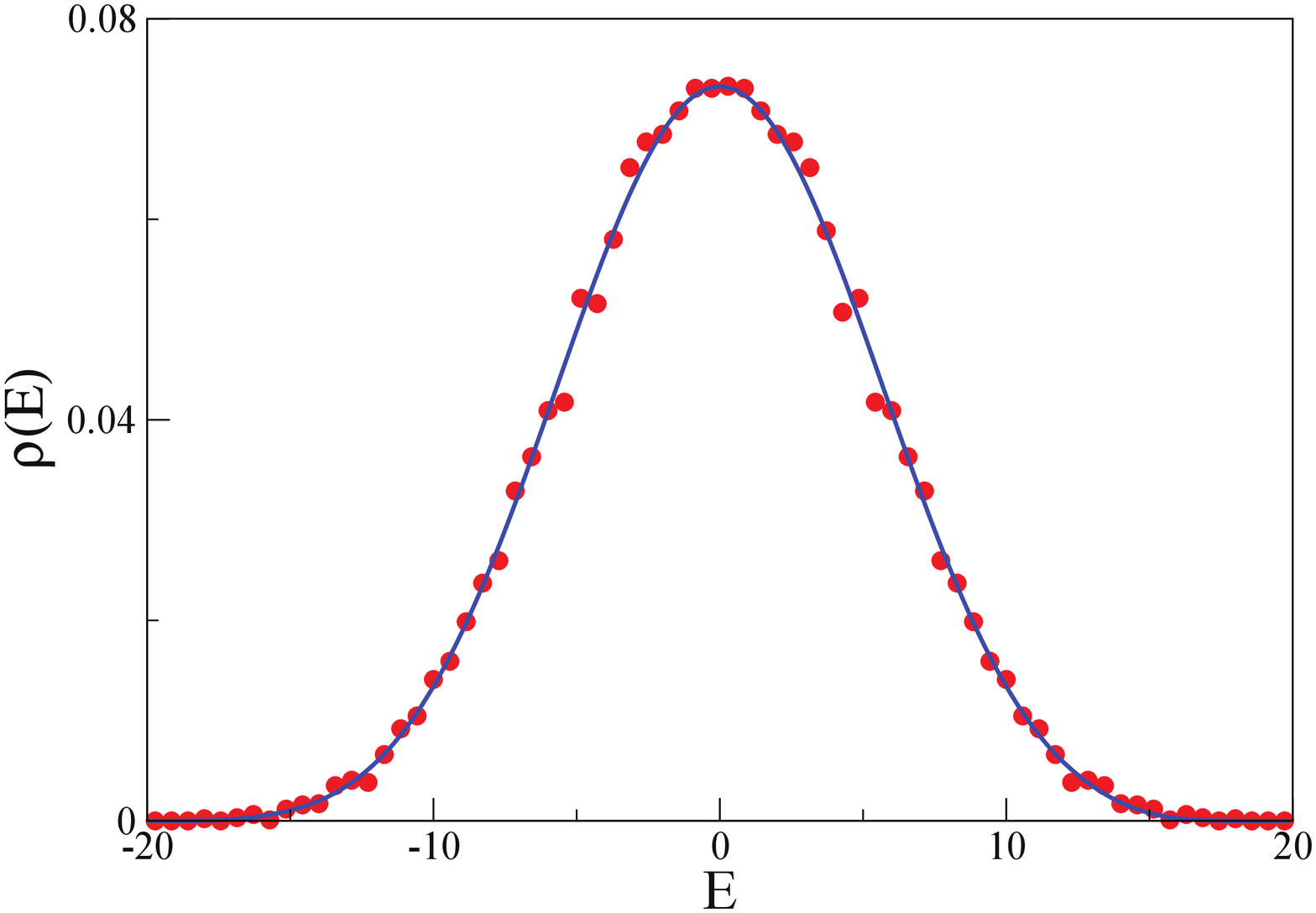}\\
(a)
\end{center}
\end{minipage}
\begin{minipage}{.48\linewidth}
\begin{center}
\includegraphics[width=.8\linewidth, angle=-90,clip]{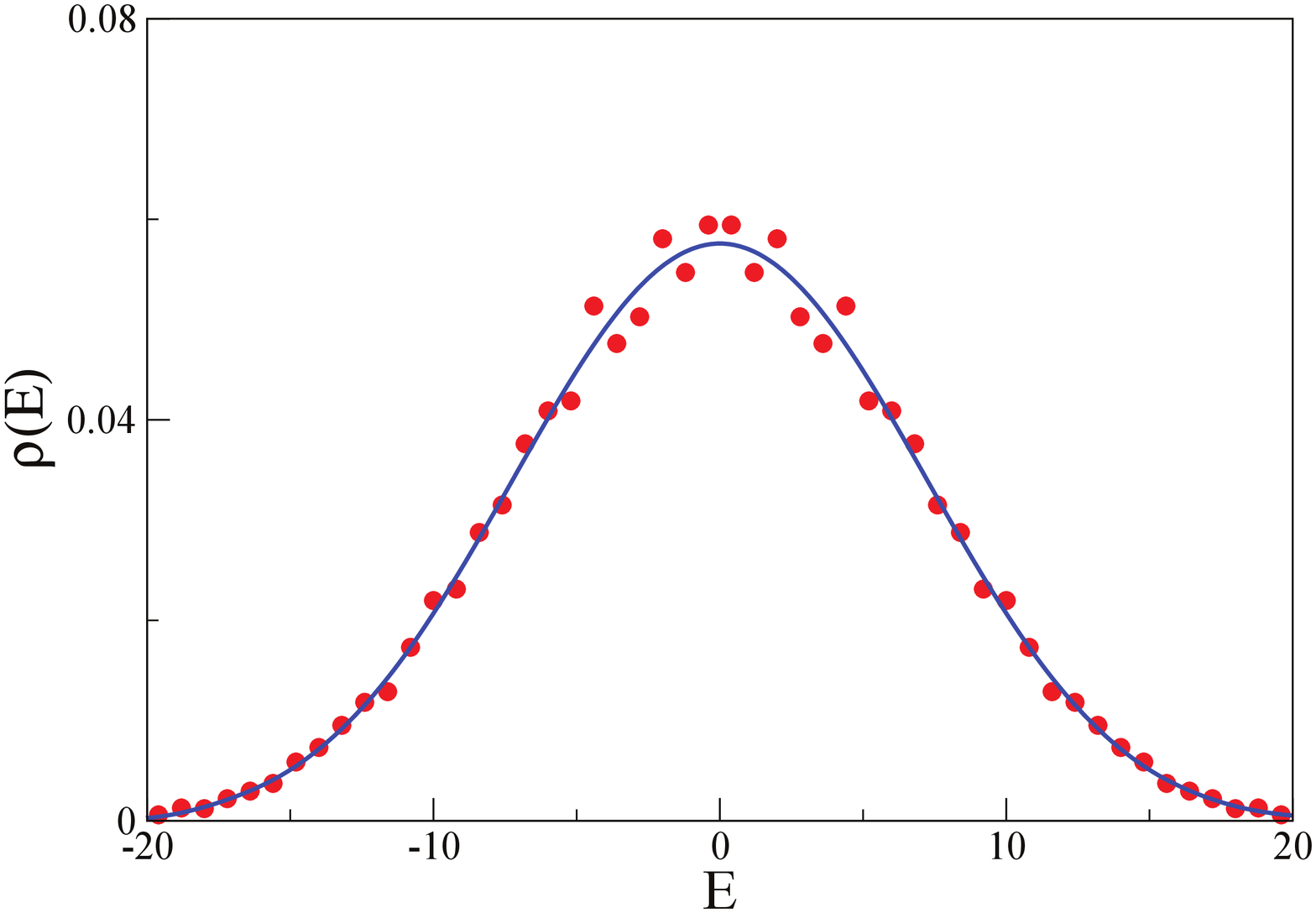}\\
(b)
\end{center}
\end{minipage}

\begin{minipage}{.48\linewidth}
\begin{center}
\includegraphics[width=.8\linewidth, angle=-90,clip]{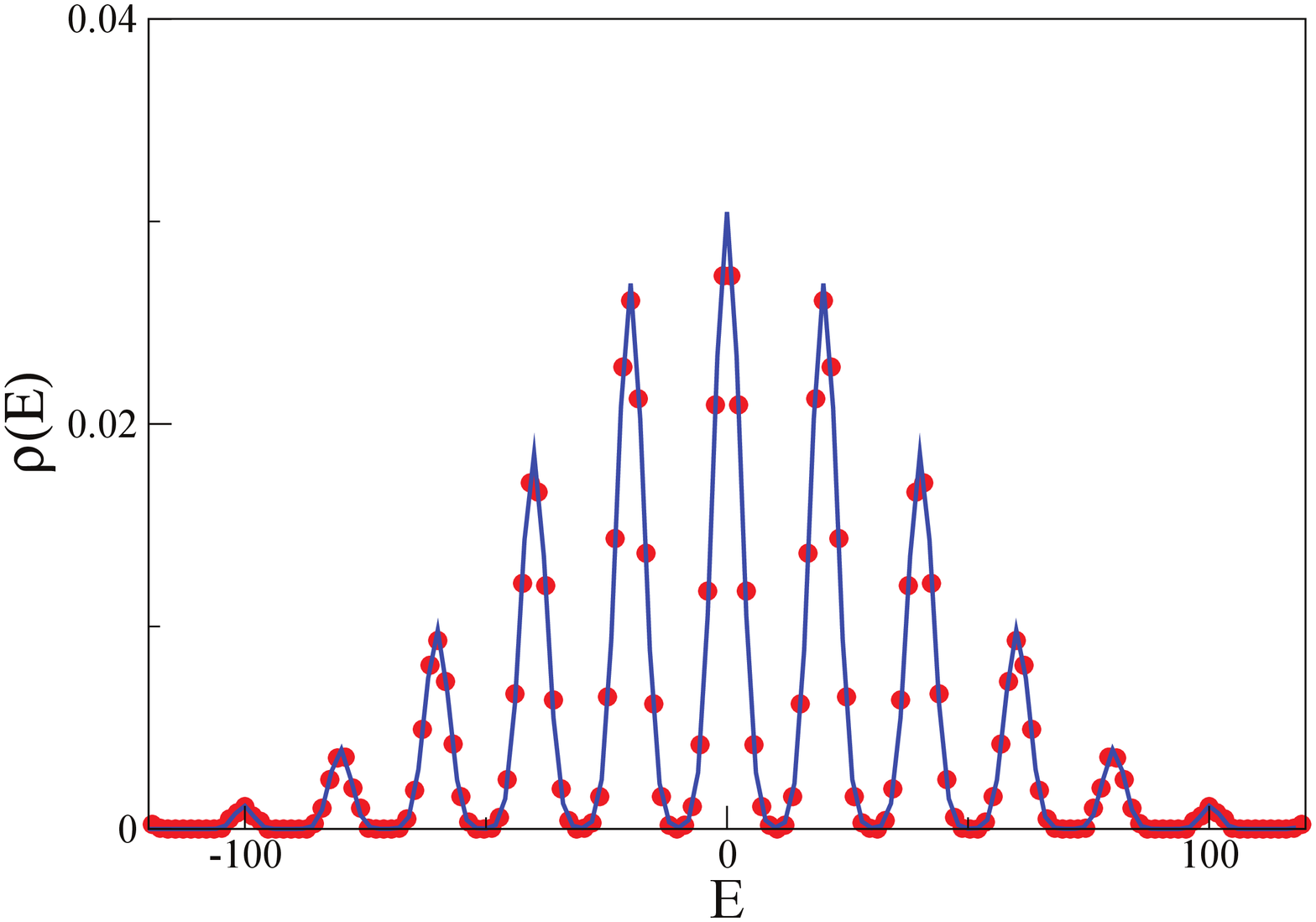}\\
(c)
\end{center}
\end{minipage}
\begin{minipage}{.48\linewidth}
\begin{center}
\includegraphics[width=.8\linewidth, angle=-90,clip]{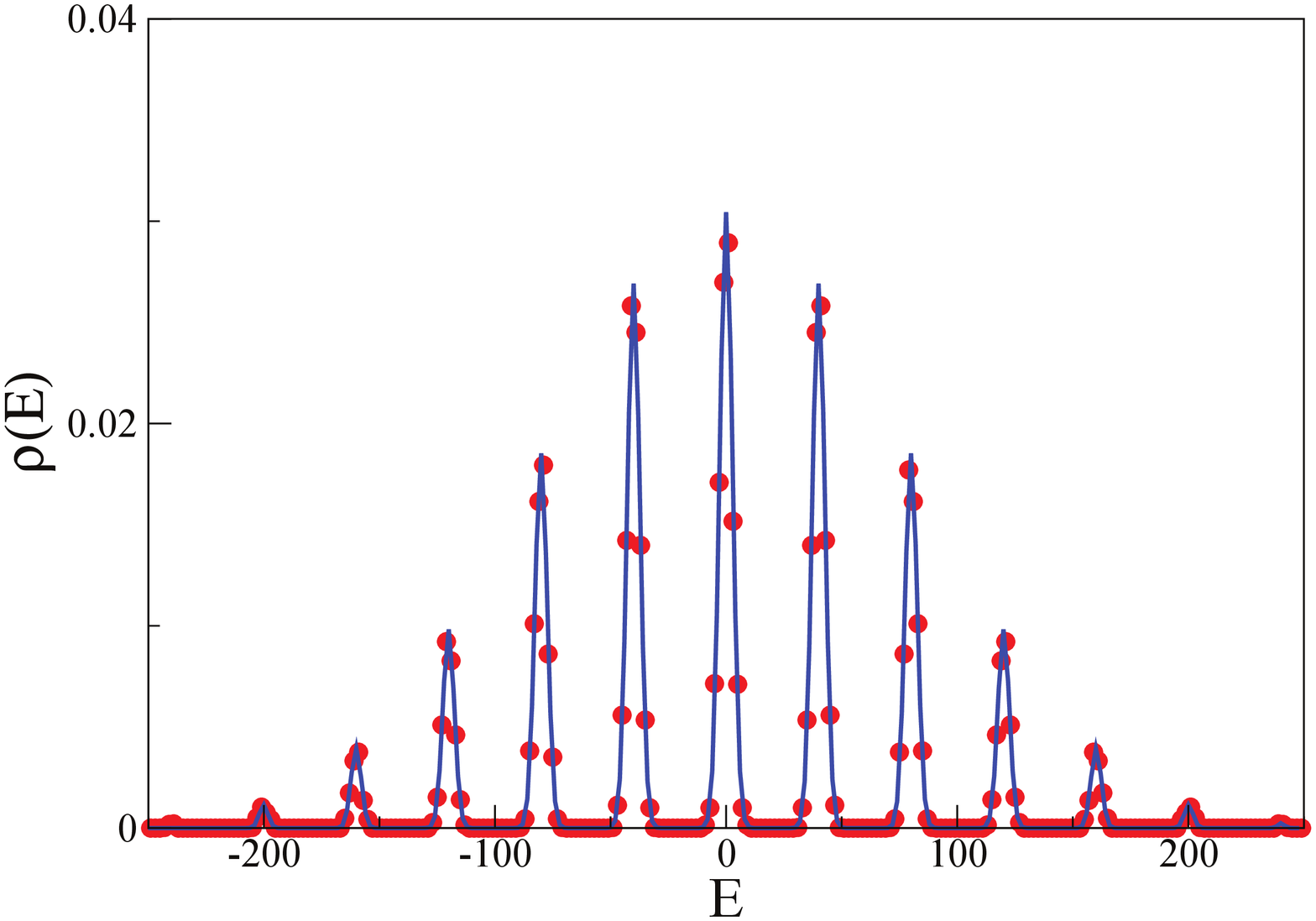}\\
(d)
\end{center}
\end{minipage}
\caption{The same as in figure~\ref{Ising_less_1} but with $\lambda\geq 1$: 
(a) $\lambda=1$, (b) $\lambda=1.5$, (c) $\lambda=10$. (d) $\lambda=20$.  Blue thick lines indicate the multi-Gaussian approximation \eqref{rho_lambda_bigger_1}.  }
\label{Ising_greater_1}
\end{figure}

The exact asymptotic spectral density \eqref{saddle_point_ising} for the Ising model in transverse field and its Gaussian approximation \eqref{Ising_Gaussian} are valid in the limit when the number of spins tends to infinity, $N\to\infty$. Nevertheless for large but finite values of $N$ the situation may and often will  be different. For example, in figure~\ref{Ising_less_1} and \ref{Ising_greater_1} we present the results of numerical calculations for the spectral density at different values of coupling constants. Notice that the density has clear peaks which are not described by the above asymptotic formulae. 

The origin of these peaks is evident. At very small or very large coupling constant the one-particle energy levels \eqref{one_particle_levels} are strongly degenerated and for all $\phi_j$
\begin{equation}
e(\phi_j)\underset{\lambda\to 0}{\longrightarrow} 2\ ,\qquad e(\phi_j)\underset{\lambda\to \infty}{\longrightarrow} 2\lambda\ .
\end{equation}

At finite coupling constant the interaction removes degeneracies and leads to overlapping peaks. 
As for the model considered  exact results are available (see \eqref{energy_exact}), the calculation of the spectral density at finite $N$ is straightforward. Indeed, it is plain that peaks correspond to states with fixed number of fermions in \eqref{energy_exact} and each peak is related with sum of all possibilities to have fixed sum of  occupation numbers
\begin{equation}
n=\sum_{j=1}^N n_j\ .
\end{equation}   
This restriction can be taken into account by considering the generating function given by \eqref{generation_function}. When $N\to\infty$ it takes the form 
\begin{equation}
Z(\beta,x)=\exp\left [\frac{N}{2\pi}  \int_0^{2\pi} \ln \left (  \frac{1+x}{2}\cosh(\beta e(\phi)/2)+\frac{1-x}{2}\sinh(\beta e(\phi)/2)\right )  \right ]\mathrm{d}\phi\ .
\label{asymptotic_generation_function}
\end{equation}
The extraction of the term with fixed total occupation number, $n$, is achieved by the integration in the complex $x$-plane around a contour encircling the origin 
\begin{equation}
Z_n(\beta)=\frac{1}{2\pi \mathrm{i}}\oint  \frac{Z(\beta,x)}{x^{n+1}}\mathrm{d}x\ ,
\end{equation}
Finally the spectral density is the sum over all $n$  densities of states with fixed total occupation number (for $\lambda>1$)
\begin{equation}
\rho(E)\equiv \frac{1}{2^N}\sum_{n}C_N^n \rho_n(E)\ .
\end{equation}
Here $C_N^n$ is the binomial coefficient which counts the total number of states with fixed $n$ and $\rho_n(E)$  is given by the double integral
\begin{equation}
\rho_n(E) =-\frac{1}{4\pi^2 }\int_{c-\mathrm{i}\infty}^{c+\mathrm{i} \infty}\mathrm{d}\beta\, \mathrm{e}^{\beta\, E}\oint \frac{Z(\beta, x)}{x^{n+1}} \mathrm{d}x\ .
\end{equation}
The both integrations can be performed by the saddle point method as above. Calculations are straightforward but the results are tedious, not transparent, and we shall not present them here.  

We found that for our purposes the Gaussian approximation is sufficient. To settle it, one has to calculate the lowest moments of the Ising Hamiltonian in a sub-space with fixed $n$. The simplest way to find the necessary quantities is to calculate the two first terms of expansion of the generating function \eqref{asymptotic_generation_function} into series of $\beta$ and then find the term with $x^n$ power. The direct calculations lead to the following results
 \begin{equation}
\langle E\rangle_n=(N-2n)\langle e\rangle, \qquad  \langle e\rangle = \frac{1}{2N}   \sum_j e_j\approx \frac{1}{2\pi}\int_0^{2\pi} \sqrt{1-2\lambda\cos \phi+\lambda^2}\ \mathrm{d}\phi\ ,
\label{mean_n}
\end{equation}   
and 
\begin{equation}
\sigma_n^2\equiv \langle E^2\rangle_n-\langle E\rangle_n^2= 4\frac{n(N-n)}{N-1}\left ( \langle e^2\rangle-\langle e\rangle^2\right ),\qquad 
\langle e^2\rangle  =\frac{1}{4N} \sum_j e_j^2= 1+\lambda^2\ .
\label{variance_n}
\end{equation}
It means that the mean density of states with fixed $n$ corresponding to each individual peak can well  be approximated by the Gaussian 
\begin{equation}
\rho_n(E)\approx \frac{1}{\sqrt{2\pi \sigma_n^2}}\exp \left (-\frac{(E-\langle E\rangle_n)^2}{2\sigma_n^2}\right )\ ,
\end{equation}
and the total mean density is the sum over all allowed $n$. 

The limiting values of $\langle e \rangle$ and of $\langle e^2\rangle-\langle e \rangle^2$ are the following:   
\begin{equation}
\langle e \rangle =  \lambda, \;  \langle e^2\rangle-\langle e \rangle^2=1,\; \mathrm{when}\;  \lambda\to\infty \ ,\qquad 
\langle e \rangle  = 1, \; \langle e^2\rangle-\langle e \rangle^2 = \lambda^2,\; \mathrm{when}\;  \lambda\to 0\ .
\end{equation}

Taking into account different counting for different $\lambda$ one gets the following multi-Gaussian formulae for the spectral density for the Ising model in transverse field at large but finite $N$.

For $|\lambda|>1$
\begin{equation}
\rho_{\mathrm{mG}}(E)=\frac{1}{2^N} \sum_{n=0}^{N} C_N^n\rho_n(E) \ .
\label{rho_lambda_bigger_1}
\end{equation}

For $|\lambda|<1$ 
\begin{equation}
\rho_{\mathrm{mG}}(E)=\frac{1}{2^{N-1}} \sum_{n=0}^{N/2} C_N^{2n}\rho_{2n}(E)\ . 
\label{rho_lambda_less_1}
\end{equation}
Figures~\ref{Ising_less_1}-\ref{Ising_greater_1} demonstrate that these simple formulae are in a good agreement with numerical results for different coupling constants. 

The domain of visibility of peaks in the spectral density can roughly be estimated from the condition 
\begin{equation}
|E_{n+1}-E_n|>2\sigma_n 
\label{criterium}
\end{equation}
In the center of the spectrum, $n=N/2$ and the above formulae show that the peaks will be observable provided the number of spins, $N< N_{max}$ where
\begin{equation}
N_{max}= 2\lambda^2,\; \;\lambda\to\infty,\qquad  N_{max}= \frac{8}{\lambda^2},\; \;\lambda\to 0
\end{equation}
For $N>N_{max}$ peaks will be strongly overlapped and the descriprion by simple Gaussian formula \eqref{Ising_Gaussian}   becomes adequate. 
 

\section{Ising model in two fields}\label{two_fields}

The Ising model in the transverse field is integrable but when a second  longitudinal field is added, the resulting model whose Hamiltonian is given by \eqref{ising_2} is not integrable (but see \cite{zamolodchikov} for low lying states). Recently certain properties of such model have been investigated experimentally \cite{experiment}.

Eigenvalues of this problem cannot be calculated analytically and its spectral statistics is well described by the standard GOE distribution \cite{atas}. Nevertheless, this model belongs to the class of translational invariant  systems with  short-range interactions. It is well known   (see e.g. \cite{griffiths}) that for such problems the free energy, energy, and entropy are extensive functions of the number of spins (physically it means that the surface energy for short-range interaction systems  is much smaller that the volume energy). In particular, the full entropy of $N$-spins, $S(E,N)$ which determines the full spectral density according to the formula     
\begin{equation}
\rho(E)\sim \mathrm{e}^{S(E,N)}
\end{equation}
has the following form (in weak sense, in general,)
\begin{equation}
S(E,N)=NS(E/N)+\mathcal{O}(1).
\end{equation}
It is also well known (and can be proved e.g. by calulating moments by saddle point method) that the $N$-th power of any convex function $f(e)$ when $N\to\infty$ can be approximated in the bulk  by the Gaussian centered at the point of maximum, $e_0$ ($f^{\prime}(x_0)=0$) 
\begin{equation}
f^N(e)\approx \left (f(e_0)+\frac{(e-e_0)^2}{2}f^{\prime \prime}(e_0)\right )^N\approx f^N(e_0) \mathrm{e}^{-N(e-e_0)^2/2 \sigma^2},\qquad \sigma^2=-\frac{f(e_0)}{f^{\prime \prime}(e_0)}\ .  
\end{equation}
Combining these two statements together leads to the known conclusion that the spectral density of many-body systems with short-range interactions should be  well described in the bulk by the Gaussian approximation which is fixed  by the knowledge of only two lowest momenta of the Hamiltonian. For random spin systems, similar result has been proved in \cite{keating} by a different method. 

\begin{figure}
\begin{center}
\includegraphics[angle=-90, width=.6\linewidth,clip]{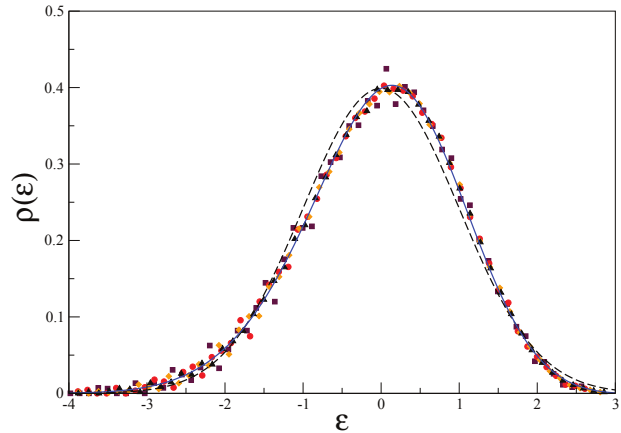}
\end{center}
\caption{Spectral density for the quantum Ising model in two fields with $\lambda=1$ and $\alpha=1$ versus rescaled energy \eqref{rescaled_energy}. Different symbols indicate results of direct diagonalization   with different number of spins. Black squares: $N=13$, Red circles: $N=14$, Orange diamond: $N=15$. Blue triangles: $N=16$. Blue solid line shows prediction \eqref{Gaussian_two_fields} for $N=16$. Dashed black line indicates the pure  Gaussian formula for $N=16$ without the cubic correction term. }
\label{gaussian_ising_two_fields}
\end{figure}

The first moments of the Hamiltotian \eqref{ising_2} (as in \eqref{traces}) can  be calculated straightforwardly. The result is 
\begin{align}
\langle \mathcal{H} \rangle &=0,
\qquad 
\langle \mathcal{H}^{2} \rangle =  N(1+\lambda^2+\alpha^2),\qquad 
\langle \mathcal{H}^{3} \rangle = -6N\alpha^2, \label{third} \\
\langle \mathcal{H}^{4} \rangle &=3N^2(1+\lambda^2+\alpha^2)^2+N( 24\alpha^2-2\alpha^4-2-2\lambda^4 -8\lambda^2 -4\lambda^2\alpha^2)
\end{align}
These expressions lead to the following formula for the spectral density in the bulk of the Ising model in two fields
\begin{equation}
\rho(\varepsilon)=\frac{1}{\sqrt{2\pi}}\mathrm{e}^{-\frac{\varepsilon^2}{2}}\left [1-\frac{\alpha^2}{\sqrt{N}(1+\lambda^2+\alpha^2)^{3/2} } (\varepsilon^3-3 \varepsilon  )\right ]\ ,
\label{Gaussian_two_fields}
\end{equation} 
where $\varepsilon$ is rescaled energy
\begin{equation}
\varepsilon=E/\sqrt{N(1+\lambda^2+\alpha^2)}\ .
\label{rescaled_energy}
\end{equation}
The term in the square brackets is introduced to take into account the third moment of the Hamiltonian \eqref{third}. If necessary, it is easy to incorporate  the forth moment and a  few higher moments as well by using well known Gram–Charlier and (or) Edgeworth series  in Hermite polynomials. Notice that the convergence of the spectral density to the pure Gaussian form is slow. The first correction term is of the order of $1/\sqrt{N}$ where $N$ is the number of spins. It agrees with the error term obtained in \cite{keating} for random spin chain. In figure~\ref{gaussian_ising_two_fields} we present an example of a good agreement of the above formula with the results of direct numerical calculations of spectral density for the Ising model \eqref{ising_2} with $\lambda=\alpha=1$. In particular, this figure demonstrates that the correction term in \eqref{Gaussian_two_fields} gives non-negligible contribution at accessible   number of spins.  

\subsection*{Strong fields}

\begin{figure}
\begin{minipage}{.48\linewidth}
\begin{center}
\includegraphics[width=.8\linewidth, angle=-90,clip]{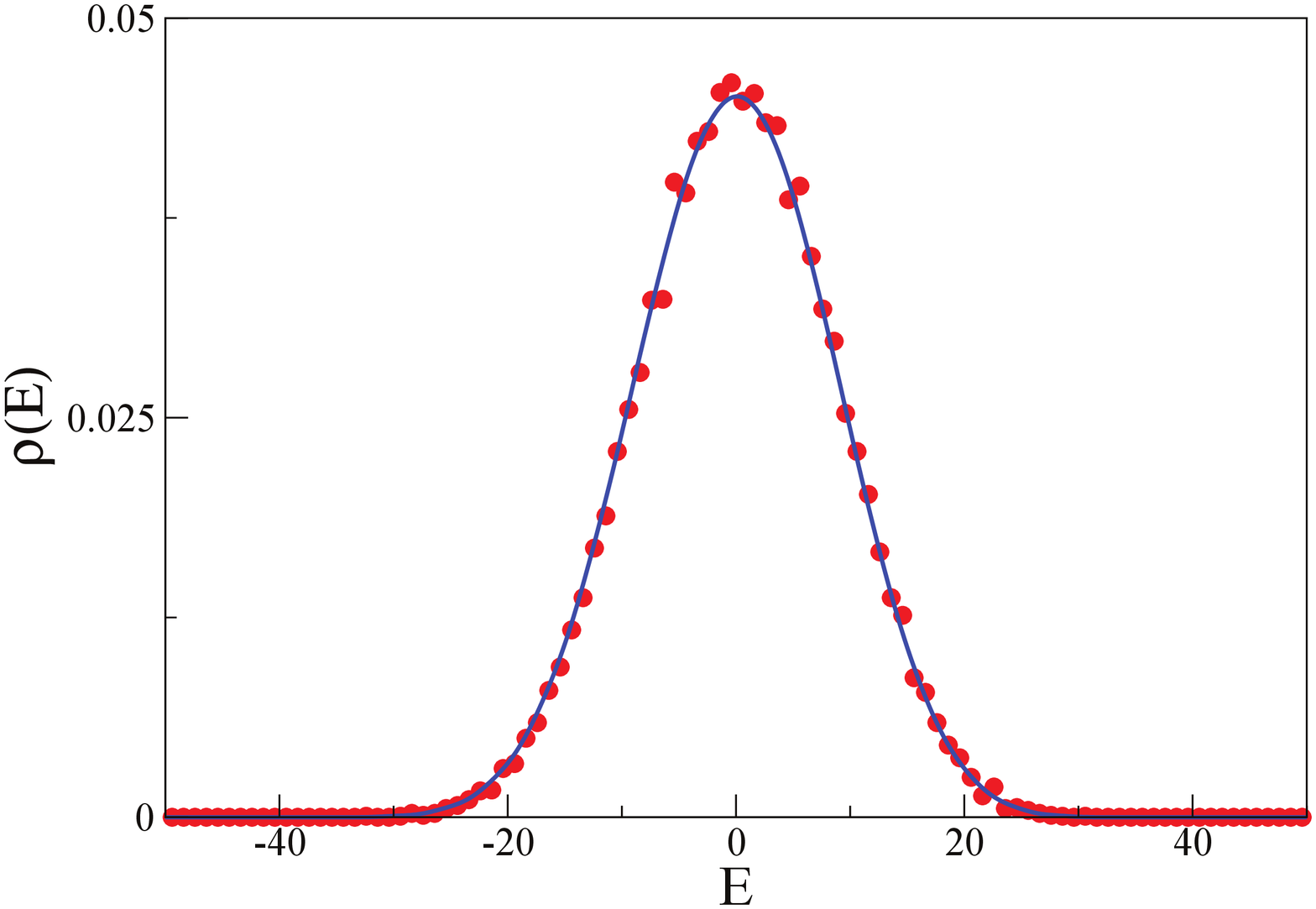}\\
(a)
\end{center}
\end{minipage}
\begin{minipage}{.48\linewidth}
\begin{center}
\includegraphics[width=.8\linewidth, angle=-90,clip]{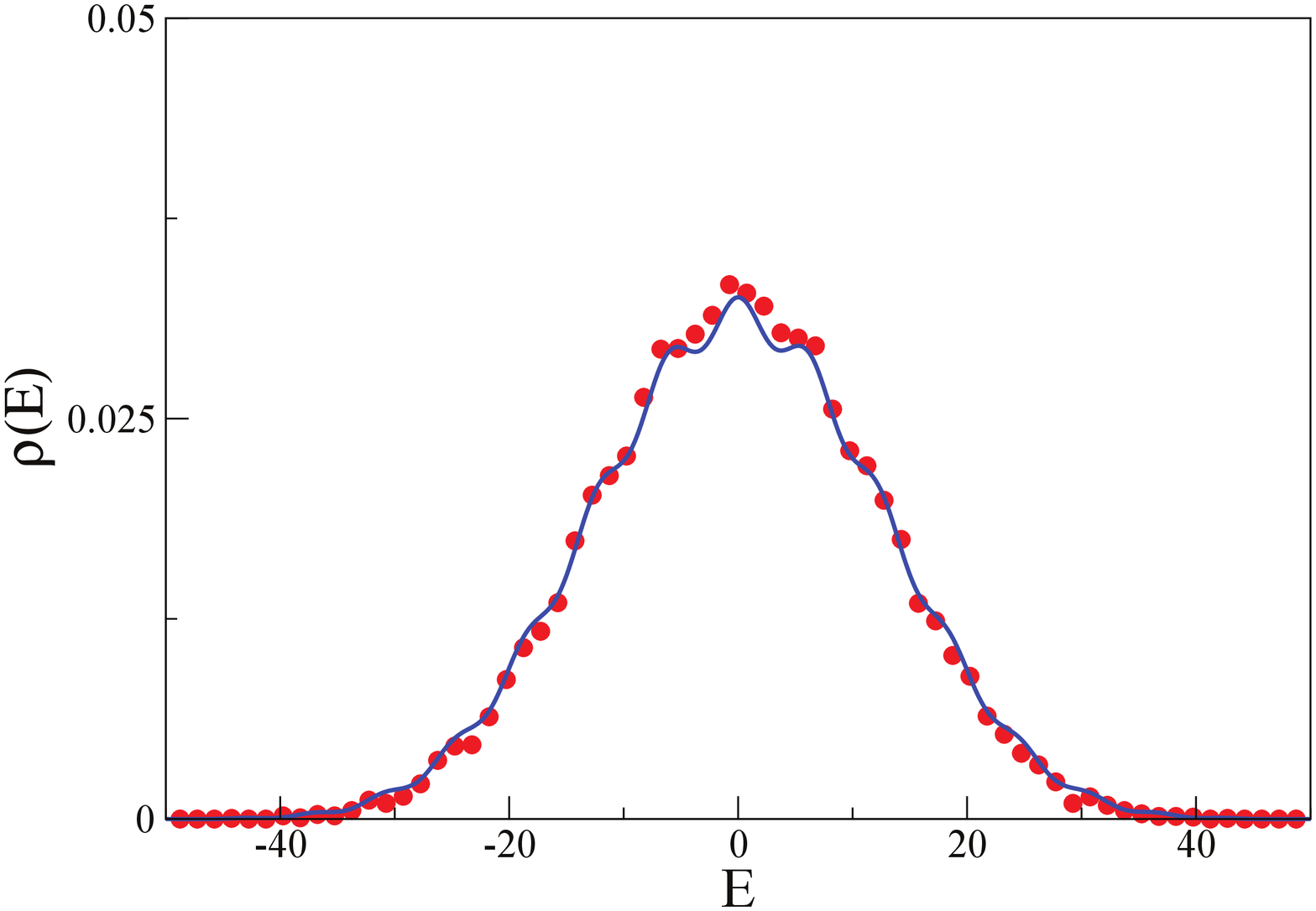}\\
(b)
\end{center}
\end{minipage}

\begin{minipage}{.48\linewidth}
\begin{center}
\includegraphics[width=.8\linewidth, angle=-90,clip]{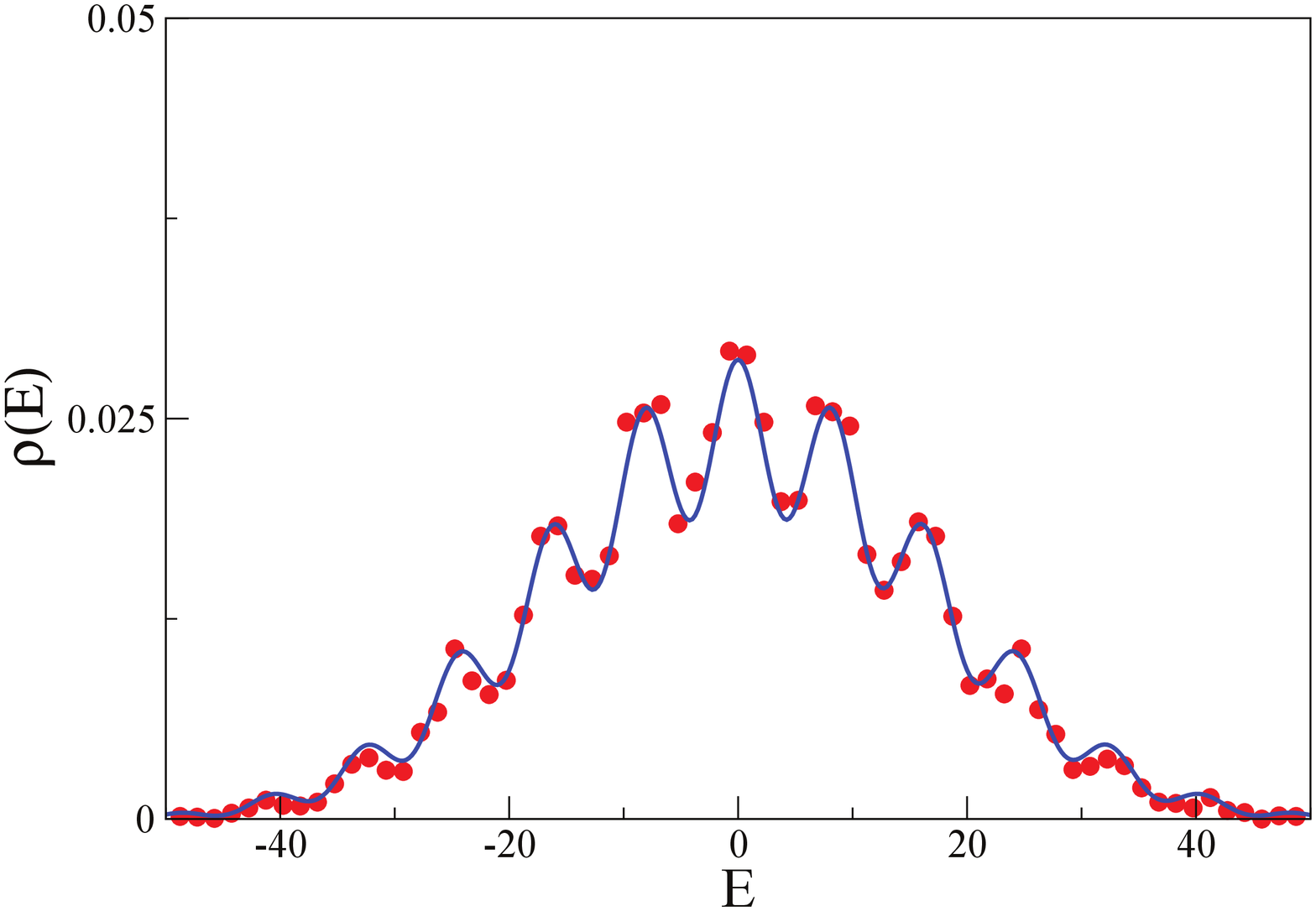}\\
(c)
\end{center}
\end{minipage}
\begin{minipage}{.48\linewidth}
\begin{center}
\includegraphics[width=.8\linewidth, angle=-90,clip]{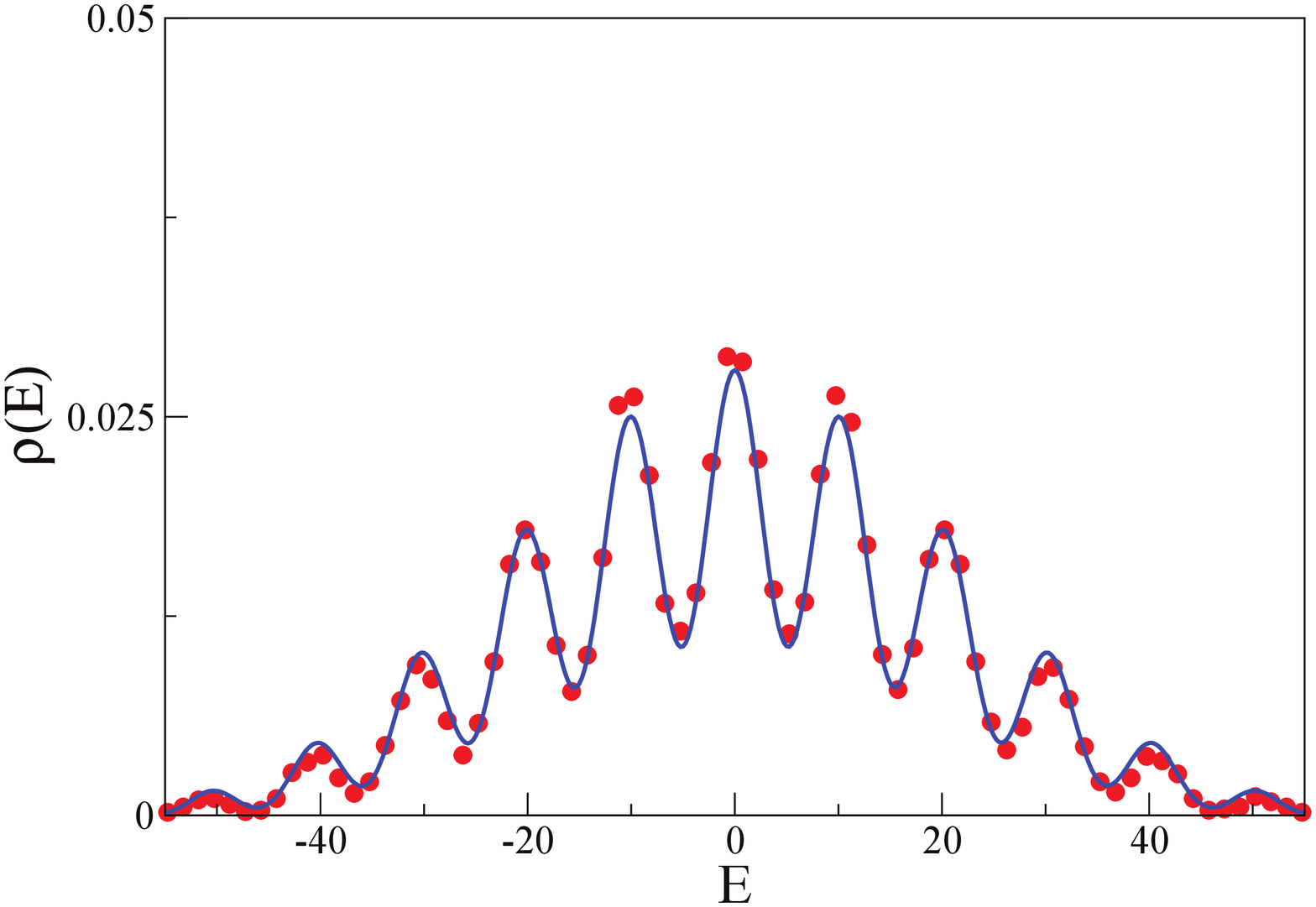}\\
(d)
\end{center}
\end{minipage}
\caption{Spectral density for the quantum Ising model in two fields with $\alpha=0.5$ and $N=14$ for different values of $\lambda$:  (a) $\lambda=2$, (b) $\lambda=3$, (c) $\lambda=4$. (d) $\lambda=5$. Red  circles  are numerically calculated densities. Blue thick lines indicate the multi-Gaussian approximation \eqref{density_two_fields}.  }
\label{multi_density}
\end{figure}

As presented above, there exist general arguments explaining why the density of states for $N$-body short-range interaction models  should be well approximated asymptotically in the bulk by the Gaussian function (see \eqref{Gaussian_two_fields}). However, the numerical calculations for large but finite $N$ clearly demonstrate that it is not always the case (see  figure~\ref{multi_density}). 

As for the case of quantum Ising model in transverse field discussed in the previous Section, the existence of such peaks in the spectral density is related with  degeneracy of energy levels at strong fields.   For the non-integrable case of the Ising model in two fields it is possible to construct a kind of perturbation series to determine the splitting of such degenerate states. 

Let us start with the case of large  $\lambda$ and $\alpha$ where  non-perturbed Hamiltonian is the sum of one-spin terms 
\begin{equation}
\mathcal{H}_0=-\sum_n [\lambda\, \sigma^z_n+\alpha\,  \sigma^x_n]\ .
\end{equation}
It can be diagonalized by the rotation
\begin{equation}
\mathcal{H}_0=-\sum_n[ \lambda\, \sigma^z_n+\alpha\,  \sigma^x_n]=-\sqrt{\lambda^2+\alpha^2}\, \sum_n  \hat{\sigma}^z_n
\label{H_0}
\end{equation}
where 
\begin{equation}
\hat{\sigma}^z_n=\cos \phi \ \sigma^z_n+\sin \phi \ \sigma^x_n,\qquad \cos \phi=\frac{\lambda}{\sqrt{\lambda^2+\alpha^2}},\; \sin \phi=\frac{\alpha}{\sqrt{\lambda^2+\alpha^2}}\ . 
\label{hat_sigma}
\end{equation} 
The other component of the new spin operator is 
\begin{equation}
\hat{\sigma}^x_n=-\sin  \phi \ \sigma^z_n+\cos \phi \ \sigma^x_n\ . 
\end{equation}
It is plain that the unperturbed spectrum of this Hamiltonian is
\begin{equation}
E_n^{(0)}=\sqrt{\lambda^2+\alpha^2}(N-2n)
\label{unperturbed_E}
\end{equation}
where $n$ is the total number of (new) spins up and the degeneracy of each level is $C_N^n$.  

The remaining Hamiltonian is the sum of nearest-neighbour interaction terms  
\begin{equation}
\mathcal{H}_1=\sum_n \sigma^x_n \sigma^x_{n+1}\ .
\end{equation}
After the rotation to new spins  it takes the form
\begin{align}
\notag \mathcal{H}_1&=-\sum_n (\sin  \phi \ \hat{\sigma}_{n}^z+\cos \phi \ \hat{\sigma}_{n}^{x})(\sin  \phi \ \hat{\sigma}_{n+1}^z+\cos \phi \ \hat{\sigma}_{n+1}^{x})\\
 &=- \sin^2 \phi \sum_n \hat{\sigma}^z_n \hat{\sigma}^z_{n+1} -\cos^2 \phi \sum_n \hat{\sigma}^x_n \hat{\sigma}^x_{n+1}-\sin \phi \cos \phi \sum_n \, 
 \hat{\sigma}^x_n( \hat{\sigma}^z_{n+1}+ \hat{\sigma}^z_{n-1})\ .
 \label{H_1}
\end{align}
The standard way of splitting the degeneracies consists in the diagonalization of the interaction  Hamiltonian $H_1$  in the sub-space of all states with the same energy (which in our case corresponds to the fixed number of spin up). Due to exponentially large dimension of such sub-space for the Ising model considered here, this direct approach is not effective.

We propose a different method. As the initial Hamiltonian belongs to the class of short-range interaction, the Hamiltonian projected to the sub-space of states with fixed number of spins up will, necessarily also contain only short-range interactions. As it was done above, one can argue that the spectral density of such projected Hamiltonian is well described by the first two moments of this Hamiltonian.  The calculation of these momenta is performed in \ref{ap_1} and the results are 
\begin{eqnarray}
E_n&\equiv &\langle \mathcal{H} \rangle_{n}= \sqrt{\lambda^2+\alpha^2}(N-2n)-\left (N-\frac{4n(N-n)}{N-1}\right ) \frac{\alpha^2}{\lambda^2+\alpha^2},\label{mean_energy_two_fields} \\ 
\sigma_n^2 &= & \frac{2n(N-n)}{N-1} \frac{\lambda^4}{(\lambda^2+\alpha^2)^2}.
\end{eqnarray}  
It means that the normalised density of states with fixed number of spins up is the Gaussian with the above parameters
\begin{equation}
\rho_n(E)=\frac{1}{\sqrt{2\pi\sigma_n^2}}\exp \left (-\frac{(E-E_n)^2}{2\sigma_n^2}\right )
\end{equation} 
and the total spectral density is the sum over all $n$
\begin{equation}
\rho_{\mathrm{mG}}(E)=\frac{1}{2^N}\sum_{n=0}^NC_N^n \rho_n(E)
\label{density_two_fields}
\end{equation}
Figure~\ref{multi_density} shows that this formula agrees well with results of numerical calculations for different values of $\lambda$.  

In \ref{a p_3} it is demonstrated that for the Ising model in two fileds the Hamiltonian projected on the sub-space with fixed number of spins up reduces to the well known XX model which has an exact solution. The spectrum  of the XX model corresponds to the fermionic filling as in \eqref{energy_exact}. Therefore,  in the bulk the XX spectral density can be well approximated by the Gaussian function whose parameters agree with \eqref{mean_energy_two_fields}. It gives a direct proof that in sub-space of fixed unperturbed energy (i.e. fixed $n$) spectral density is Gaussian. 

The criterium  \eqref{criterium} implies that peaks in strong fields are visible if $N<N_{max}$ where
\begin{equation}
N_{max}=\frac{2(\lambda^2+\alpha^2)^3}{\lambda^4}.
\end{equation}


\subsection*{Small transverse field}

Another interesting limit of the Ising model in two fields \eqref{ising_2}  is  when  $\alpha$ is fixed and $\lambda \to 0$. To investigate this case it is convenient first to diagonalize  the one-spin part as it is done in \eqref{H_0} and \eqref{H_1}. In particular, the diagonal energy in the new $z$-spin representation is 
\begin{equation}
E_{n,k}(\alpha,\lambda)=\sqrt{\alpha^2+\lambda^2}(N-2n)-(N-4k) \sin^2\phi,\qquad \sin \phi=\frac{\alpha}{\sqrt{\alpha^2+\lambda^2}}
\label{E_shifted}
\end{equation}  
Here as above $n$ is the total number of spins up and $k$ is the number of groups of spins in the same directions (cf. figure~\ref{structure}).

When $\lambda=0$ the unperturbed energy takes the form
\begin{equation}
E_{n,k}^{(0)}=\alpha(N-2n)+4k-N
\label{alpha_E_0}
\end{equation}  
We assume that parameter $\alpha$ is of order of $1$, therefore this energy is a function  of  two integers, $n$ and $k$. The existence of strong peaks in the spectral density depends crucially on exact degeneracies of this energy. 

\begin{figure}
\begin{minipage}{.32\linewidth}
\begin{center}
\includegraphics[width=.9\linewidth, angle=-90,clip]{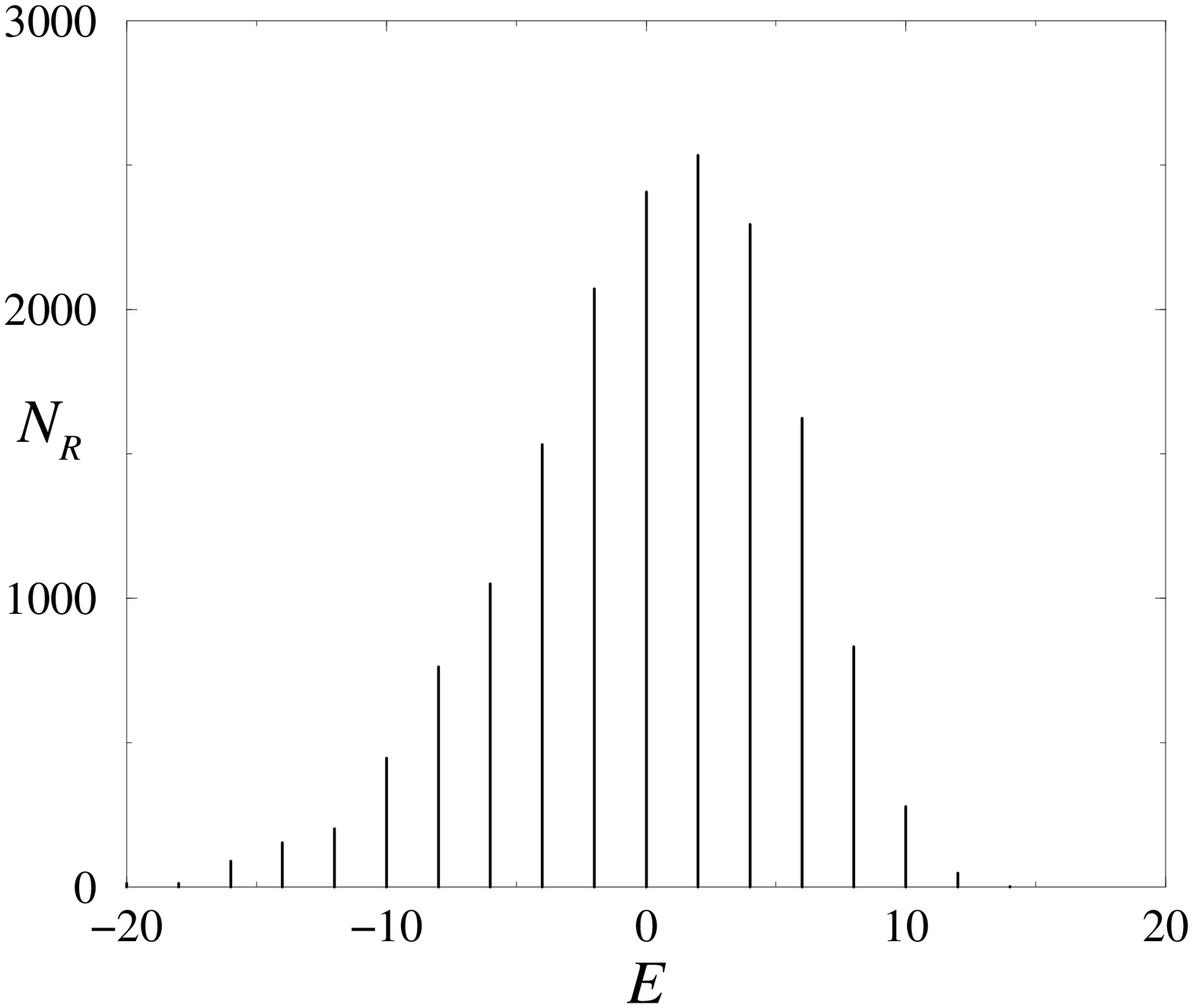}\\
(a)
\end{center}
\end{minipage}
\begin{minipage}{.32\linewidth}
\begin{center}
\includegraphics[width=.9\linewidth, angle=-90,clip]{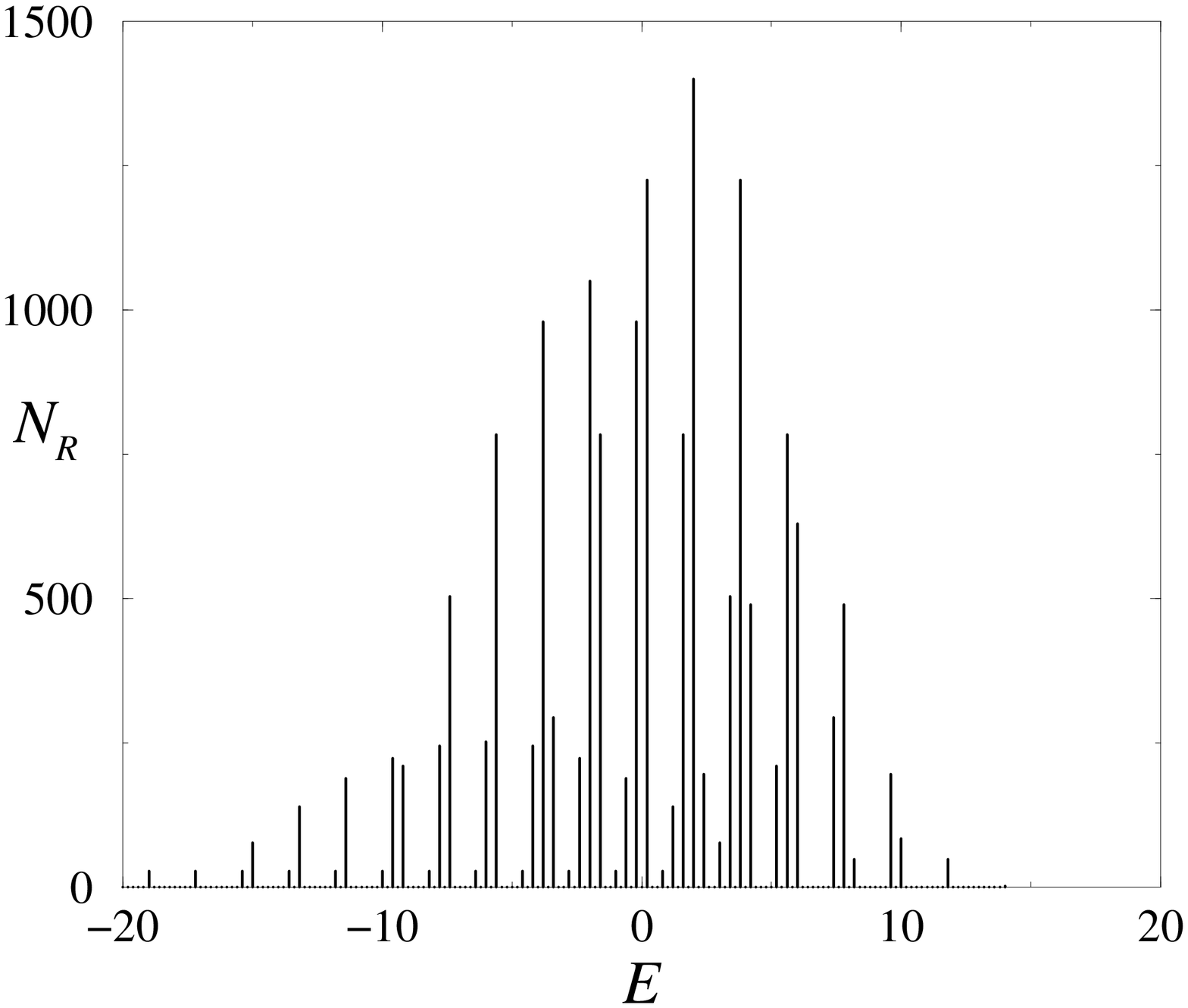}\\
(b)
\end{center}
\end{minipage}
\begin{minipage}{.32\linewidth}
\begin{center}
\includegraphics[width=.9\linewidth, angle=-90,clip]{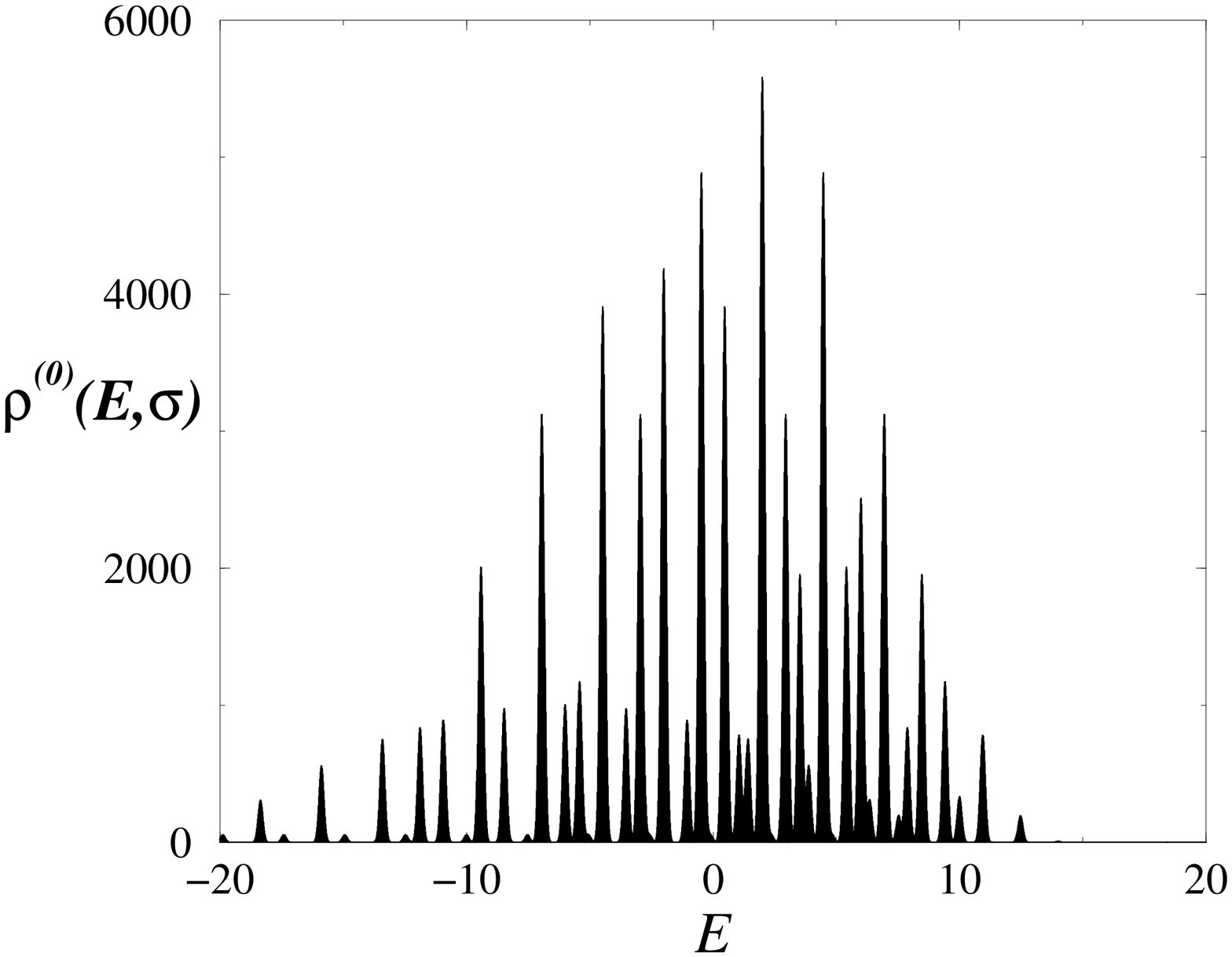}\\
(c)
\end{center}
\end{minipage}
\caption{Degeneracies of the unperturbed spectrum of the Ising model in two fields with $N=14$ and $\lambda=0$: (a) $\alpha=1$, (b) $\alpha=0.9$, and (c) $\alpha=\sqrt{5}-1$. }
\label{degeneracy}
\end{figure}

We consider first a characteristic example with integer $\alpha=1$. In this case, unperturbed energy is simply an even integer, $E_{n,k}^{(0)}=2R$ where 
\begin{equation}
R=2k-n
\label{E_0}
\end{equation}   
It means that the sub-space of exactly degenerated states consists of all combinations of spins up and down such that integer $R\equiv 2k-n$ takes a constant value. From \eqref{k_configurations} derived in \ref{ap_1}, it follows that the dimension of each degenerate sub-space is 
\begin{equation}
\mathcal{N}_R=\sum_{2k-n=R} \frac{N}{k}C_{n-1}^{k-1}C_{N-n-1}^{k-1}
\label{N_R}
\end{equation}
These degeneracies are plotted in figure~\ref{degeneracy}a. 

Our second example is $\alpha=0.9$. In this case unperturbed energy is $E_{n,k}^{(0)}=-0.1N +R/5$ where $R$ is an integer 
\begin{equation}
R=20k-9n
\end{equation}   
In this case the exact degeneracies are calculated by the formula similar to \eqref{N_R}
\begin{equation}
\mathcal{N}_R=\sum_{20k-9n=R} \frac{N}{k}C_{n-1}^{k-1}C_{N-n-1}^{k-1}\ .
\end{equation}
These values are presented in figure~\ref{degeneracy}b. Notice that the combination of number theory and combinatorics leads to peaks with irregular heights.  

When longitudinal field $\alpha$ is irrational, the unperturbed spectrum \eqref{alpha_E_0} 
is formally non-degenerated. Nevertheless the density of such spectrum is not constant and has clear peaks related with the continued fraction expansion of $\alpha$. To illustrate it, we present in figure~\ref{degeneracy}c the density of unperturbed spectrum represented as the double sum over $n$ and $k$ for $\alpha=\sqrt{5}-1\approx 1.24$
\begin{equation}
\rho^{(0)}(E,\sigma)=\sum_{n=0}^N\sum_{k=1}^{n-1}f(n,k)\frac{1}{\sqrt{2\pi \sigma^2}}\exp\left ( -\frac{(E-\alpha(N-2n)+(N-4k))^2}{2\sigma^2}\right )\ .
\end{equation} 
with $\sigma$ chosen arbitrary, $\sigma=0.1$. $f(n,k)$ as above is the total number of symbols with fixed $n$ and $k$ given by \eqref{k_configurations}.  

Peaks of different heights and irregular positions are clearly visible in this picture which has to be compared with figure~\ref{degeneracy}a for $\alpha=1$ and  figure~\ref{degeneracy}b for $\alpha=0.9$. Notice that figure~\ref{degeneracy}a and figure~\ref{degeneracy}b show true integer degeneracies but figure~\ref{degeneracy}c is only the density of unperturbed states smoothed by $\sigma$. For smaller $\sigma$ more peaks in this figure become visible.

At non-zero but  small values of $\lambda$ each degenerate peak is  shifted and acquires a finite width. If peaks are well separated,  the peak shift is determined by the first moment the Hamiltonian $\mathcal{H}$ projected into sub-space with fixed $R$ and the peak width is related with the second moment of the projected Hamiltonian. When unperturbed peaks are very close to each other such simple approximation is questionable and one has to add more perturbation terms related with transitions to almost degenerated states. 

In the above examples the most regular peaks correspond to integer $\alpha=1$ (cf. figure~\ref{degeneracy}a). In this case  the contribution to the first moment of the the Hamiltonian $\mathcal{H}_0$ is given by the expression
\begin{align}
E_R&\equiv  \langle \mathcal{H}\rangle_R=\frac{1}{\mathcal{N}_R}\sum_{2k-n=R} E_{n,k}(1,\lambda)\\
&=2R \sqrt{1+\lambda^2} +N\left (\sqrt{1+\lambda^2}-\frac{1}{1+\lambda^2}\right)\left [1-\frac{4}{\mathcal{N}_R }\sum_{2k-n=R} C_{n-1}^{k-1}C_{N-n-1}^{k-1} \right ]\ .
\label{E_R}
\end{align}
Of course, this formula is valid at small $\lambda$. In the limit $\lambda\to 0$ it coincides with \eqref{E_0}. To take into account all second order terms proportional to $\lambda^2$ one should add the contribution due to transitions into states with different energies. They can be calculated within the usual perturbation series. The details of the calculations in the second order in $\lambda$ are presented in \ref{ap_2}. The result is 
\begin{equation}
\Delta E_{R}=\frac{1}{\mathcal{N}_R}\sum_{2k-n=R}\Delta E_{n,N-n;k}\ , 
\end{equation}
where 
\begin{equation}
\Delta E_{n,m;k}=\frac{2 \lambda^2 N}{(1+\lambda^2)^2}\left [ \left (\frac{2k-n}{3}+2k-m\right ) \frac{1}{k}\, C_{n-1}^{k-1}\, C_{m-1}^{k-1}\right .
+\left .\frac{2}{3} \left ( C_{n-2}^{k-2}\, C_{m-1}^{k-1}- C_{n-1}^{k-1}\, C_{m-2}^{k-2}\right ) \right ]\ .
\end{equation} 
Up to the second order in $\lambda$ the mean energy is the sum $E_R+\Delta E_{R}$. We stress that we ignore here the mixing of  different transitions. The discussion of higher order terms is beyond the scope of the paper and will be given elsewhere. 

To calculate the variances of Hamiltonian \eqref{H_1} for $\alpha=1$ in sub-spaces with fixed value of $R$  it is necessary to select terms which do not change   the energy \eqref{E_0}. Hamiltonian \eqref{H_1} consists of two terms, the first one consists in the inversion of two-nearby spins,
\begin{equation}
\mathcal{H}_1^{(1)}=-\cos^2 \phi \sum_n \hat{\sigma}^x_n \hat{\sigma}^x_{n+1}
\label{H_1_1}
\end{equation} 
and the second changes the direction of only one spin
\begin{equation}
 \mathcal{H}_1^{(2)}=-\sin \phi \cos \phi \sum_n \, \hat{\sigma}^x_n( \hat{\sigma}^z_{n+1}+ \hat{\sigma}^z_{n-1})
\end{equation}
The later term always corresponds to  transitions with $n\to n \pm 1$ and $k\to k$ or $k\to k\pm 1$. It changes the energy of states and can be ignored when we are interesting in splitting of degenerated states.  The calculation of corresponding  variances of $\mathcal{H}_1^{(1)}$ in the sub-space can be done by straightforward combinatorics. The details are presented in \ref{ap_4}. The final result is 
 \begin{equation}
\sigma^2_R\equiv \langle \mathcal{H}^2\rangle_R=\frac{\lambda^4}{(1+\lambda^2)^2\, \mathcal{N}_R}\sum_{2k-n=R} \left [ N_a(n,k)+N_b(n,k)+N_c(n,k) \right ]
\label{sigma_R}
\end{equation}
with 
\begin{eqnarray}
N_a(n,k)&=&N\,  C_{n-1}^{k-1}\, C_{N-n-3}^{k}\ ,\qquad 
N_b(n,k)=N\,  C_{n-3}^{k-2}\,C_{N-n-1}^{k-1}\ ,\label{N_abc}\\
N_c(n,k)&=&2N\left [ C_{n-2}^{k-1}\, C_{N-n-2}^{k-2}+C_{N-n-2}^{k-1}\, C_{n-2}^{k-2}\right ]\ . \nonumber 
\end{eqnarray} 
Here subscripts $a$, $b$, and $c$ indicate different transitions with constant $R$:  (a) $n\to n-2$ and $k\to k-1$,   (b)  $n\to n+2$ and $k\to k+1$, (c)  $n\to n$ and $k\to k$.  Transition indicated by subscript $b$ are the inverse of transitions $a$, so  their total contributions to the variance are the same. 
 
\begin{figure}
\begin{minipage}{.48\linewidth}
\begin{center}
\includegraphics[width=.8\linewidth, angle=-90,clip]{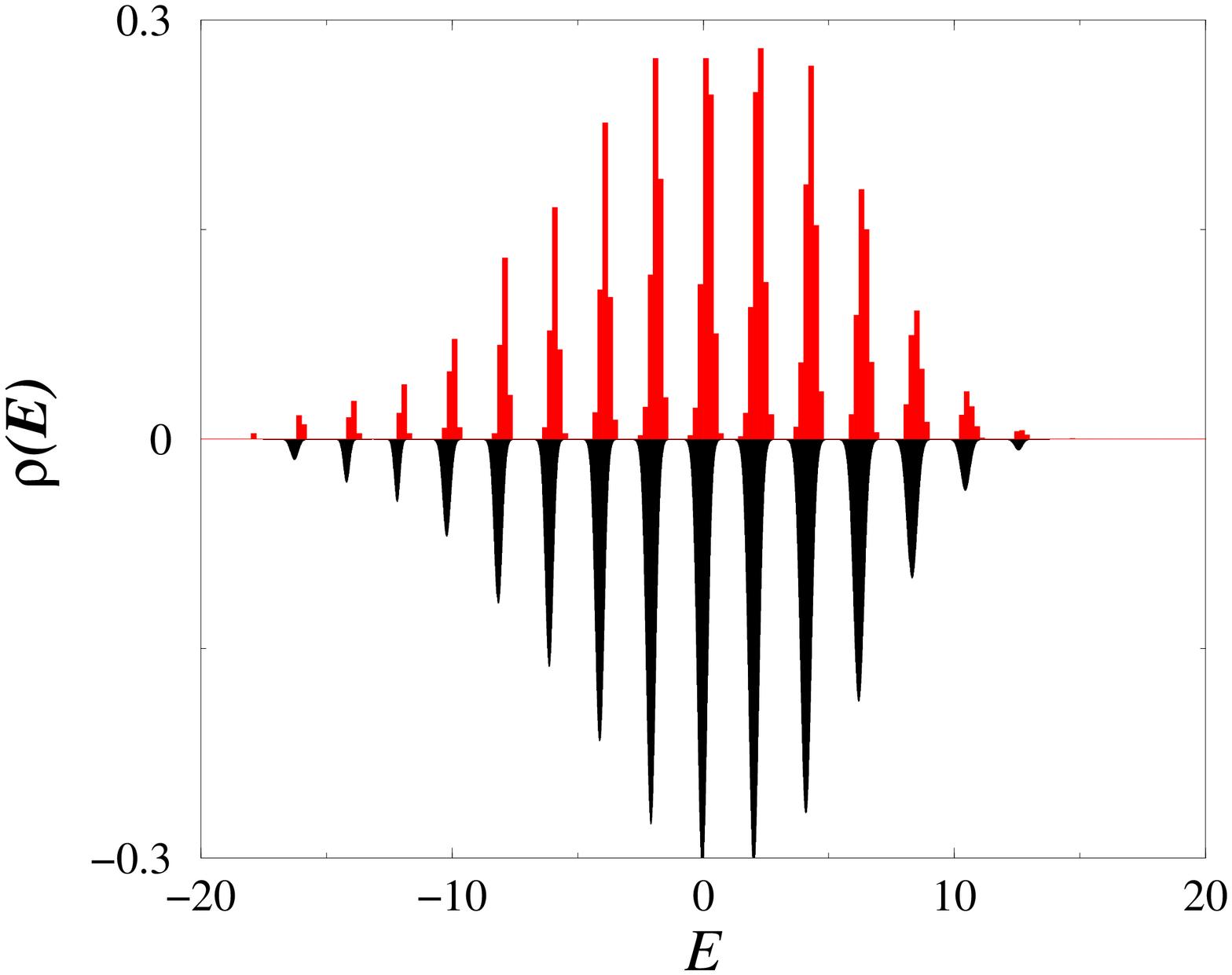}\\
(a)
\end{center}
\end{minipage}
\begin{minipage}{.48\linewidth}
\begin{center}
\includegraphics[width=.8\linewidth, angle=-90,clip]{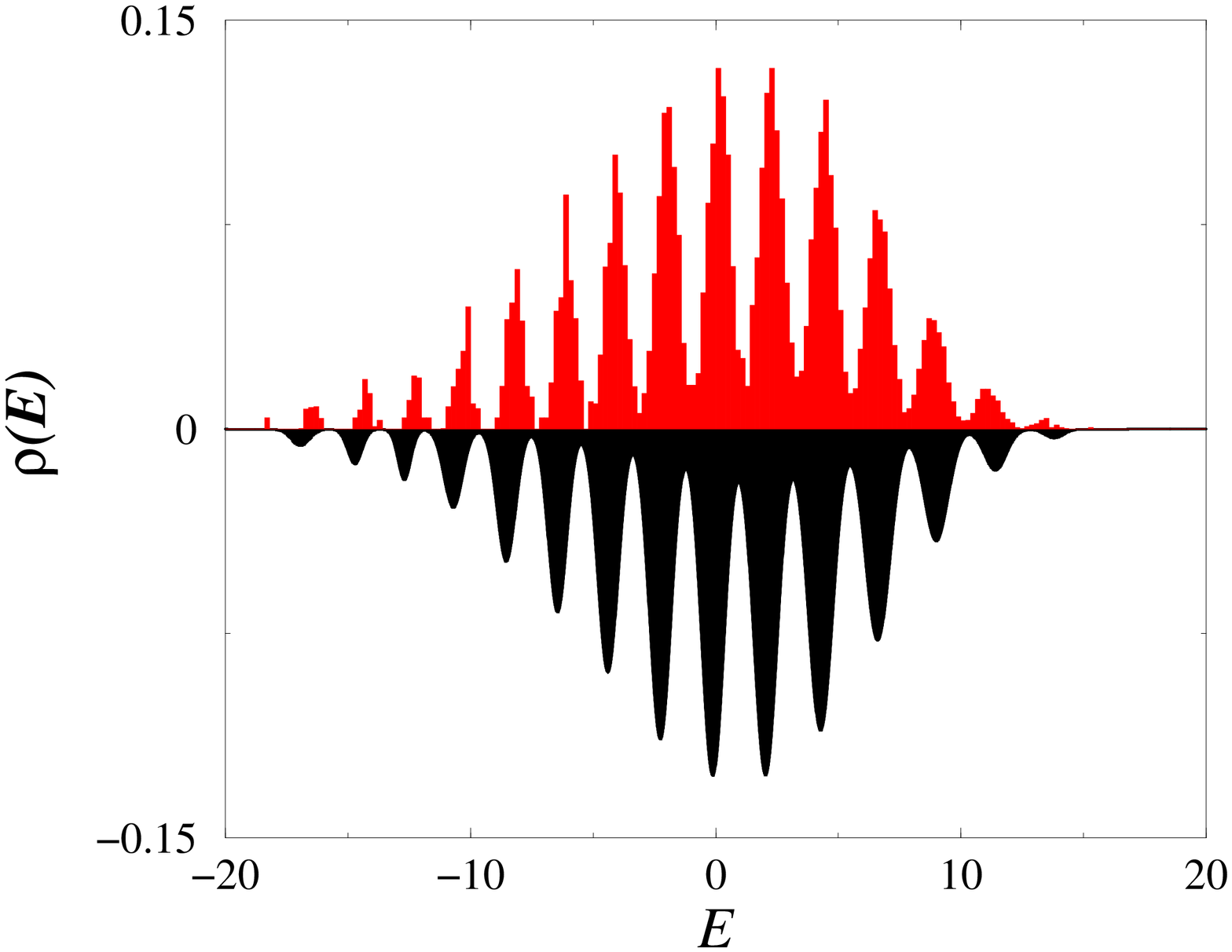}\\
(b)
\end{center}
\end{minipage}

\begin{minipage}{.48\linewidth}
\begin{center}
\includegraphics[width=.8\linewidth, angle=-90,clip]{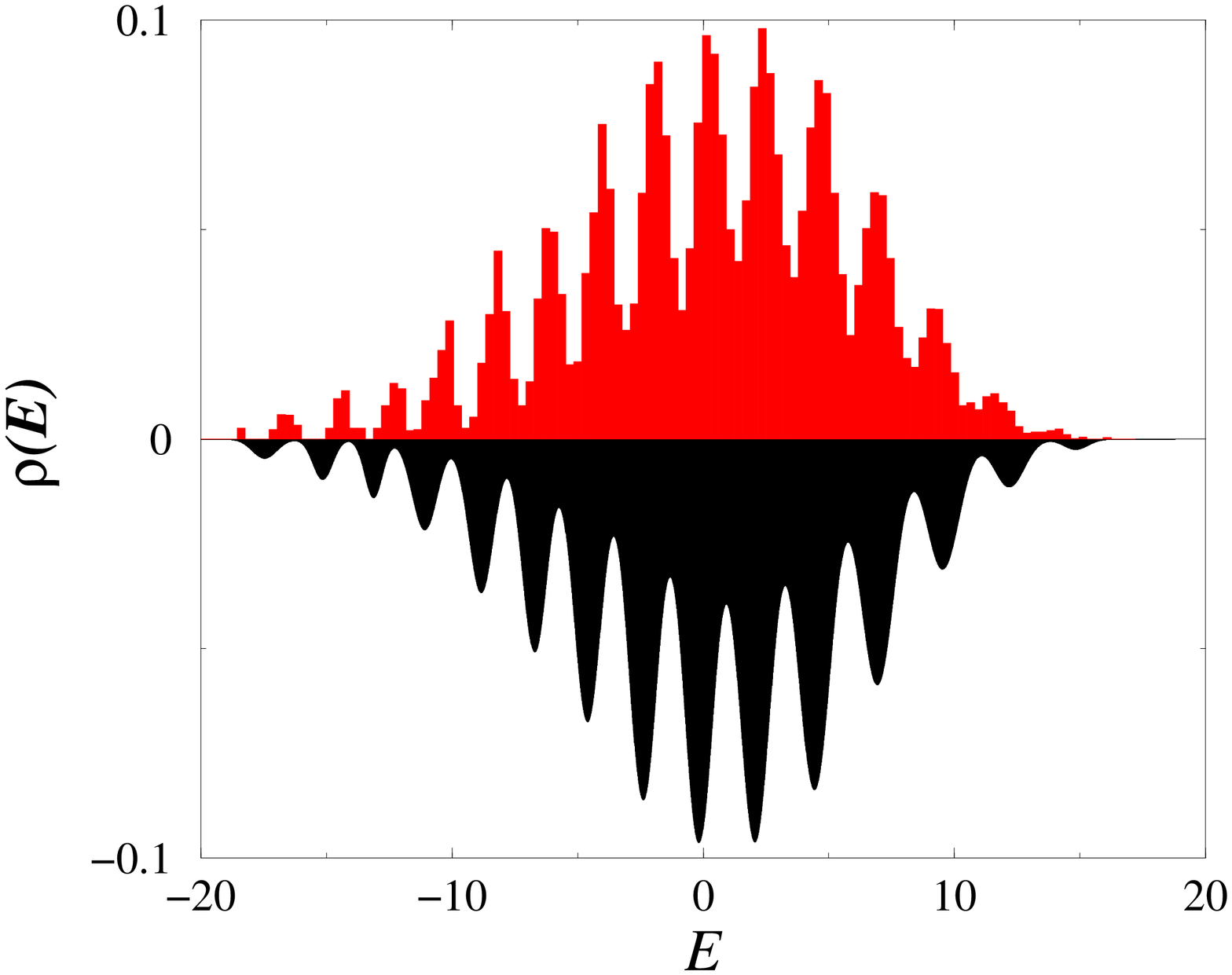}\\
(c)
\end{center}
\end{minipage}
\begin{minipage}{.48\linewidth}
\begin{center}
\includegraphics[width=.8\linewidth, angle=-90,clip]{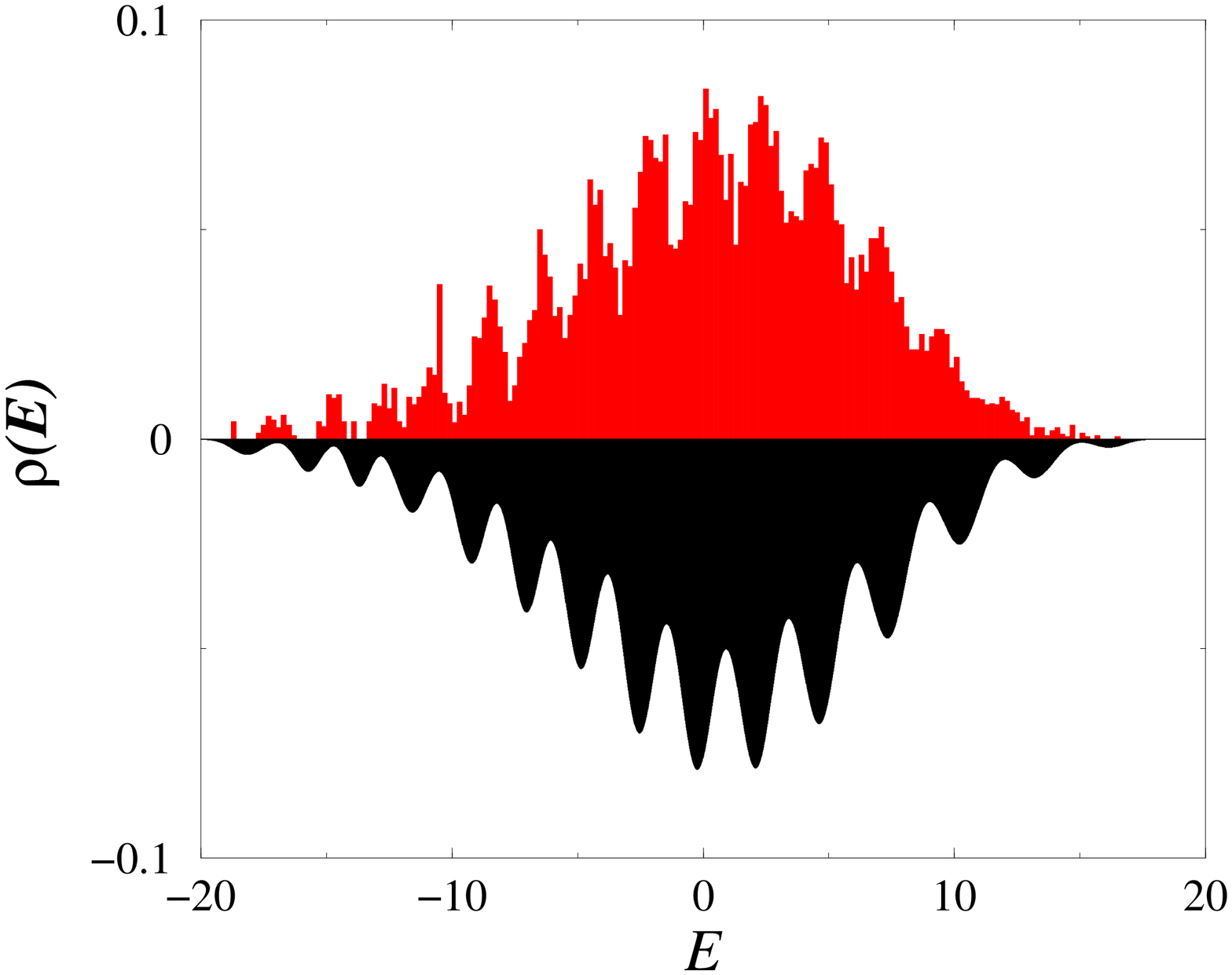}\\
(d)
\end{center}
\end{minipage}
\caption{Mean spectral density for the quantum Ising model in two fields with $\alpha=1$, $N=14$, and different $\lambda$. (a) $\lambda=0.3$, (b) $\lambda=0.5$, (c) $\lambda=0.6$, (d) $\lambda=0.7$. Upper red lines are histograms of numerically calculated densities. Lower black thick lines indicate the multi-Gaussian approximation \eqref{multi_gaussian_small}.  }
\label{fig_a_1_Ising}
\end{figure}

In the approximation of well separated peaks $R$ serves as a good quantum number and the spectral density in sub-space with fixed $R$ is the Gaussian
\begin{equation}
\rho_R(E)=\frac{1}{\sqrt{2\pi\sigma^2_R}}\exp \left ( -\frac{(E-E_R)^2}{2\sigma^2_R}\right )\ .
\end{equation}
In the leading order the total spectral density is the sum over all such Gaussians
\begin{equation}
\rho_{\mathrm{mG}}(E)=\frac{1}{2^N}\sum_R \mathcal{N}_R\,  \rho_R(E)\ .
\label{multi_gaussian_small}
\end{equation}
In figure~\ref{fig_a_1_Ising} the comparison between the spectral density of the Ising model with $\alpha=1$ and different values of $\lambda$ calculated numerically by direct diagonalization and by the muti-Gaussian approximation formula \eqref{multi_gaussian_small} is presented. The agreement is very good at small $\lambda$. Even at $\lambda=0.7$ main features of numerics are well reproduced by the simple formula \eqref{multi_gaussian_small}. Notice that from \eqref{E_R} and \eqref{sigma_R} it follows that for $\alpha=1$ peaks are visible when $N<N_{max}$ where $N_{max}$ is calculated from \eqref{criterium}
\begin{equation}
N_{max}\sim \frac{1}{\lambda^4}
\end{equation}
which explains the appearance of very prononced peaks in this case. 

For non-integer values of $\alpha$ different peaks in unperturbed density may come close to each other and one needs a careful calculation of a few terms perturbation series to get a good agreement with true spectral density at small $\lambda$. We consider here a slightly different approximation suitable for the both rational and irrational values of $\alpha$. 

Instead of representing the density as a simple sum over integer $R$ as in \eqref{multi_gaussian_small}, we propose to represent it as a double sum over $n$ and $k$
\begin{equation}
\rho_{\mathrm{mG}}(E)=\sum_{n=0}^N\sum_{k=1}^{n-1}f(n,k)\frac{1}{\sqrt{2\pi \sigma^2(n,k)}}\exp\left ( -\frac{(E-E(n,k))^2}{2\sigma^2(n,k)}\right )\ ,
\end{equation}
where $E(n,k)$ and $\sigma^2(n,k)$ are the first and the second moments of Hamiltonians \eqref{H_0} and \eqref{H_1} in the state with fixed $n$ and $k$.   We use unperturbed value \eqref{alpha_E_0} for $E(n,k)$:
\begin{equation}
E(n,k)=\alpha(N-2n)+4k-N.
\end{equation} 
The second moment can be calculated as in \ref{ap_4}. The difference is that for generic $\alpha$ one has to take into account transitions which conserve the both $n$ and $k$. In the notation of  \ref{ap_4} they corresponds to the case (c). From \eqref{N_c} together with \eqref{k_configurations} one finds
\begin{equation}
\sigma^2(n,k)= \frac{2\lambda^4 }{(\alpha^2+\lambda^2)^2 }\frac{k(k-1)(N-2k)}{(n-1)(N-n-1)} \approx \frac{2\lambda^4 }{(\alpha^2+\lambda^2)^2 }\frac{k^2(N-2k)}{n(N-n)}\ .
\end{equation}
The last expression coresponds to large values of $n$ and $k$. It is this formula that we use in calculations.    
 
\begin{figure}
\begin{minipage}{.48\linewidth}
\begin{center}
\includegraphics[angle=-90, width=.9\linewidth]{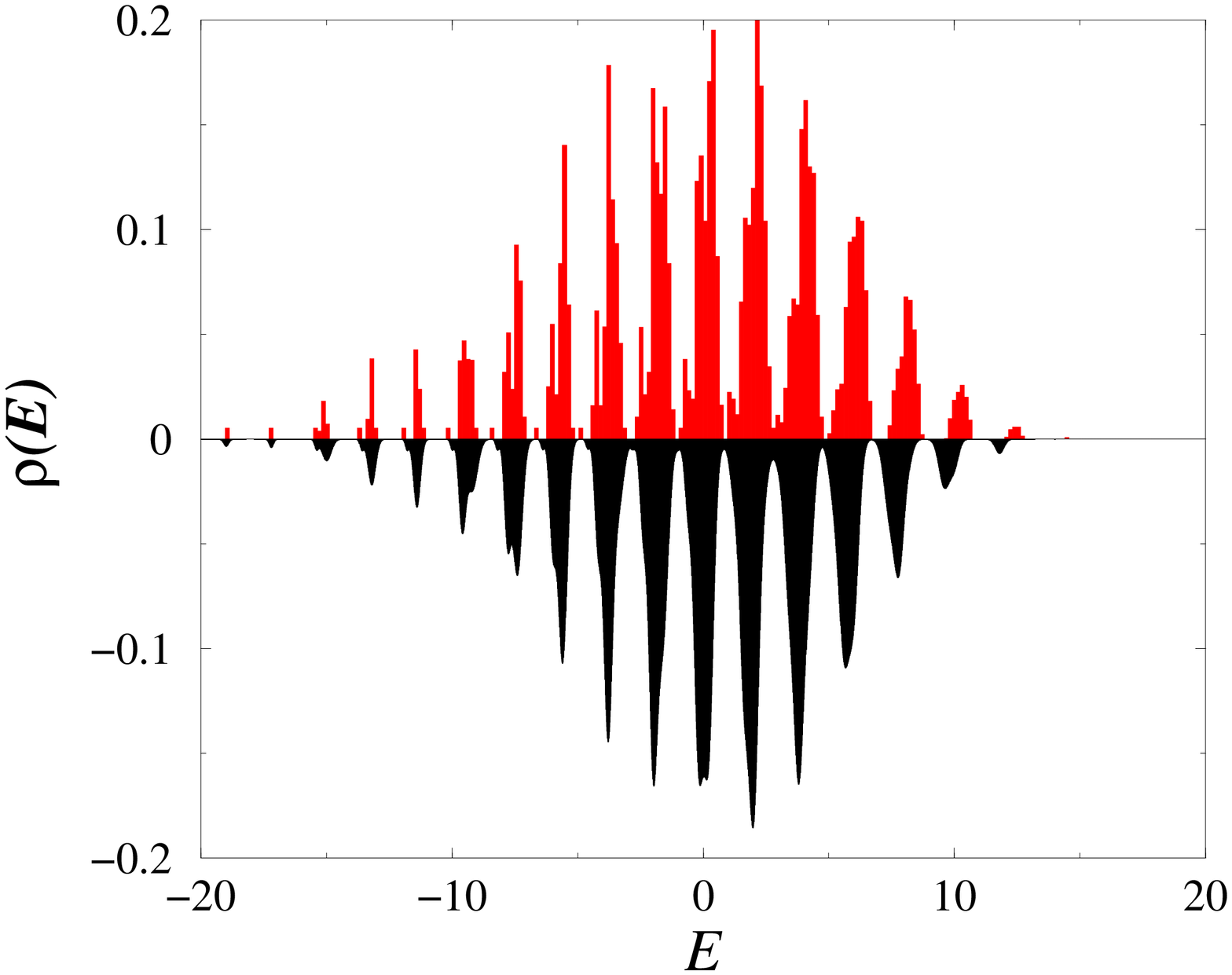}\\
(a)
\end{center}
\end{minipage}
\begin{minipage}{.48\linewidth}
\begin{center}
\includegraphics[angle=-90,  width=.9\linewidth]{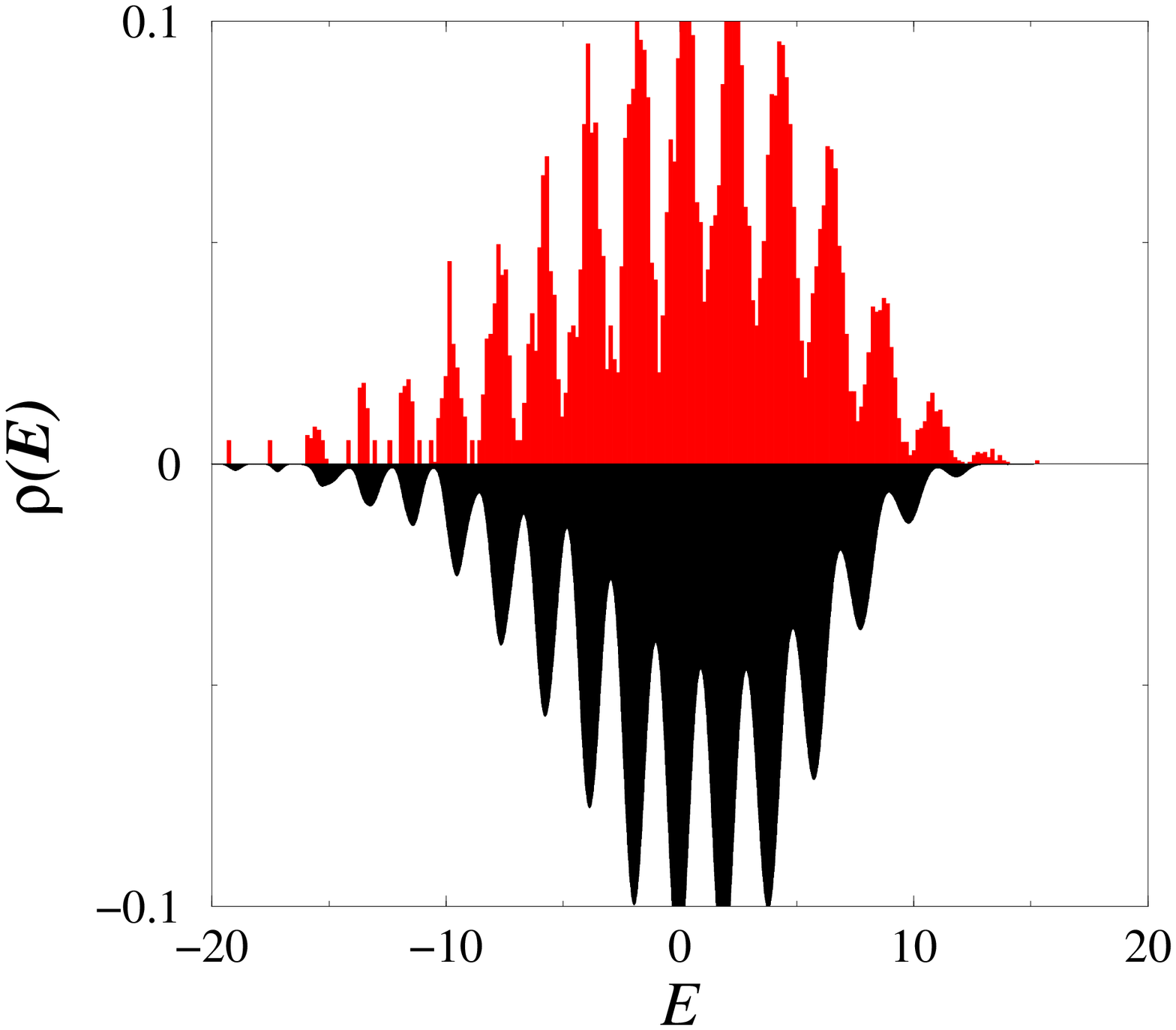}\\
(b)
\end{center}
\end{minipage}

\begin{minipage}{.48\linewidth}
\begin{center}
\includegraphics[angle=-90, width=.9\linewidth]{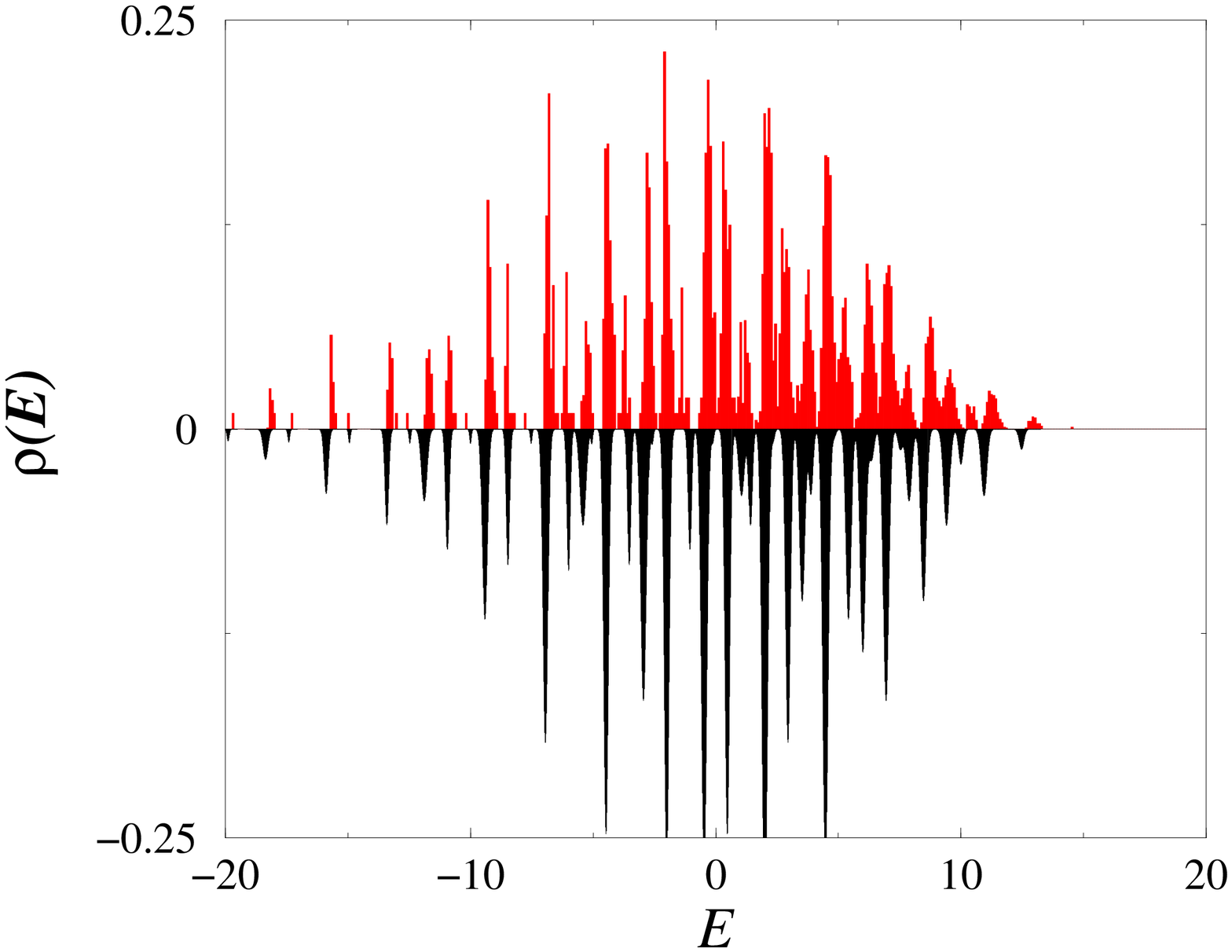}\\
(c)
\end{center}
\end{minipage}
\begin{minipage}{.48\linewidth}
\begin{center}
\includegraphics[angle=-90,  width=.9\linewidth]{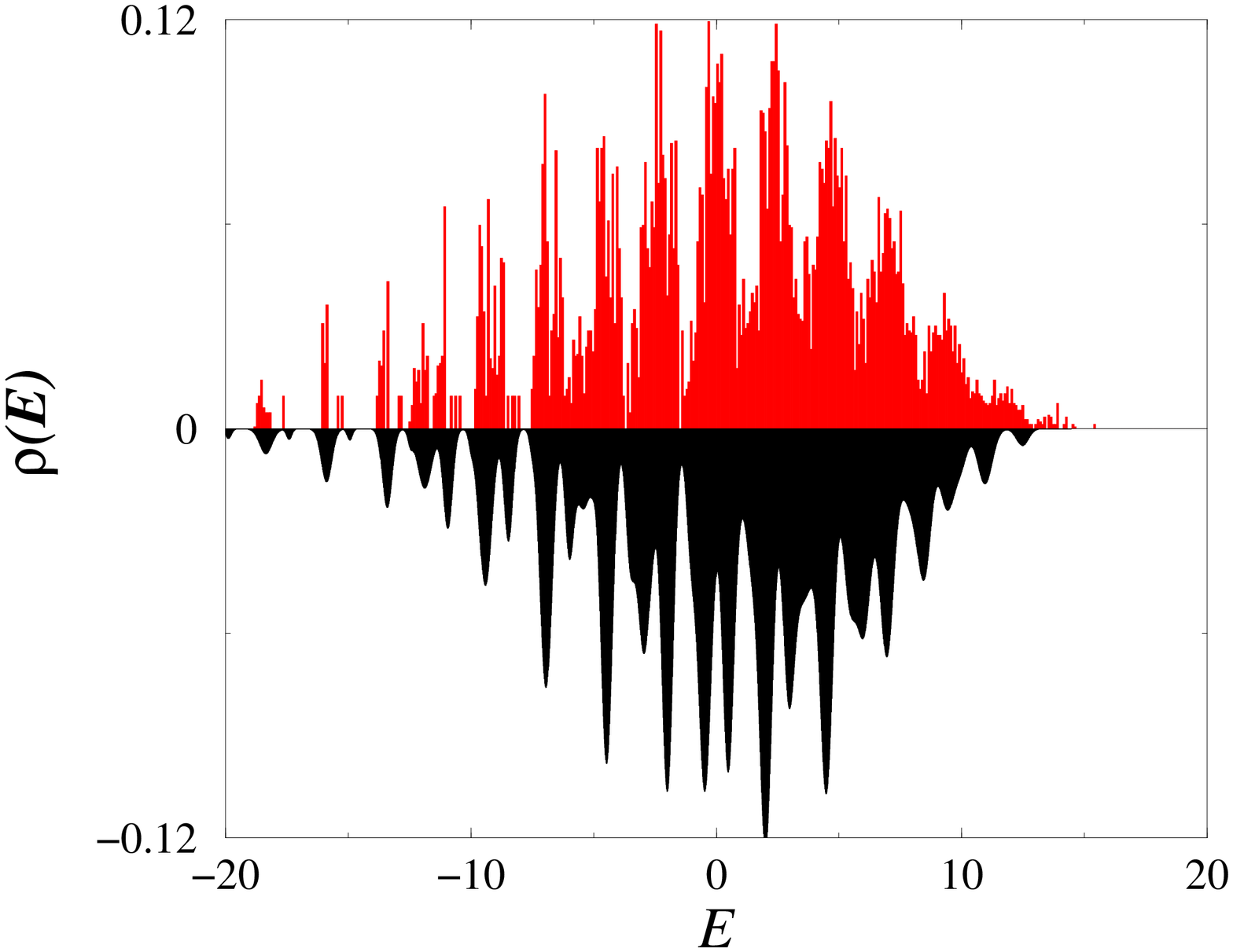}\\
(d)
\end{center}
\end{minipage}
\caption{The same as in figure~\ref{fig_a_1_Ising} but with different values of $\alpha$ and $\lambda$:  (a) $\alpha=0.9$, $\lambda=0.3$; (b)  $\alpha=0.9$, $\lambda=0.5$; (c)  $\alpha=\sqrt{5}-1$, $\lambda=0.3$; (d) $\alpha=\sqrt{5}-1$, $\lambda=0.5$.  }
\label{golden}
\end{figure}


\section{Conclusion}\label{conclusion}

We investigate spectral density of one-dimensional spin chains. As a typical example we consider the quantum Ising model in transverse and longitudinal fields. When all coupling constants are of the same order, the spectral density in the bulk is well described by the Gaussian whose parameters are calculated from the knowledge of the two first moments of the full Hamiltonian. This Gaussian shape of the density is typical for many-body models with short-range interactions in thermodynamic limit when the number of particles tends to infinity. For large but finite number of particles, there are two main types of corrections to such asymptotic result. The first corresponds to power corrections which can be calculated from the third and higher moments of the full Hamiltonian. The second corrections are of a different nature. They are related with strong degeneracies of unperturbed spectrum in certain limits of coupling constants. For a small deviations from these limits the degeneracies are lifted and finite-width peaks appear. We demonstrate that in the leading approximation these peaks can be described by a superposition of different Gaussians functions with parameters computed from the first and the second moments of the Hamiltonian projected to certain  sub-spaces.  The resulting formulae are simple but agree well with numerical calculations in the bulk. The reason of peaks existence is the application of perturbation theory for many-body systems. Usually (see e.g. \cite{deutsch_1}, \cite{deutsch_2}) one argues that for many-body systems with arbitrary small interactions perturbation series cannot be applied due to exponentially dense spectrum of unperturbed states. For systems considered this conclusion is not correct because by construction these models may have strong degeneracies.      

\begin{figure}
\begin{minipage}{.48\linewidth}
\begin{center}
\includegraphics[angle=-90, width=.9\linewidth]{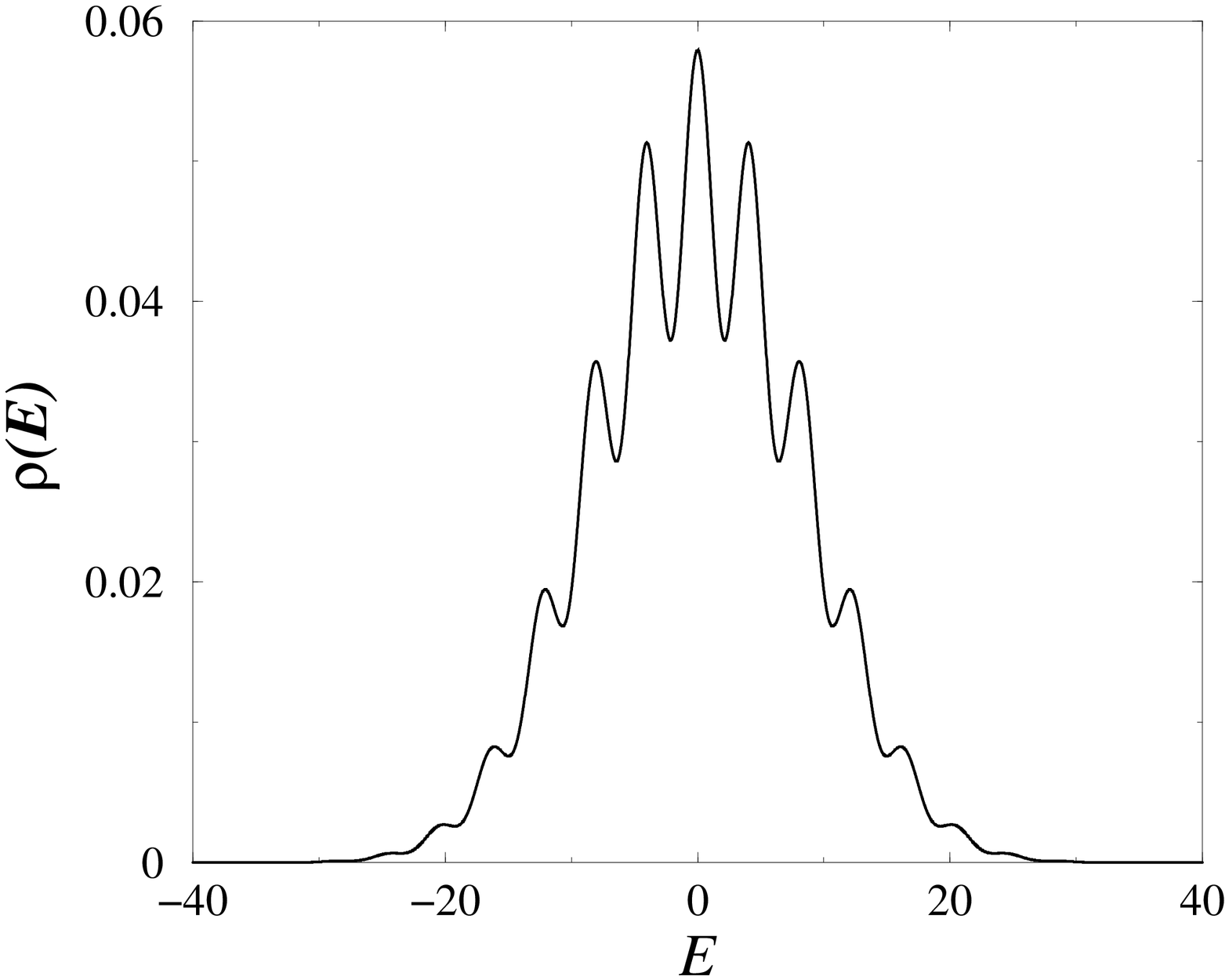}\\
(a)
\end{center}
\end{minipage}
\begin{minipage}{.48\linewidth}
\begin{center}
\includegraphics[angle=-90,  width=.9\linewidth]{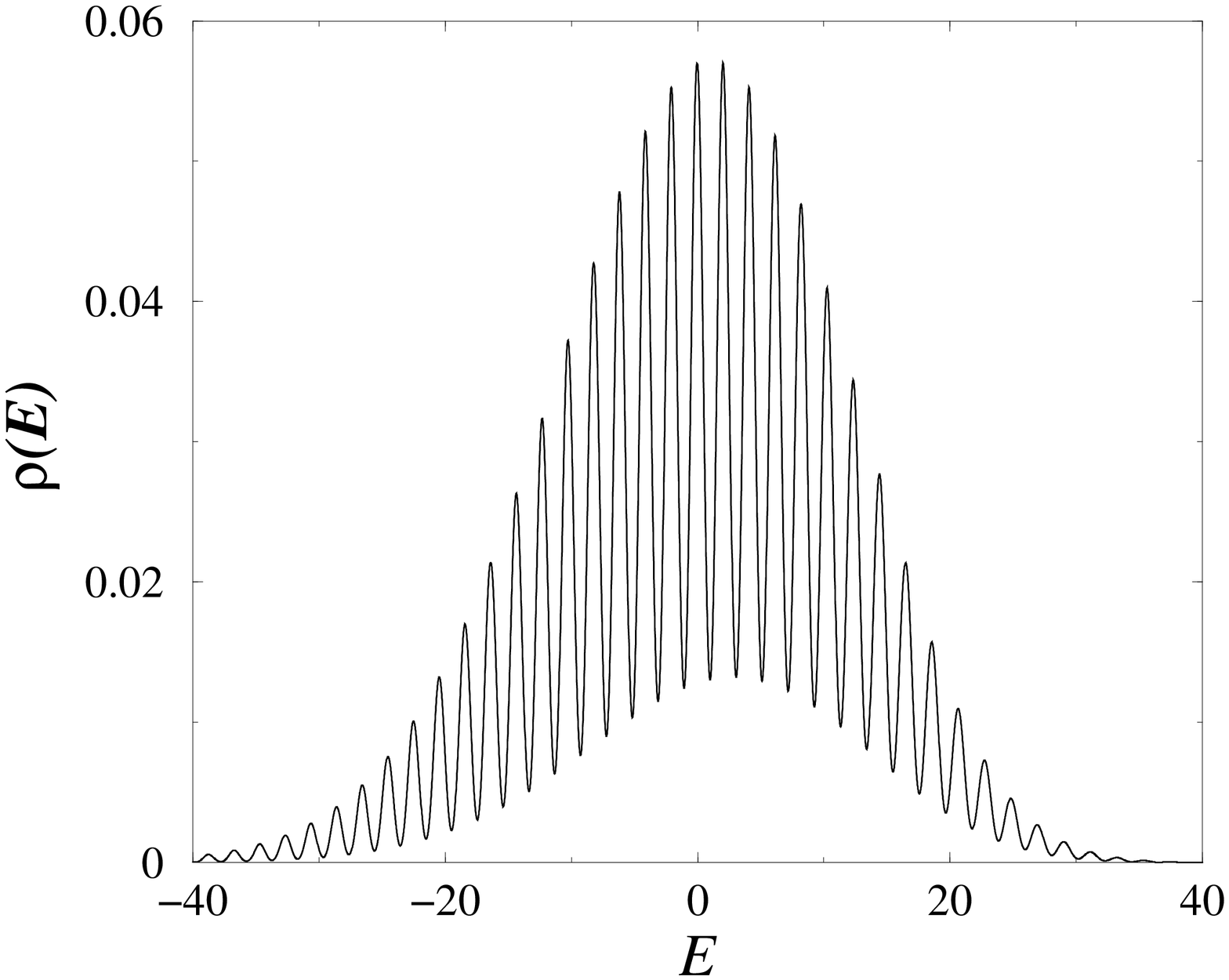}\\
(b)
\end{center}
\end{minipage}
\caption{Predictions for spectral density for the quantum Ising model with $N=64$ spins and: (a) formula (\ref{rho_lambda_less_1}) with $\alpha=0$, $\lambda=\tfrac{1}{4}$. (b) Formula (\ref{multi_gaussian_small}) with  $\alpha=1$, $\lambda=\tfrac{1}{3}$.  }
\label{64}
\end{figure}

The existence of many peaks in spectral density is quite robust phenomenon. Though for very large number of spins these peaks disappear, for number of spins accessible in today numerical calculations and reasonable coupling constants the density does contain clear peaks. For example, in figure~\ref{64} we present predictions for spectral densities of the Ising model with $N=64$ spins which is far beyond  dimensions reachable in direct diagonalization.  It is clearly seen that density is not a smooth Gaussian-type function but, instead, contains numerous identifiable peaks. 

Despite the fact that we consider only the Ising model in two fields our results should be valid for a large class of models with local interactions e.g. the Bose-Habbard model where spectral density also have pronounced peaks at certain limits of coupling constant \cite{roux}. In general,  peaks in spectral densities may and  will influence many physical (and statistical) properties of models considered. In particular,  they may alter deduction of large $N$ behaviour from finite $N$ calculations.   We think that further investigation of such mesodynamics (as opposite to thermodynamics) phenomena are of importance.          

\ack
The authors are  grateful to Guillaume Roux for numerous discussions and for pointing out Ref.~\cite{roux}. YYA is supported by the CFM foundation.


\appendix

\section{The spectrum of the Ising model }\label{ap_0}

The purpose of this Appendix is to discuss in detail the construction of energy levels $\left\lbrace E_{p},\; 0\leq p \leq 2^{N}-1 \right\rbrace$ for the Ising model in transverse field (\ref{H_Ising}) with finite value of the number of spins $N$.  This construction is rarely discussed in literature mostly due to the interest in the thermodynamic limit $N \to \infty$. 

As it is well known, the Ising model in transverse field  is integrable by the Jordan-Wigner transformation which maps the spin system into a quadratic fermion problem (see e.g.  \cite{lieb}, \cite{mattis})
\begin{equation}
a_n=\Big (\prod_{j=1}^{n-1} \sigma_j^z \Big )\sigma^{+}_n\ , \quad a_n^{\dag}=\Big (\prod_{j=1}^{n-1} \sigma_j^z \Big )\sigma^{-}_n \ ,\quad \sigma_n^z=1-2a_n^{\dagger}a_n \ ,\quad \sigma_n^{\pm}=\tfrac{1}{2}(\sigma_n^x\pm \mathrm{i}\sigma_n^y)\ .
\end{equation}
 Here $a_n$ and $a_n^{\dagger}$ are fermionic annihilation and creation   operators (spin up corresponds to the absence of fermion).
  
When one imposes periodic boundary conditions for spin operators  ($\sigma_{N+1}^{\alpha}=\sigma_{1}^{\alpha}$, $\alpha=x,y,z$),  boundary conditions for fermions have to be fixed as follows
\begin{equation}
a_{N+1}=-\mathcal{P}a_1\ ,
\end{equation}
where $\mathcal{P}$ is conserved spin-parity operator
\begin{equation}
 \mathcal{P}=\prod_{n=1}^{N}\sigma_{n}^{z}=(-1)^{\mathcal{N}_f}
 \label{parity_number_op_relation}
\end{equation}
and $\mathcal{N}_f$ is the total number of $a$-fermions
\begin{equation}
\mathcal{N}_f=\sum_{n=1}^Na_n^{\dagger}a_n\ .
\end{equation}
To diagonalize the resulting fermionic Hamiltonian one has first to perform the Fourier transform of $a_n$, i.e. introduce new fermionic operators, $b_k$
\begin{equation}
b_k=\frac{1}{\sqrt{N}}\sum_{n=1}^N\mathrm{e}^{\mathrm{i}k\,n}a_n\ .
\end{equation}
From boundary conditions it follows that values of $k$ have to be quantized,
\begin{equation}
k_j=\frac{\pi}{N}\left \{ \begin{array}{cc} 2j, & \mathcal{P}=-1\\ 2j+1,&\mathcal{P}=1\end{array}\right .\ , \qquad j=0,1,\ldots, N-1 \ .
\label{momenta}
\end{equation}
Notice that $\mathcal{N}_f=\sum_{k}^Nb_k^{\dagger}b_k$.

After such Fourier transform the Ising Hamiltonian takes the form
\begin{equation}
\mathcal{H}=2\sum_{k} (\lambda-\cos k)\left (b_k^{\dag}b_{k}-\tfrac{1}{2}\right) 
+\mathrm{i} \sum_{k} ( b_{k}^{\dag}b_{-k}^{\dag} -b_{-k} b_{k})\sin k \ .
\label{fourier}
\end{equation}
The last step consists in performing  a Bogoliubov transformation which mixed $b_k$ and $b_{-k}^{\dagger}$ operators ($b_{-j}\equiv b_{2N-j-1}$)
\begin{equation}
\eta_k = \cos \theta_k\, b_k-\mathrm{i}\sin \theta_k\,  b_{-k}^{\dagger},\qquad \tan 2\theta_k=\frac{\sin k}{\cos k-\lambda}\ .
\label{bogoliubov}
\end{equation}
In terms of these $\eta_k$ operators the Hamiltonian is diagonalized,
\begin{equation}
\mathcal{H}=\sum_{j} e_j\left( n_j-\frac{1}{2}\right)\ ,\qquad n_j=\eta_{k_j}^{\dagger}\eta_{k_j}\ ,
\end{equation}
and $e_j$ are determined in \eqref{one_particle_levels}. The spectrum of the Ising chain is obtained by adding excitations with different impulsions to the ground state energy 
\begin{equation}
E_{0}=-\frac{1}{2}\sum_{k} \Lambda_{k}.
\end{equation}  
As the parity is an integral of motion, the number of excitations added has to be choosen carefully in order to fulfill the condition (\ref{parity_number_op_relation}).

\subsection*{Anti-periodic boundary conditions}

Anti-periodic boundary conditions mean that momenta of excitations are  proportional to odd integers (see \eqref{momenta}).
From the above discussion it follows that the spin-parity $\mathcal{P}=1$.  Then for anti-periodic boundary conditions the total number of excitations has to be even.  

By definition, the Bogoliubov vacuum, $| \Psi_{0}^{\mathrm{ap}}\rangle$, is given by the condition that $\eta_{k}| {\Psi_{0}^{\mathrm{ap}}}\rangle =0$. As $\sin k \neq 0 $ for all these allowed values of $k$, the Bogoliubov vacuum in terms of $b$ operators has the form
\begin{equation}
| \Psi_{0}^{\mathrm{ap}}\rangle =\prod_{k>0}\left(\cos \theta_{k}+\mathrm{i}\sin \theta_{k} \,
b_k^{\dagger}b_{-k}^{\dagger}\right)| 0\rangle \ . 
\end{equation} 
where   $| 0 \rangle $ is the $b$ vacuum, i.e $b_{k}| 0 \rangle =0$. Therefore $|{\Psi_{0}^{\mathrm{ap}}}\rangle $ is composed with an even number of $b$ particles created by pairs $b_{k}^{\dagger}b_{-k}^{\dagger}$ of opposite non-zero momenta. Consequently, in order to fulfil the relation (\ref{parity_number_op_relation}) one has to consider addition of all combination of  even number of excitations to $E_{0}$ to reconstruct half of the eigen-energies. The other half comes from the sector of parity $\mathcal{P}=-1$ which corresponds to periodic boundary conditions. 

\subsection*{Periodic boundary conditions}

Excitations momenta are now given by even integers as in \eqref{momenta} and the total number of excitations has to be odd (for even $N$). 

In this sub-space $\sin k=0$  for $k=0$ and $\pi$, i.e. for $j=0$ and $j=N/2$ (for even $N$). 
Hamiltonian \eqref{fourier} is diagonal for $b$ fermions with these two values of $k$ and their energies  are
\begin{equation}
E_0 =2(\lambda-1)(b_0^{\dagger}b_0-\tfrac{1}{2}),\qquad E_{\pi}=2(1+\lambda)(b_{\pi}^{\dagger}b_{\pi}-\tfrac{1}{2})\ .
\end{equation}
Therefore,  the Bogoliubov transformation \eqref{bogoliubov} has to be performed only over all other values of momenta. 

If $\lambda>1$, the number of  excitations of momenta $0$ and $\pi$ have to be zero in the lowest energy state and the Bogoliubov vacuum is 
\begin{equation}
| {\Psi_{0}^{\mathrm{p}}}\rangle = \prod_{\substack{k>0\\ k\neq \pi }}\left(u_{k}+\mathrm{i}v_{k}b_{k}^{\dagger}b_{-k}^{\dagger}\right)| {0}\rangle \ .
\end{equation}
This vacuum  contains even number of $b$ particles and only odd number of excitations (i.e. vacuum energy  itself is not in the spectrum) is allowed. 

If $\lambda<1$, energy of the  excitation with momentum $0$ is negative, therefore it should not be included in  the ground state (with $\eta_0=b_0^{\dagger}$). The vacuum now has the form  
\begin{equation}
| {\Psi_{0}^{\mathrm{p}}}\rangle \equiv b_{0}^{\dagger} \prod_{\substack{k>0\\ k\neq \pi}}\left(u_{k}+\mathrm{i}v_{k}b_{k}^{\dagger}b_{-k}^{\dagger}\right)| {0}\rangle \ . 
\end{equation}
Such vacuum contains odd number of $b$ fermions and one can add only even number of excitations.

Finally, for positive spin parity one has to add odd number of excitations when $|\lambda|>1$ and even number of excitations when $|\lambda|<1$. 
 

\section{Calculation of the first two moments of the Ising Hamiltonian in the fixed-$n$ sub-space }\label{ap_1}

The full Hamiltonian of the Ising model in two fields \eqref{ising_2} is by construction  the sum of two Hamiltonians 
\begin{equation}
\mathcal{H}=\mathcal{H}_0+\mathcal{H}_1\ ,
\end{equation}
where $\mathcal{H}_0$ is determined by \eqref{H_0} and $\mathcal{H}_1$ by \eqref{H_1}.
 
\begin{figure}
\begin{center}
\includegraphics[width=.7\linewidth]{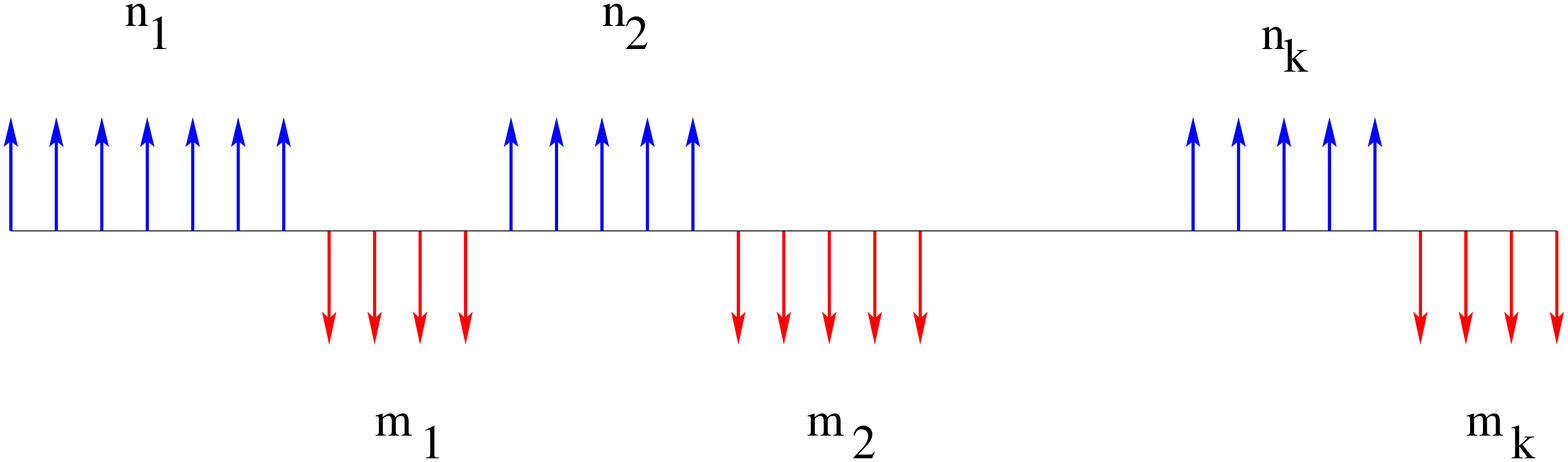}
\end{center}
\caption{Typical structure of a state with different directions of spins. Blue arrows up indicate spins with $\sigma^z=1$ and red arrow down indicate spins with $\sigma^z=-1$. }
\label{structure}
\end{figure}

Our purpose is to calculate the first and the second moments of this Hamiltonian in the sub-space of states with fixed number of spins up. The total number of such states is $C_N^n$ where  $n$ is the number of spins up.  Any of such states can be represented as a sequence of $k$ groups with $n_j$ spins up followed by $m_j$ spins down (see figure~\ref{structure}). 
We indicate such state by $\Psi_{n,k}$. Notice that there exist many states with fixed values of $n$ and $k$.  
 
By construction all $n_j$ and $m_j$ are bigger or equal to $1$, $n_j\geq 1$, $m_j\geq 1$, and their sum is fixed  
\begin{equation}
\sum_{j=1}^kn_j=n,\qquad \sum_{j=1}^km_j=m,\qquad n+m=N \ .
\end{equation}
Hamiltonian $\mathcal{H}_0$ is diagonal in this basis and in $\mathcal{H}_1$ (see \eqref{H_1}) only the term  $\sin^2 \phi \sum_n \hat{\sigma}^z_n \hat{\sigma}^z_{n+1}$ gives non-zero contribution to the first moment since other terms flip spins. Therefore in any state with fixed $n$ and $k$
\begin{equation}
\langle \mathcal{H} \rangle_{n,k}=\sqrt{\lambda^2+\alpha^2}(N-2n)-(N-4k) \sin^2\phi\ ,
\label{mean_two_fields}
\end{equation}
where $\phi$ is as in \eqref{hat_sigma} and $\langle \cdots \rangle_{n,k}$ denotes $ \langle\Psi_{n,k}| \cdots |\Psi_{n,k}\rangle$. 

The next step is to calculate the variance in the subspace with fixed $n$. It means that in the intermediate states (i.e. in $\langle \Psi_{\sigma^{\prime}}|\mathcal{H}_1|\Psi_{\sigma}\rangle  $) one has to take into account only those states $\Psi_{\sigma^{\prime}}$  which have exactly the same number $n$ of spins up as in $\Psi_{\sigma}$. By examining $\mathcal{H}_1$ it appears that only the term $\hat{\sigma}^x_n \hat{\sigma}^x_{n+1}$ may conserve $n$. It is evident that only the flipping of boundary spins separating blocks of spins down and up does not change energy. The number of such transition is $2k$ where $k$ is the number of groups of spins in the same directions. Therefore, 
\begin{equation}
\langle \mathcal{H}^2 \rangle_{n,k}-\langle \mathcal{H} \rangle_{n,k}^2\equiv \sigma_{n,k}^2= 2k \cos^4\phi .
\label{sigma_two_fields}
\end{equation}
The last step consists in finding the number of states with fixed values of $n$ and $k$ which we denote by $f(n,k)$. It can be calculated as follows. 

The number of cases when $n_1+n_2+\ldots+n_k=n$ and $n_j\geq 1$ are well known (and can easily be obtained for instance  by a generating function method)
\begin{equation}
\mathcal{N}(n,k)=C_{n-1}^{k-1}.
\end{equation}
The same is true for $m_j$ 
\begin{equation}
\mathcal{N}(m,k)=C_{m-1}^{k-1}.
\end{equation}
Total number of possibilities is the product of these two expressions. But it corresponds to a fixed initial point. Summing over all points is equivalent to multiplication by $N$. Doing this we get cyclic permuted configurations $k$ times, therefore, the result has to be divided by $k$. It gives 
\begin{equation}
f(n,k)=\frac{N}{k}C_{n-1}^{k-1}C_{N-n-1}^{k-1}.
\label{k_configurations}
\end{equation}
This formula is valid provided $k\neq 0$. There is two cases with $k=0$. One with $n=N$ when all spins are up and one with $n=0$ when all spins are down. When one uses this formula for calculating the sum over all $k$ of a polynomial of $k$ the constant term gives $C_N^n$ which is the correct answer for $n=0$  and $n=N$. Other terms are powers of $k$ and are absent when $k=0$. Therefore one can sum over $k$ from 1 to $n$ and then sum over all $n$ from $0$ to $N$.  
  
The above formulae has been calculated for states with fixed value of $n$ and $k$. But the unperturbed energy \eqref{unperturbed_E} depends only on $n$, so to get the first two moments of the Hamiltonian in the sub-space of fixed $n$ one has to sum over all possible $k$
\begin{equation}
\langle \mathcal{H} \rangle_{n}=\frac{1}{C_N^n}\sum_{k=0}^n f(n,k)\langle \mathcal{H} \rangle_{n,k},\qquad \sigma^2_n =\frac{1}{C_N^n}\sum_{k=0}^n f(n,k)\sigma_{n,k}^2
\end{equation} 
Using the  relations 
\begin{equation}
\sum_{k=1}^nf(n,k)=C_N^n,\qquad \sum_{k=1}^n kf(n,k)=\frac{n(N-n)}{N-1}C_{N}^{n}
\end{equation}
one finds that the first moment of the Ising Hamiltonian and its variance calculated in the sub-space with $n$ spins up are
\begin{eqnarray}
\langle \mathcal{H} \rangle_{n}&=& \sqrt{\lambda^2+\alpha^2}(N-2n)-(N-4\bar{k}(n) ) \sin^2\phi \label{mean_two}\\
\sigma_n^2 & =  &2\bar{k}(n) \cos^4\phi 
\label{sigma_n_two_fields}
\end{eqnarray}
where $\bar{k}(n)$ is the mean value of $k$ over states with fixed $n$
\begin{equation}
\bar{k}(n)\equiv \frac{1}{C_N^n}\sum_{k=1}^n kf(n,k)=\frac{n(N-n)}{N-1}
\end{equation} 


\section{More careful approach by using the XX model}\label{a p_3}

The Hamiltonian of the Ising model in two fields contains terms of different nature. There exists a part of this Hamiltonian which does not change the total number of spins up, $n$, and the rest which changes this number. When we are interested in the splitting of initially degenerated levels, these two parts act differently. The latter can be taken into account in the usual perturbation series as it has been done in \ref{ap_2}. The principal problem is the $n$ conserving part as no simple perturbation expansion is possible. The standard approach is to diagonalise this part to find correct energies and eigenfunctions in the zeroth order. What we did is to avoid the direct diagonalisation. Our main point is that, though we do not know in general how eigenvalues split exactly, we do know the first and the second moment of $n$-conserving Hamiltonian and as has been discussed in previous Sections the distribution of  eigenvalues of projected Hamiltonian has to be close to the Gaussian function. To check it analytically it is necessary to calculate  high moments of this projected Hamiltonian in the limit $N\to\infty$ and see if they correspond to the Gaussain values. We are unaware of direct combinatorial proof of this statement. But for the Ising model in two fields it can be done by using the XX-model. Indeed, for the Ising model the projected Hamiltonian  in the leading order has the form  (cf. \eqref{H_0} and \eqref{H_1})
\begin{equation}
\mathcal{H}_n=-\sqrt{\lambda^2+\alpha^2}\sum_{p}\hat{\sigma}_p^z-\cos^2\phi \sum_{p}\left [\hat{\sigma}_p^x \hat{\sigma}_{p+1}^x\right ]_n
\label{n_conserving_H}
\end{equation}
where $[ \ldots ]_n$ denotes the $n$-conserving part of the corresponding expression.  But it is easy to check that
\begin{equation}
[\sigma_p^x \sigma_{p+1}^x  ]_n=\frac{1}{2}[\sigma_p^x \sigma_{p+1}^x+\sigma_p^y\sigma_{p+1}^y]
\end{equation}
under conditions that the total $z$-spin is $n$. 

The last model is the well known XX-model (see e.g. \cite{lieb})
\begin{equation}
\mathcal{H}_n=-\cos^2\phi \left (\lambda^{\prime}\sum_{p}\sigma_p^z+\sum_{p}\frac{1}{2}\left [\sigma_p^x\sigma_{p+1}^x +\sigma_p^y\sigma_{p+1}^y\right ]\right ),
\qquad \lambda^{\prime}=\frac{\sqrt{\lambda^2+\alpha^2}}{\cos^2\phi}
\end{equation}
which has an exact solution. It conserves the total $z$-spin and can be solved by the Wigner-Jordan transformation. The resulting energies are
\begin{equation}
E=\cos^2\phi \sum_{n_j}2e_j\left (n_j-\frac{1}{2}\right )
\end{equation}
where for $\lambda^{\prime} >1$, $e_j=\lambda^{\prime}-\cos \phi_j$ and $n_j=0,1$. Of course, one has to fix the total number of excitations, $\sum_j n_j=n$

The discussion can be performed as above and using \eqref{mean_n} and \eqref{variance_n} one concludes that, indeed, the moments of the Hamiltonian \eqref{n_conserving_H} are Gaussian with the first moment
\begin{equation}
\langle E\rangle= \sqrt{\lambda^2+\alpha^2}(N-2n)
\end{equation}
and the variance  
\begin{equation}
\sigma_n^2\equiv \langle E^2\rangle-\langle E\rangle^2= \frac{2n(N-n)}{N-1}\cos^4\phi 
\end{equation}
which  are the same as what we have used above. 


\section{Calculations of variances for the Ising model in two fields with $\alpha=1$ and small $\lambda$}\label{ap_4} 

\begin{figure}
\begin{minipage}{.3\linewidth}
\begin{center}
\includegraphics[width=.8\linewidth]{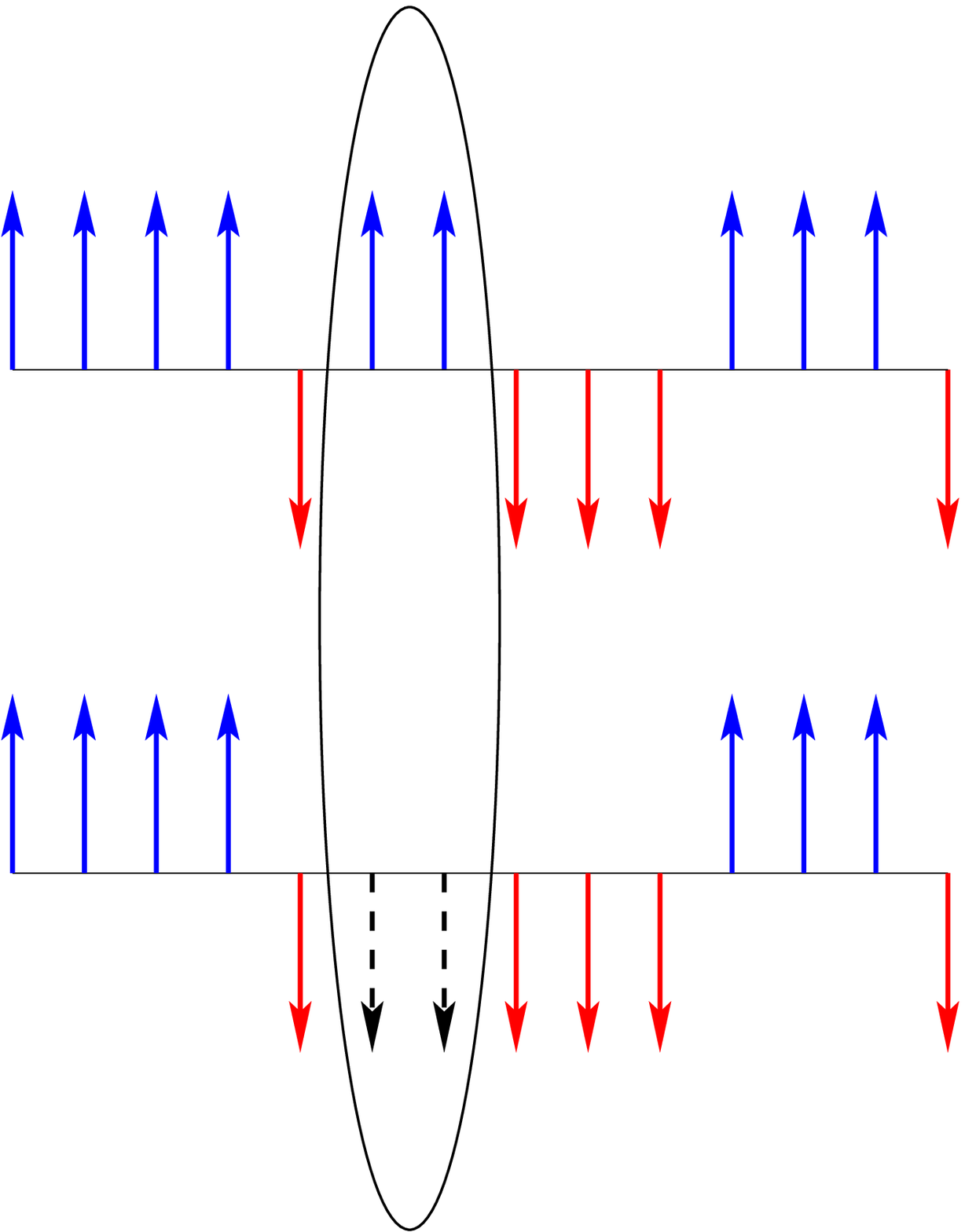}\\
(a)
\end{center}
\end{minipage}
\begin{minipage}{.3\linewidth}
\begin{center}
\includegraphics[width=.8\linewidth]{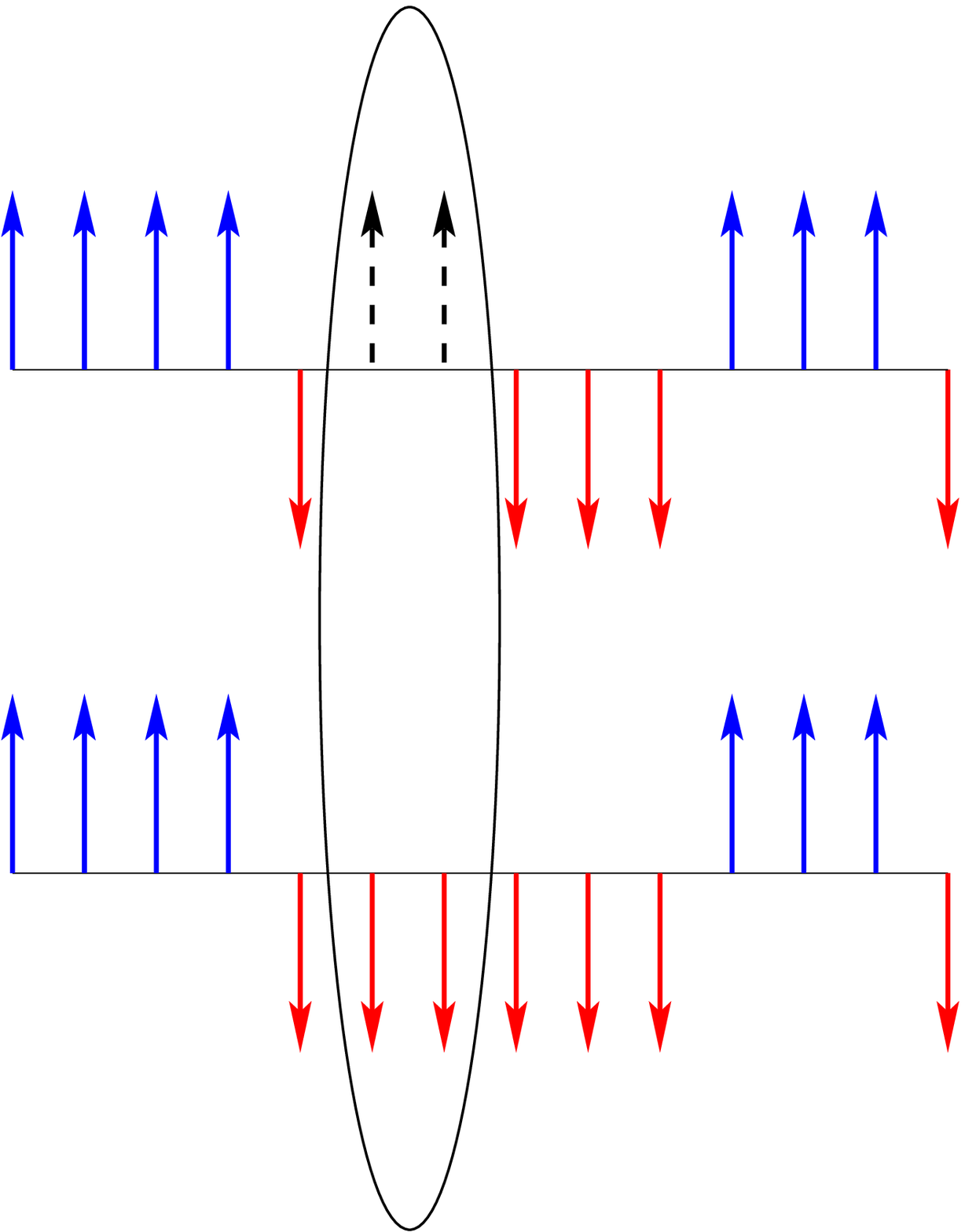}\\
(b)
\end{center}
\end{minipage}
\begin{minipage}{.3\linewidth}
\begin{center}
\includegraphics[width=.8\linewidth]{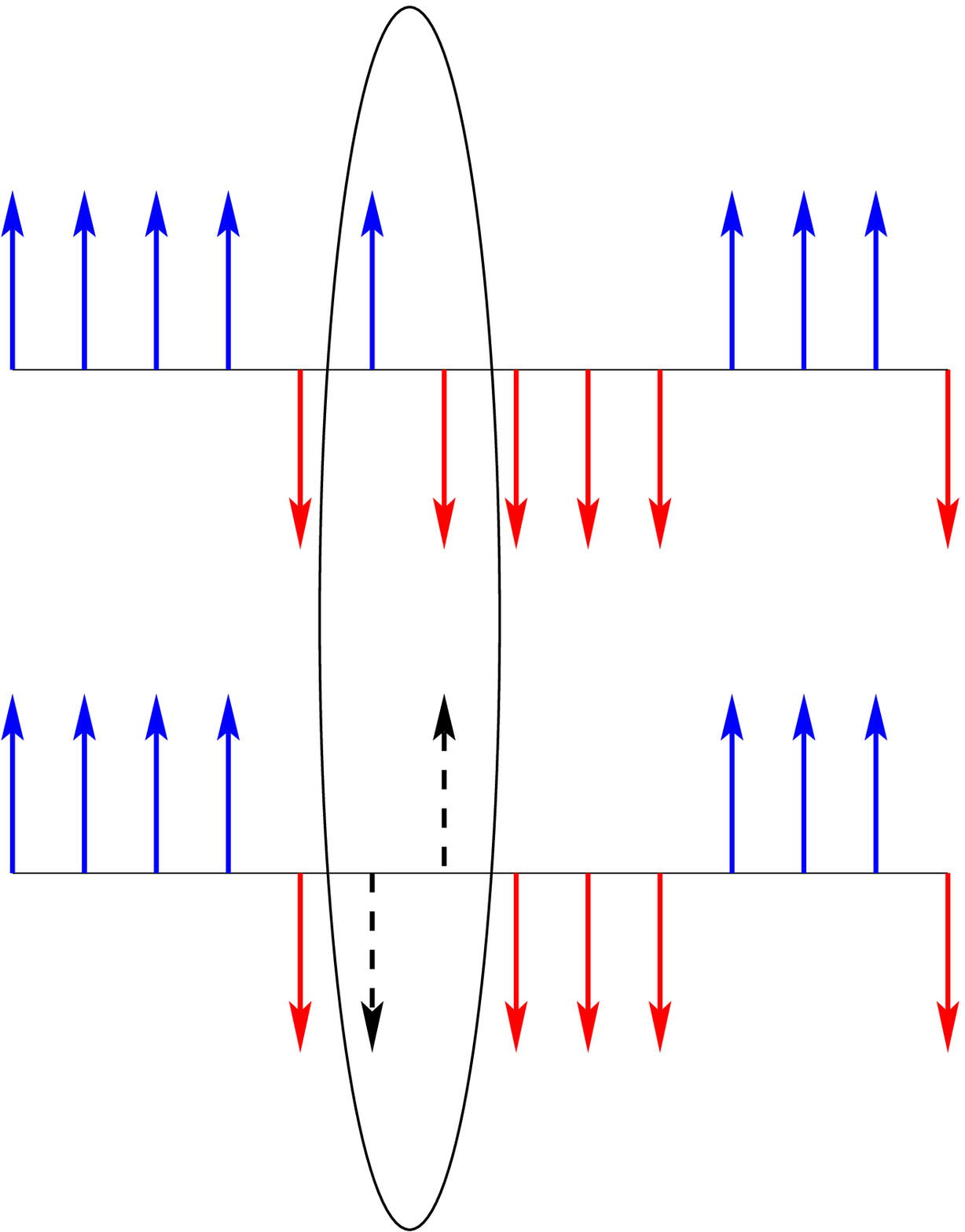}\\
(c)
\end{center}
\end{minipage}
\caption{Two-spins flip transitions which do not change unperturbed energy \eqref{E_0}: (a) $n\to n-2$ and $k\to k-1$,   (b)  $n\to n+2$ and $k\to k+1$, (c)  $n\to n$ and $k\to k$. For clarity the inverted spins are encircled.}
\label{alpha_1_spin_exchange}
\end{figure}

Hamiltonian $\mathcal{H}_1^{(1)}$ given by \eqref{H_1_1} induces transitions with $n\to n$, $n\to n\pm 2$, $k\to k$, and $k\to k\pm 1$. Direct inspections show that there exist 3 possible type of transitions which do not change energy \eqref{E_0}. They correspond to (a)  $n\to n+2$ and $k\to k+1$, (b) $n\to n-2$ and $k\to k-1$, and (c) $n\to n$ and $k\to k$. These possibilities are schematically indicated in figure~\ref{alpha_1_spin_exchange}.   

The simplest way of finding  the corresponding  variances is the use of generating function formalism. We present in details the calculations only for the  transition $n\to n-2$ and $k\to k-1$ indicated in figure~\ref{alpha_1_spin_exchange}a. Other cases are  similar  and we shall give only the final results. 

The transition $n\to n-2$ and $k\to k-1$ appears  any times when $n_j=2$ (see figure~\ref{alpha_1_spin_exchange}a). Therefore, it is necessary to find the total number of groups of only two nearby spins in a sequence of $k$ groups and $n_j$ spins up such that
\begin{equation}
\sum_{j=1}^k n_j=n,\qquad n_j\geq 1\ .
\label{partition}
\end{equation} 
The total number of all partitions satisfying \eqref{partition} is $C_{n-1}^{k-1}$.  

Consider partitions \eqref{partition}  when there exist exactly $p$ groups of nearby two spins, $n_j=2$ up and all other groups have either $n_i\geq 3$ or $n_i=1$. Let $F_p(x)$ be the generating function of such events
\begin{equation}
F_p(x)=\sum_{n_1,n_2,\ldots,n_k}x^{n_1+n_2+\ldots+n_k}\ ,
\end{equation}
where the summation is done by imposing all required restrictions. To find it one has to choose first $p$ indices from $k$ indices and put for them $n_j=2$. All remaining $k-p$ indices have to be either $1$ or any number greater or equal $3$. It gives  
\begin{equation}
F_p(x)=C_k^p\, x^{2p}\, \left (x+\frac{x^3}{1-x}\right )^{k-p}\ .
\end{equation}
The number of states with $p$ groups of two spins up equals  the coefficient, $A_p(n)$, in front of $x^n$ term in series on $x$ of this function, 
\begin{equation}
F_p(x)=\sum_{n=0}^{\infty}A_n(p)x^n
\end{equation}
which can be expressed through the binomial coefficients. 

The total contribution of transitions $n\to n-2$ and $k\to k-1$ is, thus,  the sum over all $p$
\begin{equation}
N_2(n,k)=\sum_{p=0}^k pA_n(p)\ .
\end{equation}
To find this sum it is convenient to first calculate  the sum
\begin{equation}
\sum_{p=0}^k \mathrm{e}^{sp}C_k^p\, x^{2p}\, \left (x+\frac{x^3}{1-x}\right )^{k-p}=\left ( x^2\mathrm{e}^{s}+x+\frac{x^3}{1-x} \right )^k\ ,
\end{equation}
and then differentiate it by $s$, and put in the result $s=0$. It leads to the statement that $N_2(n,k)$ is the coefficient of $x^n$ of the function
\begin{equation}
F_1(x)=k x^2 \left ( \frac{x}{1-x}\right )^{k-1}\ .
\end{equation}
Using the relation
\begin{equation}
C_{-m}^r=(-1)^rC_{m+r-1}^{m-1}
\end{equation}
one obtains 
\begin{equation}
N_2(n,k)=k\, C_{n-3}^{k-2}\ .
\label{N_2}
\end{equation}
The full answer is the product of this quantity and $N/k\, C_{m-1}^{k-1}$ (cf. \ref{ap_1})
\begin{equation}
N_a(n,m;k)=N C_{n-3}^{k-2}\,C_{m-1}^{k-1}\ .
\label{N_a}
\end{equation} 
Transition $n\to n+2$ and $k\to k+1$  correspond to the inversion of two nearby spins when  there exist a group of spins down such that $m_j\geq 3$ as in figure~\ref{alpha_1_spin_exchange}b. This transition is the inverse of the transition $n\to n-2$ and $k\to k-1$ discussed above (cf. figure~\ref{alpha_1_spin_exchange}a  and figure~\ref{alpha_1_spin_exchange}b).  The total number of such events can be calculated as above by using the corresponding generating function. The result is  
\begin{equation}
N_b(n,m;k)=N_a(n+2,m-2;k+1)=N\,  C_{n-1}^{k-1}\, C_{m-3}^{k}\ .
\end{equation}
Transitions $n\to n$ and $k\to k$ are possible in 4 closely related  cases: (i) $m_j=1$, $n_{j+1}\geq 2$, (ii)  $m_j=1$, $n_{j}\geq 2$, (iii) $n_j=1$, $m_{j}\geq 2$, (iv) $n_j=1$, $m_{j-1}\geq 2$. The calculations are straightforward and the total contribution of all 4 above possibilities is
\begin{equation}
N_c(n,m;k)=2N\left [C_{m-2}^{k-1}\, C_{n-2}^{k-2}+ C_{n-2}^{k-1}\, C_{m-2}^{k-2}\right ]\ . 
\label{N_c}
\end{equation}


\section{Correction terms}\label{ap_2}

Expression \eqref{E_shifted} is just the dominant contributions for fixed  $\alpha$ and small $\lambda$. Correction terms include the transitions between states with different unperturbed energies and can be calculated within the usual perturbation series. In the leading order
\begin{equation}
\Delta E_{\sigma}=\sum_{f} \frac{|\mathcal{H}_{\sigma\, f}|^2}{E_{\sigma}-E_{f}}\ ,
\end{equation}
where the summation is done over all possible final states such that $E_{f}\neq E_{\sigma}$. 

The Hamiltonian $\mathcal{H}_1$  given by \eqref{H_1} contains two  terms which give contributions to the mean energy. The first one 
\begin{equation}
\mathcal{H}^{(1)}=-\frac{\lambda^2}{\alpha^2+\lambda^2}  \sum_n \hat{\sigma}^x_n \hat{\sigma}^x_{n+1}
\end{equation}
changes two near-by spins and is of the second order in $\lambda$. The second term 
\begin{equation}
\mathcal{H}^{(2)}=-\sin \phi \cos \phi \sum_n \hat{\sigma}^x_n \, ( \hat{\sigma}^z_{n+1}+ \hat{\sigma}^z_{n-1}) 
\end{equation}
flips only one spin and is of the first order in $\lambda$. Therefore up to the second order in $\lambda$ the contribution to the mean energy is connected only with $\mathcal{H}^{(2)}$. 

\begin{figure}
\begin{minipage}{.3\linewidth}
\begin{center}
\includegraphics[width=.8\linewidth]{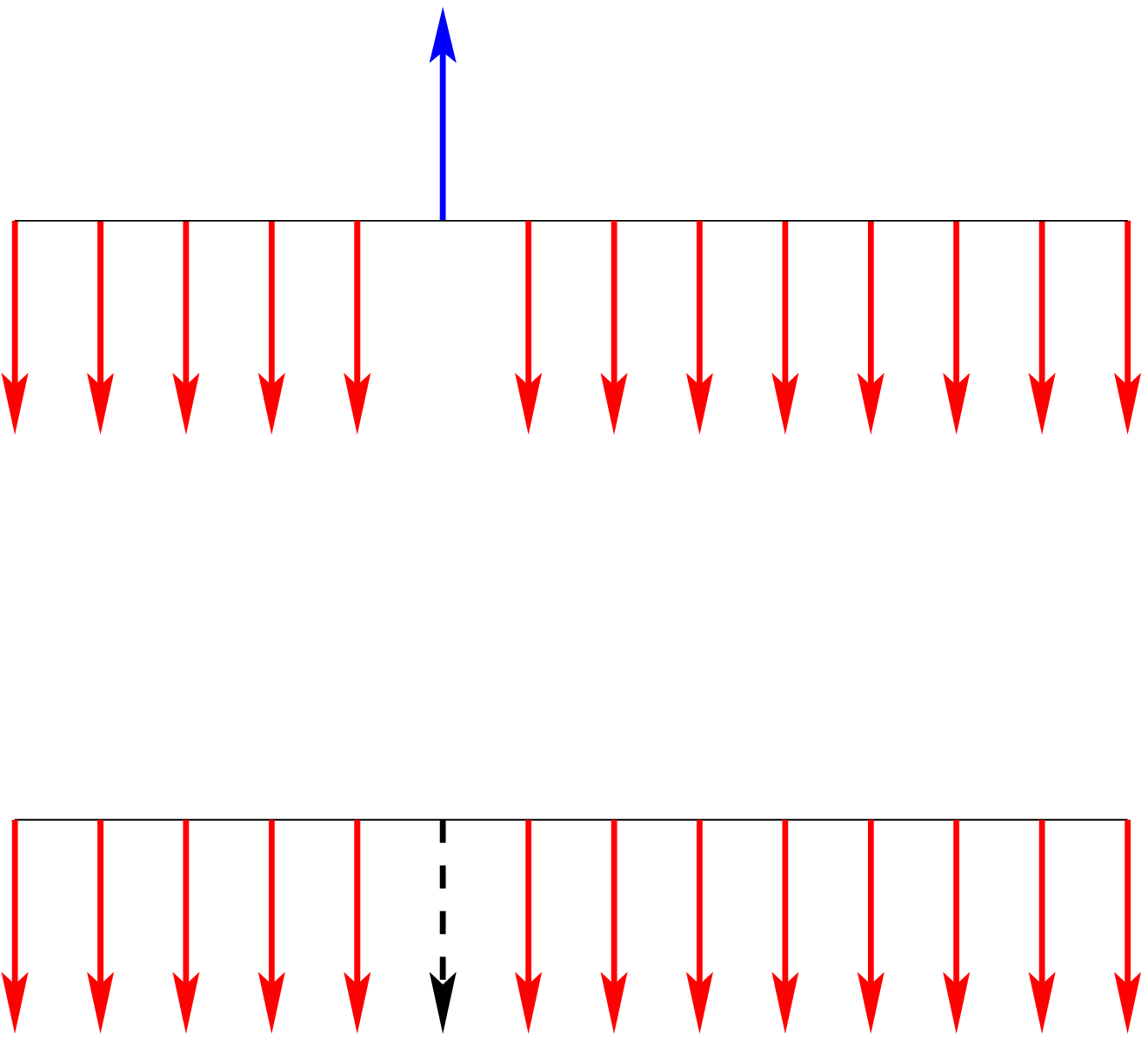}\\
(a)
\end{center}
\end{minipage}
\begin{minipage}{.3\linewidth}
\begin{center}
\includegraphics[width=.8\linewidth]{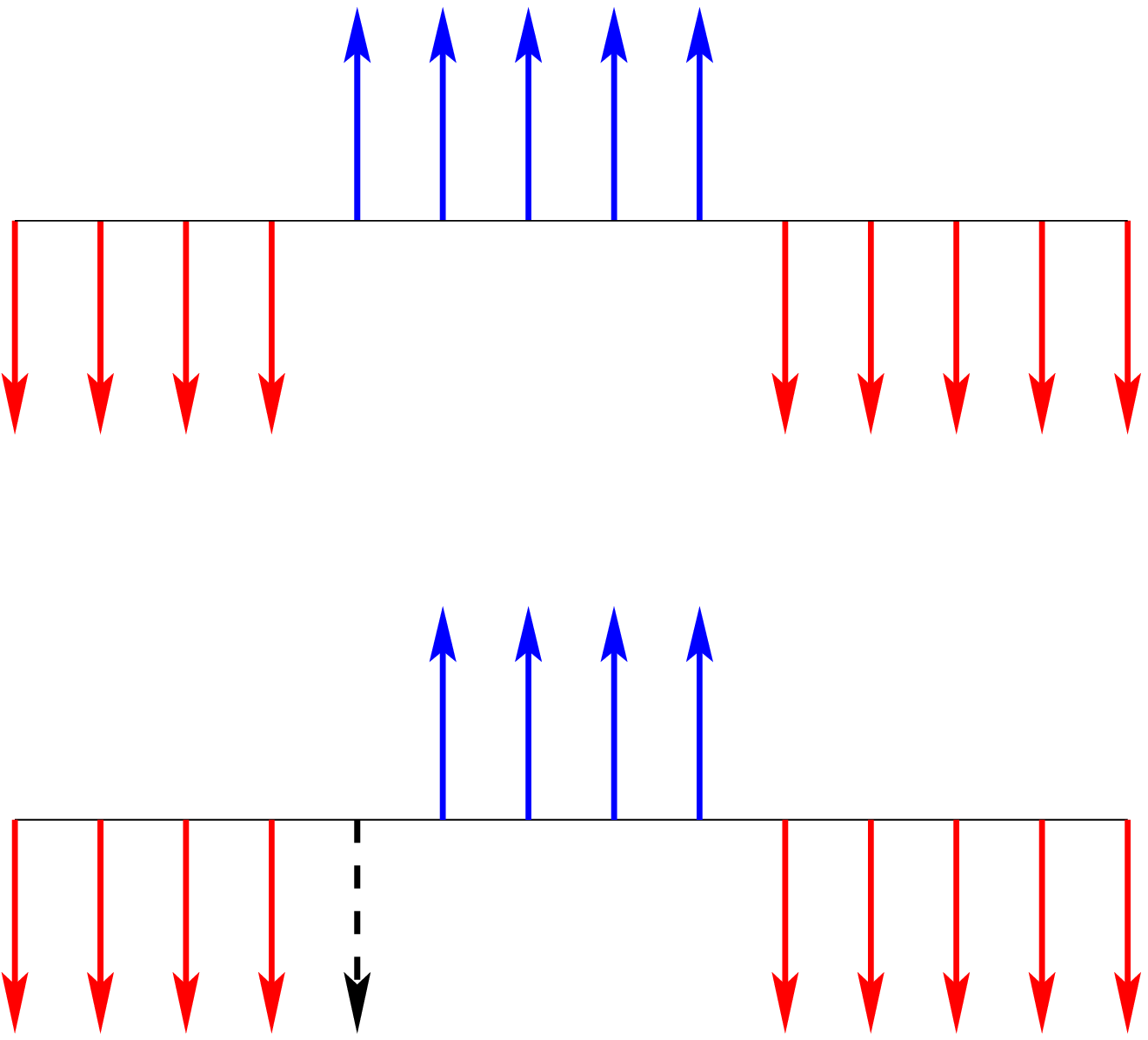}\\
(b)
\end{center}
\end{minipage}
\begin{minipage}{.3\linewidth}
\begin{center}
\includegraphics[width=.8\linewidth]{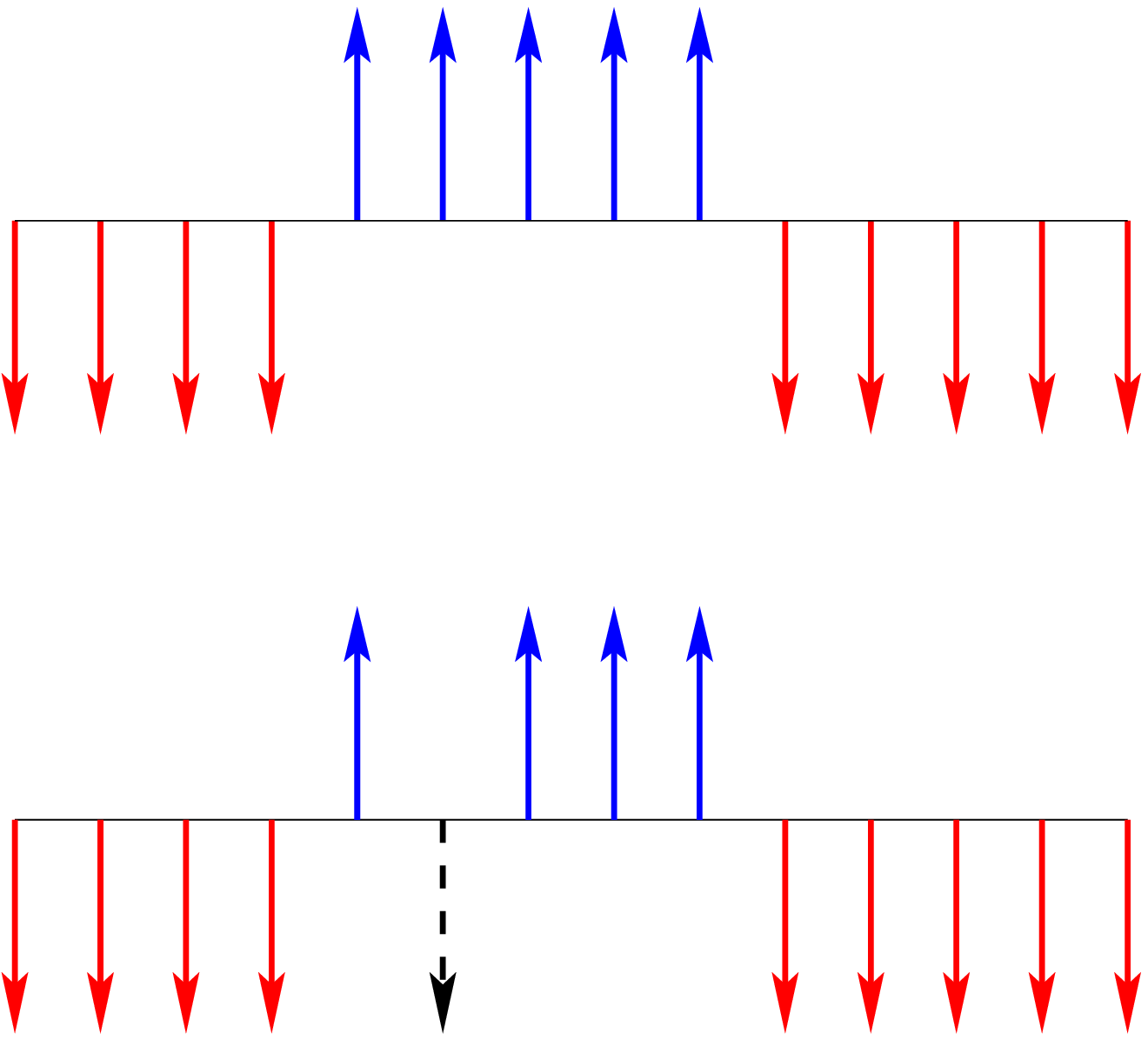}\\
(c)
\end{center}
\end{minipage}
\caption{One-spin flip transitions with $n\to n-1$ and: (a) $k\to k-1$, (b)  $k\to k$, (c)  $k\to k+1$. }
\label{one_spin_change}
\end{figure}

In figure~\ref{one_spin_change} all possible transitions with $n\to n-1$ induced by one spin flips are  indicated graphically. All these transitions change the unperturbed energy \eqref{alpha_E_0} provided that $\alpha\neq \pm 2$.

Transitions (b) vanish as the two nearby spins are opposite. 
The total contribution of transitions (a) and (c) together with the corresponding transitions related with flips of spins down is given by the following expression
\begin{align}
\notag \Delta E_{n,k}=\sin^2 \phi \cos^2\phi & \, \left [ \sum_{n_j=1}\frac{4}{E_{n,k}^{(0)}-E_{n-1,k-1}^{(0)}}  + \sum_{n_j\geq 2}\frac{4(n_j-2)}{E_{n,k}^{(0)}-E_{n-1,k+1}^{(0)}}\right.\\
&+\left . \sum_{m_j=1}\frac{4}{E_{n,k}^{(0)}-E_{n+1,k-1}^{(0)}} +\sum_{m_j\geq 2}\frac{4(m_j-2)}{E_{n,k}^{(0)}-E_{n+1,k+1}^{(0)}} 
\right ]\ .
\end{align}
Here $E_{n,k}^{(0)}=\alpha(N-2n)+4k-N$ is the unperturbed energy (for further convenience calculations here are performed at arbitrary $\alpha\neq \pm 2$).  

Calculating the denominators one gets
\begin{align}
\notag \Delta E_{n,k} &=\frac{2 \alpha^2 \lambda^2}{(\alpha^2+\lambda^2)^2} 
  \left [\sum_{n_j=1} \frac{1}{2-\alpha}  -\sum_{n_j\geq 2}\frac{n_j-2}{2+\alpha} + \sum_{m_j=1}\frac{1}{2+\alpha} -\sum_{m_j\geq 2}\frac{m_j-2}{2-\alpha}
\right ] \\
&=\frac{2 \alpha^2 \lambda^2}{(\alpha^2+\lambda^2)^2}  \left [ \frac{2\alpha}{4-\alpha^2}\Big (\sum_{n_j=1} 1-\sum_{m_j=1}1\Big ) -\frac{1}{2+\alpha} \sum_{n_j\geq 1}(n_j-2)-\frac{1}{2-\alpha} \sum_{m_j\geq 1}(m_j-2)\right ]\ .
\end{align}
The calculation of the number of one spin up, $n_j=1$, which we denoted by $N_1(n,k)$, can be performed exactly as it was done in \ref{ap_4} for the case $n_j=2$ (cf. \eqref{N_2}) and one finds that 
\begin{equation}
N_1(n,k)=k\, C_{n-2}^{k-2}\ .
\end{equation}
Therefore, in the second order in $\lambda$  the change of energy $E_{n,m;k}$ is
\begin{align}
\notag \Delta E_{n,m;k}(\alpha,\lambda)= 
\frac{2 \alpha^2 \lambda^2 N}{(\alpha^2+\lambda^2)^2}&
\left [ \left (\frac{2k-n}{2+\alpha}+\frac{2k-m}{2-\alpha} \right ) \frac{1}{k}\, C_{n-1}^{k-1}\, C_{m-1}^{k-1}\right .\\ 
&+\left .\frac{2\alpha }{4-\alpha^2} \left ( C_{n-2}^{k-2}\, C_{m-1}^{k-1}- C_{n-1}^{k-1}\, C_{m-2}^{k-2}\right ) \right ] \label{Delta} \ , 
\end{align} 
and the total contribution to the mean energy when $\alpha=1$ is
\begin{equation}
\Delta E_{R}=\frac{1}{\mathcal{N}_R}\sum_{2k-n=R}\Delta E_{n,N-n;k}(1,\lambda)\ . 
\end{equation}
where $\mathcal{N}_R$ is the total number of states with fixed $R$ given by \eqref{N_R}.\\



\begin{thebibliography}{100}
\bibitem{wigner_1} E. P. Wigner,  \textit{Characteristic vectors of bordered matrices with infinite dimensions}, Ann. of Math., \textbf{62}, 548 (1955). 
\bibitem{wigner_2}E. P. Wigner, \textit{On the distribution of the roots of certain symmetric matrices}, Ann. of Math., \textbf{67},  325  (1958).
\bibitem{mehta} M. L. Mehta, \textit{Random matrices}, Second Ed., Academic Press, (1991). 
\bibitem{free} D. Voiculescu, \textit{Limit laws for random matrices and free products}, Invent. Math. \textbf{104}, 201 (1991).
\bibitem{french_1} J. French and S. Wong, \textit{Validity of random matrix theories for many-particle systems}, Phys. Lett. B \textbf{33}, 449 (1970).
\bibitem{french_2} J. French and S. Wong, \textit{Some random-matrix  level and spacing distributions for fixed-particle-rank interactions}, Phys. Lett. B \textbf{35}, 5 (1971).
\bibitem{gervois} A. Gervois,  \textit{Level densities for random one- or two-body potentials}, Nucl. Phys. A \textbf{184}, 507 (1972)
\bibitem{brody} T. A. Brody, J. Flores, J. B. French, P. A. Mello, A. Pandey, S. S. M. Wong, \textit{Random-matrix physics: spectrum and strength fluctuations}, Reviews of Modern Physics, \textbf{53}, 385-480 (1981).
\bibitem{bohigas_1} O. Bohigas and J. Flores, \textit{Two-body random hamiltonian and level density}, Phys. Lett. B \textbf{34}, 261 (1071)
\bibitem{bohigas_2} O. Bohigas and J. Flores, \textit{Spacing and individual eigenvalue distributions of two-body random hamiltonians}, Phys. Lett. B \textbf{35}, 383 (1971).
\bibitem{tail} B. V. Bronk, \textit{Accuracy of the semicircle approximation for the density of eigenvalues of random matrices}, J. Math. Phys. \textbf{5}, 215 (1964).
\bibitem{bethe_1}  H. A. Bethe, \textit{An attempt to calculate the number of energy levels of a heavy nucleus}, Phys. Rev. \textbf{50}, 332 (1936).  
\bibitem{bethe_2} H. A. Bethe, \textit{Nuclear physics  B: nuclear dynamics, theoretical}, Rev. Mod. Phys. \textbf{9}, 69 (1937). 
\bibitem{pfeuty} P. Pfeuty, \textit{The one-dimensional Ising model with a transverse field}, 
 Annals of Physics, \textbf{57}, 79 (1970).
\bibitem{sachdev} S. Sachdev, \textit{Quantum phase transitions}  (Camb. Univ. Press, 1999).
\bibitem{lieb} E. Lieb, T. Schultz, and D. Mattis,  Annals of Physics \textbf{16}, 407 (1961).
\bibitem{mattis} D. C. Mattis, \textit{The many body problem}  (World Scientific, Singapore, 1994). 
\bibitem{zamolodchikov}   A. B. Zamolodchikov, \textit{Integrals of motion and $S$-matrix of the (scaled) $T=T_c$ Ising-model with magnetic-field}, Int. J. Mod. Phys. A \textbf{4}, 4235 (1989).
\bibitem{experiment} R. Coldea at al., \textit{Quantum Criticality in an Ising Chain: Experimental Evidence for Emergent $E_8$ Symmetry Science},  \textbf{327}, 177 (2010).
\bibitem{atas} Y. Y. Atas, E. Bogomolny, O. Giraud, G. Roux, \textit{The distribution of the ratio of consecutive level spacings in random matrix ensembles},  Phys. Rev. Lett. \textbf{110}, 084101 (2013). 
\bibitem{griffiths}  R. B. Griffiths, \textit{A proof that the free energy of a spin system is extensive},  J. Math. Phys. \textbf{5}, 1215 (1964). 
\bibitem{keating} J. P. Keating, N. Linden, and H. J. Wells, \textit{Spectra and eigenstates of spin chain models}, in preparation (2013)
\bibitem{roux} C. Kollath, G. Roux, G. Biroli and A. M. L{\"a}uchli, \textit{Statistical properties of the spectrum of the extended Bose-Hubbard model}, J. Stat. Mech. Theor. Exp., \textbf{2010}, P08011 (2010).
\bibitem{deutsch_1} J. M. Deutsch, \textit{Quantum statistical mechanics in a closed system}, Phys. Rev. A \textbf{43}, 2046 (1991).
\bibitem{deutsch_2} J. M. Deutsch, \textit{A closed quantum system giving ergodicity}, unpublished, (1991), \\ http://physics.ucsc.edu/{\raise.17ex\hbox{$\scriptstyle\sim$}}josh/publications.html . 
\end{thebibliography}
\end{document}